\documentclass[journal=jacsat,manuscript=article]{achemso}

\usepackage{chemformula} % Formula subscripts using \ch{}
\usepackage[T1]{fontenc} % Use modern font encodings

\usepackage{dsfont}
\usepackage{amsfonts}
\usepackage{upgreek}
\usepackage{subcaption}
\usepackage{algorithm}
\usepackage{algpseudocode}
\usepackage{algorithmicx}

\usepackage{tikz}
\SectionNumbersOn            % achemso switch to number sections

\author{Nils E. Strand}
\affiliation[JFI]
{James Franck Institute, University of Chicago, Chicago, Illinois, 60637, United States}
\alsoaffiliation[Stats]
{Department of Statistics, University of Chicago, Chicago, Illinois, 60637, United States}
\author{Siyao Yang}
\affiliation[Stats]
{Department of Statistics, University of Chicago, Chicago, Illinois, 60637, United States}
\alsoaffiliation[CCAM]
{Committee on Computational and Applied Mathematics, University of Chicago, Chicago, Illinois, 60637, United States}
\author{Yuehaw Khoo}
\affiliation[Stats]
{Department of Statistics, University of Chicago, Chicago, Illinois, 60637, United States}
\alsoaffiliation[CCAM]
{Committee on Computational and Applied Mathematics, University of Chicago, Chicago, Illinois, 60637, United States}
\author{Aaron R. Dinner}
\email{dinner@uchicago.edu}
\affiliation[Chem]
{Department of Chemistry, University of Chicago, Chicago, Illinois, 60637, United States}
\alsoaffiliation[JFI]
{James Franck Institute, University of Chicago, Chicago, Illinois, 60637, United States}
\alsoaffiliation[CCAM]
{Committee on Computational and Applied Mathematics, University of Chicago, Chicago, Illinois, 60637, United States}

\title[ttmetad]
  {Adaptive tensor train metadynamics for high-dimensional free energy exploration}

\usepackage{xr}
\makeatletter
\newcommand*{\addFileDependency}[1]{
  \typeout{(#1)}
  \@addtofilelist{#1}
  \IfFileExists{#1}{}{\typeout{No file #1.}}
}
\makeatother

\newcommand*{\myexternaldocument}[1]{
    \externaldocument{#1}
    \addFileDependency{#1.tex}
    \addFileDependency{#1.aux}
}
\myexternaldocument{SI}

\begin{document}

\begin{abstract}

A key challenge for molecular dynamics simulations is efficient exploration of free energy landscapes over relevant collective variables (CV).
Common methods for enhancing sampling become prohibitively inefficient beyond only a few CVs; in the case of the widely-used metadynamics method, the computational cost of evaluating and storing the bias potential grows exponentially with the number of dimensions.
Here, we introduce TT-Metadynamics, in which the accumulated sum of Gaussian functions in the original metadynamics method is periodically compressed into a low-rank tensor train (TT) representation.
The TT enables efficient memory use and prevents the computational cost of evaluating the bias potential from increasing with simulation time.
We present a ``sketching'' algorithm that allows us to construct the TT with linear scaling in the number of CVs.
Applied to benchmark systems with up to 14 CVs, the accuracy of TT-Metadynamics matches or exceeds that of standard metadynamics in long simulations, particularly in systems with high barriers.
These results establish TT-Metadynamics as a scalable and effective method for computing free energies that are functions of several CVs.

\end{abstract}

\section{Introduction}

% MD suffers from separation of time scales
Over the past few decades, advances in molecular dynamics (MD) algorithms and computer hardware have enabled the simulation of complex molecular phenomena across physics, chemistry, and biology.
These include long-timescale processes such as protein folding~\cite{shaw2010atomic,lindorff2011fast,piana2013atomic}, voltage sensing in ion channels~\cite{jensen2012mechanism,guo2024dynamics}, and drug–target binding~\cite{shan2011drug}.
Although the underlying equations of motion are simple, a central challenge in studying molecular systems is that they have many metastable states, and traversing the (free) energy barriers between them can take exorbitant amounts of time in unbiased simulations.
For the above examples, parallelization has allowed a handful of transition events to be observed in unbiased simulations \cite{shaw2010atomic,jensen2012mechanism,shan2011drug}, but these simulations do not provide sufficient statistics to accurately estimate thermodynamics and kinetics.

% Most enhanced sampling approaches bias along CVs
Enhanced sampling methods bridge separations of timescales by increasing the probabilities of rare but important states to facilitate the recovery of unbiased statistics~\cite{henin2022enhanced}.
A popular example of such an approach is metadynamics \cite{laio2002escaping}.
The idea behind metadynamics is to build a bias potential adaptively as a sum of Gaussian functions centered on positions explored by the simulation, which encourages escape from metastable states and, in turn, repeated transitions between them.
The bias potential is generally expressed as a function of one or more order parameters or collective variables (CV) that are computed from the Cartesian coordinates used for numerical integration.

% The choice of CVs is important and limited to 2-3 CVs
The particular choice of CVs has a strong influence on the amount of simulation time required for metadynamics (and enhanced sampling more generally) to converge.
For systems such as small peptides (e.g., alanine dipeptide, polyalanines), backbone dihedral angles are common choices for CVs.
Some fast-folding proteins are sufficiently well-characterized so that specific CVs that provide good control of sampling are established \cite{lindorff2011fast,qiu2002smaller,juraszek2006sampling,juraszek2008rate,piana2007bias,marinelli2009kinetic,sidky2019high,strahan2021long}.
However, as the complexity of the system increases, it becomes harder to identify useful CVs \cite{bussi2020using}.
In its original formulation \cite{laio2002escaping}, metadynamics performs well only when the number of CVs is limited to a small number (typically two or three), since the volume of the CV space to be explored grows exponentially with dimension.
This limitation presents a chicken-and-egg problem:  CVs that provide good control of sampling are usually not known before simulations are run.

% Dimensional reduction methods can be used to find a few CVs
% VAC/VAMP-like approaches yield slow modes and first few can be used as CVs
When some simulation data are available, one strategy is to use generated data to identify a small number of CVs by unsupervised dimensional reduction.
Methods include principal component analysis (PCA)~\cite{pearson1901lines}, diffusion maps~\cite{coifman2005geometric}, and autoencoders \cite{bank2023,chen2018collective,chen2018molecular,chen2023chasing}.
Perhaps the most successful and thus widely-used unsupervised dimensional reduction methods use two time points to identify slow modes \cite{amadei1993essential,molgedey1994separation,naritomi2013slow,noe2013variational,nuske2014variational,perez2013identification,lorpaiboon2020integrated}.
However, these methods can require significant amounts of data.  More fundamentally, there is no reason that unsupervised methods should identify the variables that are most important for sampling transitions of interest (i.e., that the properties they optimize align with the scientific problem).

% An alternative to finding a good low-dimesional representation is to make metadynamics work with more CVs
An alternative strategy is to modify the sampling algorithm to be able to simultaneously explore a large number of CVs with the hope that they encompass---either in tandem or as a combination---ones that accelerate sampling.
Metadynamics methods that are compatible with biasing along more than three variables at a time have thus been developed.
An early such approach, bias-exchange metadynamics, runs multiple parallel simulations, each biasing along a single variable, with configuration exchanges between replicas proposed periodically and accepted or rejected based on the Metropolis criterion \cite{piana2007bias,marinelli2009kinetic}.
A more recent approach, termed parallel-bias metadynamics, similarly biases the simulation using multiple simultaneous 1D potentials; the main difference is that it does not require the use of multiple replicas~\cite{pfaendtner2015efficient}.
Both of these methods scale more favorably with the number of CVs than standard metadynamics but remain limited by the fact that inter-variable correlations are not captured in the bias potential itself: in bias-exchange metadynamics, correlations enter only indirectly through the exchange of configurations between replicas biasing different CVs, while in parallel-bias metadynamics, the simulation switches between 1D potentials within a single replica, and correlations are similarly absent from any individual bias term.
% enter indirectly through the exchange of a replica between CVs rather than the bias potential.
%For example, in the original application of bias-exchange metadynamics to the Trp-cage miniprotein, only five CVs were used~\cite{piana2007bias}.
More fundamentally, neither method yields a free energy as a function of all biased variables jointly, but rather a collection of low-dimensional free energy projections.

% Integrated machine learning is enabling capture of the correlations between CVs
Alternatively, machine learning methods that incorporate many CVs have been integrated with metadynamics.
For example, Galvelis and Sugita~\cite{galvelis2017neural} proposed a method in which the high-dimensional bias potential is represented using artificial neural networks that are trained on the fly with the aid of nearest-neighbor density estimation; they were able to achieve converged results with up to eight CVs.
Zhang et al.~\cite{zhang2018reinforced} also represented the bias potential with neural networks and developed a training scheme using an analogy to reinforcement learning with a reward function based on statistical uncertainty; they presented examples with up to 20 CVs.
These methods show the promise of machine learning for metadynamics but also make clear that neural networks can be poorly behaved in regions with sparse sampling. 
As such, enhanced sampling based on bias potentials obtained by machine learning continues to be an active area of research.

Tensor decompositions offer ways of representing high-dimensional functions complementary to neural networks.
Although neural networks are trained iteratively through gradient descent to approximate nonlinear functions, tensor decompositions are grounded in numerical linear algebra and construct low-rank representations through well-established operations such as singular value decompositions and least-squares solves~\cite{oseledets2011tensor}.
Biasing strategies centered on tensor decompositions were previously applied to the adaptive biasing force (ABF) method~\cite{ehrlacher2022adaptive}.
In contrast to metadynamics, which constructs a history-dependent bias potential, ABF  estimates and applies the mean force along the reaction coordinates directly, with the goal of flattening the free energy landscape through force-based rather than potential-based biasing.
However, in its standard form, ABF suffers from the same limitations in the number of CVs \cite{alrachid2015long}.
Ehrlacher et al. \cite{ehrlacher2022adaptive} introduced a variant of ABF in which the free energy is approximated as a memory-efficient sum of separable tensor products of one-dimensional functions.
The authors established rigorous long-time convergence and consistency results for the resulting algorithm and demonstrated numerically that tensor approximations can capture correlations between reaction coordinates in systems with up to five dimensions.
Their study focuses primarily on theoretical guarantees and proof-of-concept numerical experiments, leaving open questions regarding algorithmic optimization, integration into large-scale molecular dynamics engines, and performance in substantially higher-dimensional settings.

Motivated by the studies above~\cite{galvelis2017neural,zhang2018reinforced,ehrlacher2022adaptive}, here, we develop a form of metadynamics that uses a compact tensor decomposition known as a tensor train (TT)~\cite{oseledets2011tensor} to represent the bias potential.
TTs have shown great success in quantum many-body physics, where they are widely known as matrix product states (MPS)~\cite{schollwock2011density,white1993density,yu2024re}, in data science~\cite{liu2012tensor,oseledets2010tt,novikov2021tensor}, and in applied mathematics~\cite{chen2023combining,dolgov2012fast}.
In the context of molecular dynamics, TTs have previously been used to approximate the dynamical evolution (Koopman) operator \cite{nuske2014variational,cao2025amuset}.
 As we elaborate, TT-based algorithms generalize matrix factorization methods such as QR and singular value decomposition (SVD) to higher-dimensional tensors.

% We start from a very simple idea, replace the grid sometimes employed
The starting point for our method is a version of metadynamics that stores the bias potential on a grid~\cite{plumed}, so that the cost of evaluating the bias potential is  constant.
However, the memory requirement for grid storage scales exponentially with dimension~\cite{plumed}.
Alternatively, one could compute the bias potential as a sum over the full list of Gaussian functions.
Because this list grows without bound, the computational cost to evaluate the bias potential increases linearly with simulation time, rendering this strategy unappealing.
The essential idea of our approach is to replace grid storage with TT storage.
With the TT, the cost of evaluating the bias potential is still constant, and the memory grows only linearly with dimension.
We achieve this improvement without significant overhead through an efficient compression algorithm that scales linearly with both the dimensionality and the number of Gaussian functions.
In addition to compressing the bias potential efficiently, the TT factorization introduces a form of regularization that accelerates convergence by preventing overfitting to localized, undersampled regions of the CV space.
We demonstrate our method by computing PMFs that are functions of up to 14 CVs for peptides.

\section{Methods}

\subsection{Background}
\label{sec:ttmetad}

\subsubsection{Metadynamics}

We begin by summarizing the standard metadynamics algorithm, as originally developed by Laio and Parrinello \cite{laio2002escaping}.
We are interested in classical systems that move according to a potential function $V\equiv V(\mathbf r)$, where $\mathbf r$ is the position of the molecule in Cartesian coordinates.
Metadynamics adaptively constructs a bias potential, which is expressed as a function of a chosen set of collective variables $(V_{\text{bias}}\equiv V_{\text{bias}}(\mathbf x))$, which are themselves functions of the positions of particles $(\mathbf x\equiv\mathbf x(\mathbf r))$.
Practically, the metadynamics kernels take the form of Gaussian functions with diagonal covariance matrices.
A standard metadynamics bias potential takes the form
\begin{align}
    \nonumber V_{\text{bias}}^{\text{MetaD}}(\textbf x)&=\sum_{i}\mathcal G^i(\textbf x)\\
    &=\sum_{i}h_i\exp\left(-\frac{|\textbf x-\textbf x_i|^2}{2\boldsymbol\sigma^2}\right),
    \label{eq:metad}
\end{align}
where $\mathbf x_i$ is the center position of Gaussian $i$, and $h_i$ is its height.
We assign the same scalar bandwidth $\boldsymbol\sigma$ to each dimension.
% constant vector of bandwidths assigned to each dimension.

We consider throughout this work the \emph{well-tempered metadynamics}~\cite{barducci2008well} format, whereby the Gaussian heights $h_i$ are tempered as a function of the accumulated bias:
\begin{equation}
    h_i=h_0\exp\left(\frac{-\beta V_{\text{bias}}^{(i-1)}(\textbf x_i)}{\gamma-1}\right),
    \label{eq:height}
\end{equation}
where $h_0 > 0$ is a user-specified initial height and $\gamma > 1$ is the bias factor.
The tempering rule ensures that Gaussian heights decrease over time: regions of CV space that have already accumulated a large bias receive progressively smaller increments, while under-explored regions continue to be filled at a higher rate.
This self-correcting mechanism drives the bias toward a fixed point at which the rate of accumulation is uniform across CV space.
Denoting the free energy by $\mathcal{F}(\mathbf{x})$, at that fixed point, the biased sampling distribution $p_{\text{bias}}(\mathbf{x}) \propto \exp(-\beta[\mathcal{F}(\mathbf{x}) + V_{\text{bias}}(\mathbf{x})])$ requires
\begin{equation}
    V_{\text{bias}}(\mathbf{x}) \xrightarrow{t\to\infty} -\frac{\mathcal{F}(\mathbf{x})}{\gamma} + C',
    \label{eq:convergence}
\end{equation}
where $C'$ is an irrelevant additive constant~\cite{dama2014well}.
The bias thus converges to a scaled negative of the free energy.
The parameters $\gamma$ and $h_0$ are user-determined and affect the convergence rate. In practice, $\gamma$ is chosen based on the expected height of barriers, and $h_0$ is tuned empirically.

It is customary to store the bias potential in a multidimensional grid and update this grid every $\omega$ timesteps~\cite{plumed}.
The memory requirement of this strategy scales exponentially with the number of CVs, or the number of dimensions $D$.
For computations involving a single compute node with 375~GB RAM on the Midway3 cluster within the University of Chicago Research Computing Center, grid storage is infeasible for $D\geq6$.

\subsubsection{Tensor trains}
\label{sec:tt}

The central idea of the paper is that we can efficiently represent a bias potential that is a function of $D$ CVs using a tensor train (TT) representation.
In this section, we briefly review TTs.

A $D$-dimensional tensor $\mathcal{P}$ can be represented in TT format if it admits the factorization \cite{oseledets2011tensor}
\begin{equation}    \mathcal P(i_1,\dots,i_D)=\sum_{\alpha_1=1}^{r_1}\cdots\sum_{\alpha_{D-1}=1}^{r_{D-1}}G_1(i_1,\alpha_1)G_2(\alpha_1,i_2,\alpha_2)\cdots G_D(\alpha_{D-1},i_D),
    \label{eq:tt}
\end{equation}
where the cores $G_1,\dots,G_D$ are third-order tensors (except for the first and last cores, which are of second-order).
Each index $i_k \in \{1,\dots,n_k\}$ is a \emph{physical} index running over the $n_k$ degrees of freedom along dimension $k$; in our setting, these correspond to the coefficients of the basis functions defined along the $k$th collective variable.
The indices $\alpha_k \in \{1,\dots,r_k\}$ are \emph{auxiliary} indices that are summed over and exist solely to mediate correlations between adjacent cores.
The tuple $(r_1,\dots,r_{D-1})$ denotes the collection of TT ranks; $r_k$ is the dimension of the auxiliary index between the $k$th and $(k+1)$th TT cores, and the choice of $r_k$ affects the TT's ability to capture interdimensional correlations.
The ranks are typically small in practice.

\begin{figure}[bt]
    \centering
    \subfloat[Tensor train]{{\includegraphics[width=0.4\textwidth]{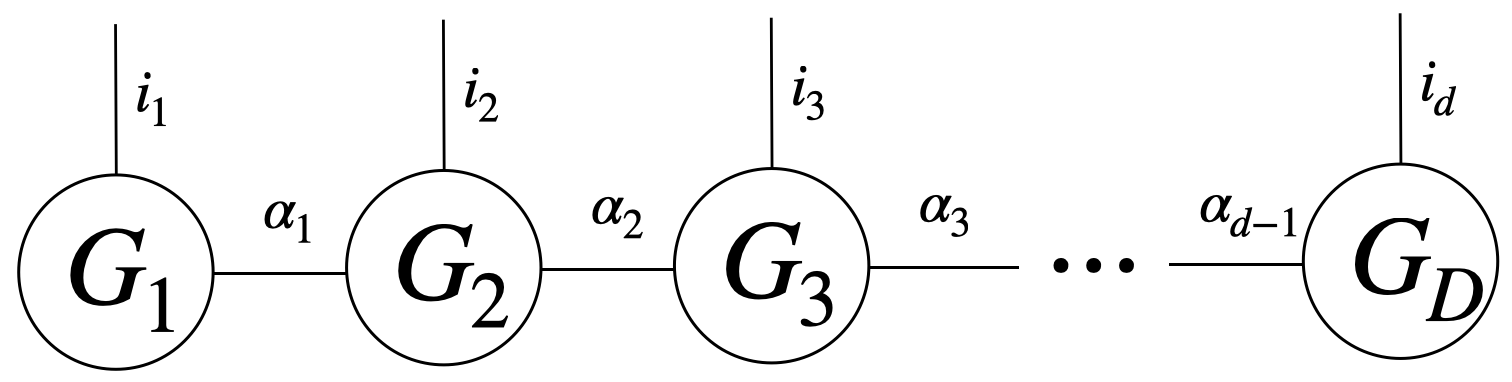}}}
    \qquad  \qquad  
    \subfloat[Functional tensor train]{{\includegraphics[width=0.4\textwidth]{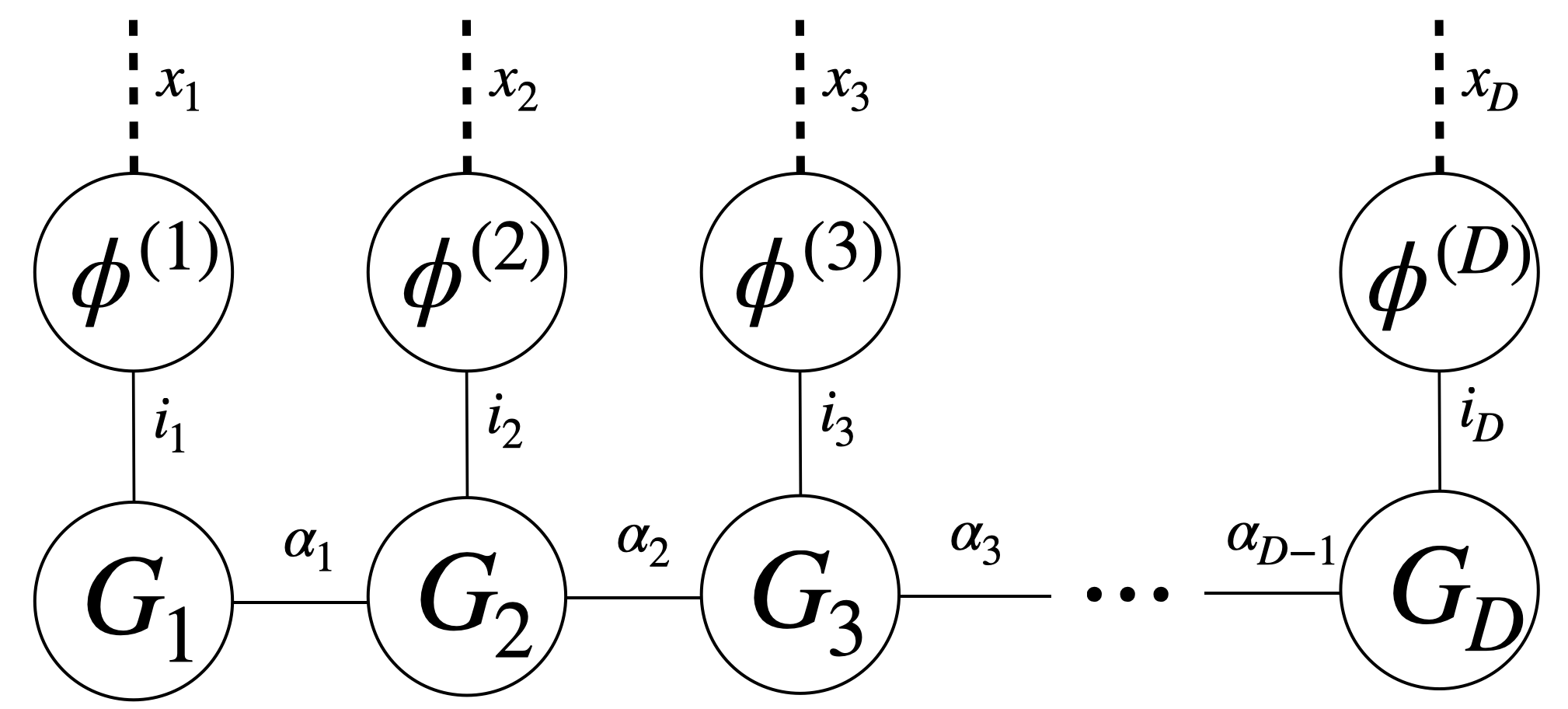}}}\\[20pt]
    \caption{Tensor diagrams illustrating tensor trains (TTs).
    (a) TT representation of the $D$-dimensional tensor defined in Eq.~\eqref{eq:tt}.
    Each vertical solid leg represents a (discrete) coefficient index $i_k$, and each horizontal solid leg represents an (discrete) auxiliary index $\alpha_k$.
    A horizontal connection between two cores indicates contraction over the associated auxiliary index.
    (b) Functional TT representation for the bias potential, as in Eq.~\eqref{functional_TT}, showing the contraction of the coefficient tensor with a product of univariate basis functions $\phi_{i_k}^{(k)}(x_k)$ for each collective variable $x_k$.
    Each vertical dashed leg represents a (continuous) coordinate $x_k$, and correlations between different dimensions are mediated by the horizontal solid auxiliary indices.  
    }
    \label{fig:tt-diagram}   
\end{figure}

The low-rank TT format allows for linear computational cost of tensor evaluations and various fundamental algebraic operations such as summations, Hadamard products, inner products, and integration.
In this format, the full tensor value $\mathcal P(i_1,\dots,i_D)$ is computed via \emph{contractions} between the TT cores.
Specifically, a contraction refers to a summation over a shared index between two tensors, i.e., the auxiliary index $\alpha_k$ connecting the cores $G_k$ and $G_{k+1}$.
By contracting all intermediate indices, one sequentially reduces the network of TT cores to a single scalar output.
Eq.~\eqref{eq:tt} can be represented graphically as Fig.~\ref{fig:tt-diagram}(a).
Each core \(G_k(\alpha_{k-1}, i_k, \alpha_k)\) is represented as a node labeled \(G_k\); its indices are represented by the attached edges (“legs”).
Whenever two nodes are connected by a common leg, the corresponding index is contracted.

There are several existing methods for compressing a high-dimensional tensor into a TT.
A classical approach is the TT-SVD algorithm~\cite{oseledets2011tensor}, which constructs the TT representation by repeatedly reshaping the full tensor into matrices, that is, grouping subsets of indices into rows and columns, and applying singular value decompositions (SVDs) to these matrices.
However, since tensor size grows exponentially with dimension, TT-SVD suffers from the curse of dimensionality in both storage and computation.
Another widely used method is TT-Cross\cite{oseledets2010tt}, which constructs a TT representation based on interpolation pivots.
Although effective in many settings, TT-Cross often requires multiple sweeps across dimensions to identify near-optimal pivots, making the overall computational complexity difficult to control.
In this work, we employ TT-Sketch, a method developed and refined by Khoo and co-workers~\cite{hur2023generative,chen2023combining,peng2024tensor}.
As detailed in Section~\ref{sec:ttsketch}, TT-Sketch leverages randomized ``sketching'' techniques \cite{halko2011finding,huber2017randomized} to replace full SVDs with a sequence of tractable linear algebra problems.
Unlike TT-Cross, TT-Sketch is one-shot and purely linear algebraic.
These properties lead to computational complexity that scales linearly with the number of dimensions, making TT-Sketch well-suited for high-dimensional enhanced sampling frameworks such as the one presented here.

\subsection{TT-Metadynamics}

\begin{figure}[hbt]
\centering
{\includegraphics[width=0.8\textwidth]{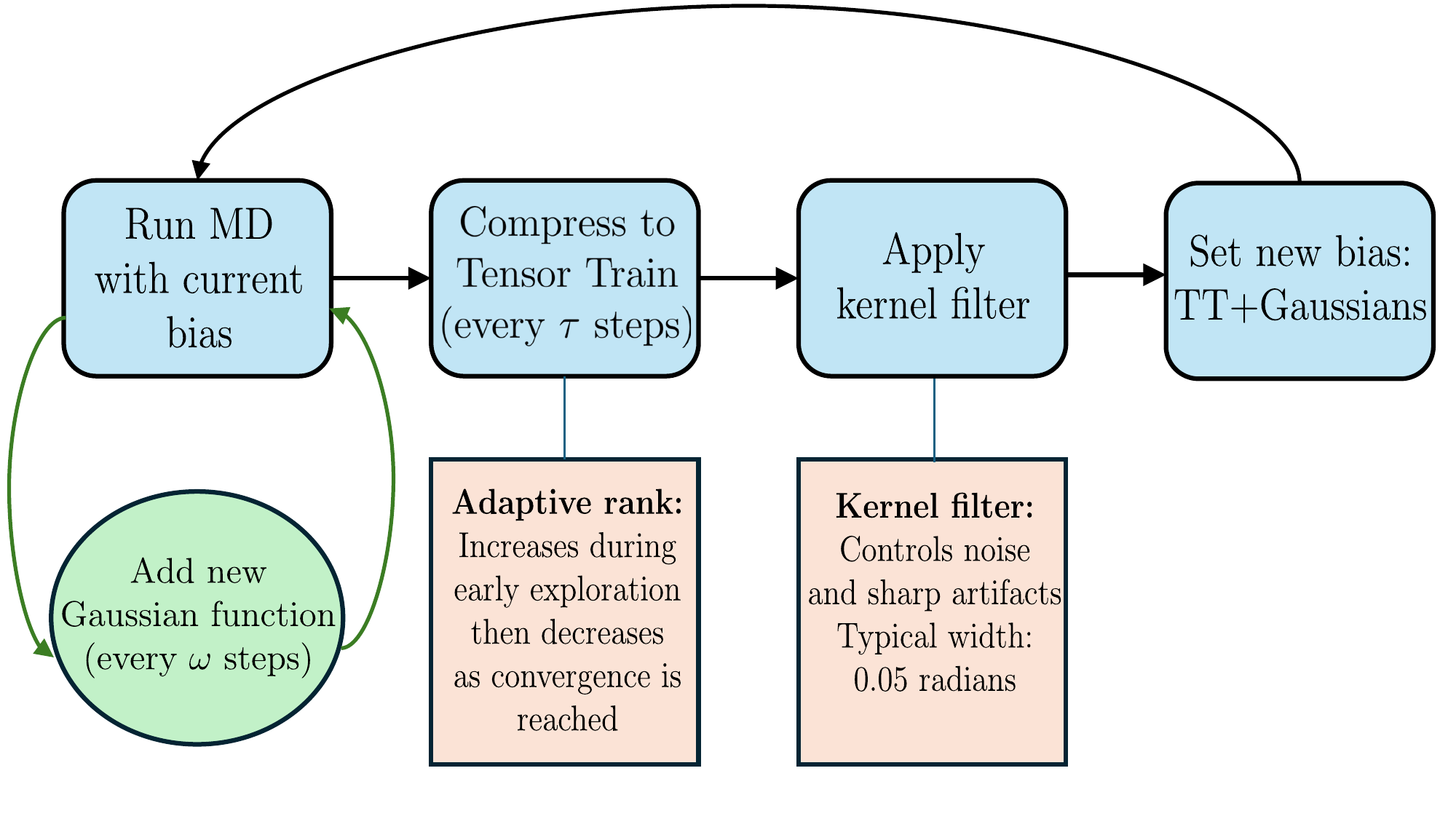}}
\caption{Workflow of the TT-Metadynamics algorithm, illustrating the steps from sampling to bias potential construction.
}
\label{fig:flowchart}
\end{figure}

The core algorithm is outlined in Algorithm~\ref{alg:full_algorithm}; its general workflow is depicted in Fig.~\ref{fig:flowchart}.
The main idea of our proposed framework is to compress the sum of Gaussian kernels, each of which is deposited every $\omega$ timesteps, into a TT.
This compression is carried out every $\tau$ timesteps.
The bias potential $V_{\text{bias}}(\mathbf x)$ is defined in a continuous domain and is represented by a product basis expansion constructed from selected univariate orthogonal basis functions $\phi_i^{(k)}(x_k)$ along each dimension $k$.
The corresponding coefficient $\mathcal P_{\text{bias}}$ is approximated by a low-rank TT as Eq.~\eqref{eq:tt}, resulting in the following functional TT representation for the bias potential:
\begin{equation} \label{functional_TT}
    V_{\text{bias}}(x_1,\dots,x_D)=\sum_{i_1=1}^{n_1}\cdots\sum_{i_D=1}^{n_D}\mathcal P_{\text{bias}}(i_1,\dots,i_D)\phi_{i_1}^{(1)}(x_1)\cdots\phi_{i_D}^{(D)}(x_D),
\end{equation}
where $i_k$ is a discrete index iterating over the $n_k$ basis functions defined along the dimension $k$.
One could, in principle, use a grid (i.e., indicator or interpolating basis functions centered at discrete points) for this expansion, as is standard in low-dimensional metadynamics~\cite{plumed}.
However, a grid-based TT would require $n_k$ to be large enough to resolve the finest features of the bias, and the cost of TT-Sketch scales with $n_k$; moreover, gradient evaluation on a grid requires finite difference derivatives, which introduce additional numerical error.
We instead use a Fourier basis; the Gaussian metadynamics kernels in this case conveniently admit closed-form inner products with the basis functions (Eq.~\eqref{coeff_tns_closed_form2}).
The smooth, global nature of Fourier modes also means that a modest number of basis functions $n_k$ suffices to represent the bias accurately, and gradients can be computed analytically.
For simplicity, we adopt the same family of Fourier basis functions for all dimensions:
\begin{align}
    \phi_i(x)=\begin{cases}(2L)^{-1/2},&\quad~~~i=1,\\
    (L)^{-1/2}\cos\left[\pi(x-a)\lfloor i/2\rfloor/L\right],&\quad\mod(i,~2)=0,\\
    (L)^{-1/2}\sin\left[\pi(x-a)\lfloor i/2\rfloor/L\right],&\quad~~~\text{otherwise},\end{cases}
    \label{eq:fourier}
\end{align}
where $L$ is half the length of the domain of $x_k$ and $a$ is the position of the midpoint of the domain.
The tensor diagram for Eq.~\eqref{functional_TT} is shown in Fig.~\ref{fig:tt-diagram}(b), where each core of the TT representation of the coefficient tensor $\mathcal P_{\text{bias}}$ connects to the chosen set of Fourier basis functions.
The correlation structure of the high-dimensional bias potential is captured by the coefficient tensor $\mathcal P_{\text{bias}}$.
After obtaining the cores in $\mathcal P_{\text{bias}}$, the bias potential can be evaluated using Eq.~\eqref{functional_TT}.
The computational cost of this evaluation is independent of the number of metadynamics kernels and scales only linearly with the number of dimensions $D$, since each evaluation amounts to a sequence of $D$ small matrix-vector products contracting the TT core; this is therefore much cheaper than evaluating the entire list of Gaussians accumulated in Eq.~\eqref{eq:metad}, whose cost grows with simulation time.

\begin{algorithm}[bt]
\caption{TT-Metadynamics}
\label{alg:full_algorithm}
\begin{algorithmic}[1]
\Require Initial $3n$-element atom position vector $\mathbf r$ and velocity vector $\mathbf v$; simulation timestep $\Delta t$; collective variables $\mathbf x(\mathbf r)$; Gaussian bandwidth $\boldsymbol\sigma$ and initial height $h_0$; bias factor $\gamma$; deposition interval $\omega$; sketching interval $\tau$; TT rank tolerance $\epsilon$; kernel smoothing bandwidth $\boldsymbol\rho$
\Ensure Final bias potential $V_{\text{bias}}(\mathbf x)$ represented as a functional TT; reweighting factors as in Eq.~\eqref{eq:reweighting}, for each collected sample
\State\textbf{initialize} $v_{\text{TT}}(\mathbf x):=0$, $\hat{v}(\mathbf x):=0$
\State\textbf{define} $V_{\text{bias}}(\mathbf x)=v_{\text{TT}}(\mathbf x)+\hat{v}(\mathbf x)$
\For{every MD step $i$}
    \If{($\mod(i,~\omega)=0$)}
    \State Compute CV values, $\mathbf x_{i/\omega}:=\mathbf x(\mathbf r)$
    \State Compute Gaussian height $h_{i/\omega}$ as Eq.~\eqref{eq:height}
        \State $\hat{v}(\mathbf x):=\hat{v}(\mathbf x)+h_{i/\omega}\exp\left(-\dfrac{1}{2}\left|\dfrac{\mathbf x-\mathbf x_{i/\omega}}{\boldsymbol\sigma}\right|^2\right)$
    \EndIf
    \If{$(\mod(i,~\tau)=0)$}
        \State $v_{\text{TT}}(\mathbf x):=\text{TT-Sketch}(V_{\text{bias}}(\mathbf x))$
        \State $\hat{v}(\mathbf x):=0$
    \EndIf
    \For{$j$ from 1 to $3n$}
        \State $F_j:=-\dfrac{\partial V(\mathbf r)}{\partial r_j}-\dfrac{\partial V_{\text{bias}}(\mathbf x)}{\partial\mathbf x}\Bigr|_{\substack{\mathbf x}}\dfrac{\partial\mathbf x(\mathbf r)}{\partial r_j}$
        \label{step:force}
        \Comment{with kernel smoothing}
    \EndFor
    \State Propagate $\mathbf r$ and $\mathbf v$ by $\Delta t$
\EndFor
\end{algorithmic}
\end{algorithm}

\begin{figure}[bt]
    \centering
    \includegraphics[width=0.9\linewidth]{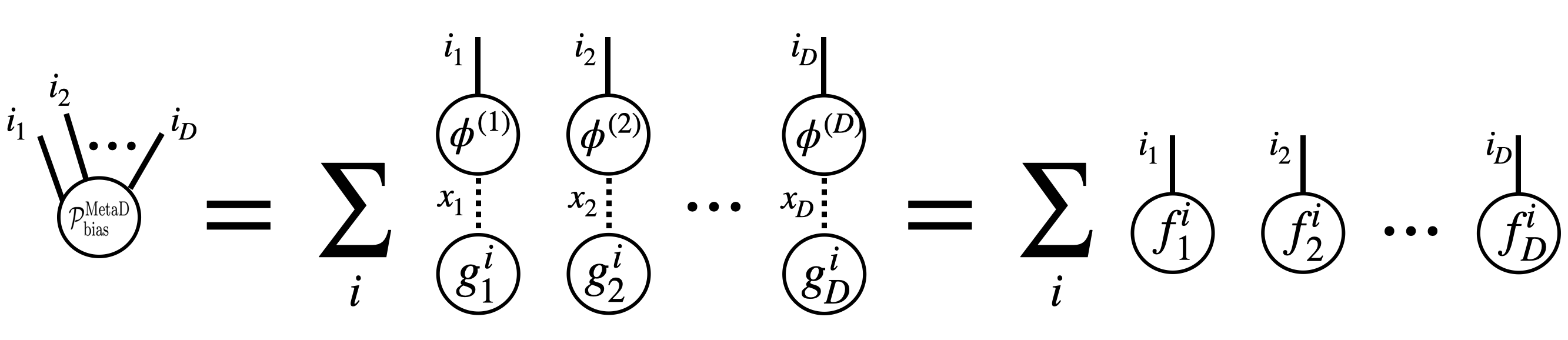}
    \caption{Tensor diagram illustrating the closed form of $ \mathcal P_{\text{bias}}^{\text{MetaD}}$ in Eq.\ \eqref{coeff_tns_closed_form2}. Here we omit the scalar weights $h_i$ in the tensor diagram for simplicity.}
    \label{fig:Pbias_closedform}
\end{figure}

The coefficient tensor can be written in the general form
\begin{equation}
\label{eq:coef}
    \mathcal P_{\text{bias}}\equiv \mathcal P_{\text{bias}}^{\text{prev}}+\mathcal P_{\text{bias}}^{\text{MetaD}},
\end{equation}
where $\mathcal P_{\text{bias}}^{\text{prev}}$ is the coefficient TT obtained from the previous TT-Sketch procedure ($\mathcal P_{\text{bias}}^{\text{prev}}$ is zero in the first TT-Sketch) and $\mathcal P_{\text{bias}}^{\text{MetaD}}$ is the contribution from the running list of Gaussian functions.
Using the orthogonality of the Fourier bases, the latter tensor admits the closed form
\begin{align}\label{coeff_tns_closed_form}
    \nonumber\mathcal P_{\text{bias}}^{\text{MetaD}} (i_1,\dots,i_D) = &\int\text{d}x_1\cdots \int\text{d}x_D~V_{\text{bias}}^{\text{MetaD}}(x_1,\dots,x_D)\\
    &\times\phi_{i_1}(x_1)\cdots\phi_{i_D}(x_D).
\end{align}
Computing the entries of $\mathcal P_{\text{bias}}^{\text{MetaD}}$ requires evaluations of inner products between the metadynamics Gaussian functions $\mathcal G^i(\textbf x)$ defined in Eq.~\eqref{eq:metad} and the Fourier basis functions defined in Eq.~\eqref{eq:fourier}.
Assuming the factorization $\mathcal G^i(\textbf x)=\prod_{k=1}^Dg^i_k(x_k)$, where $g^i_k(x_k)$ is a univariate Gaussian function defined along dimension $k$, inserting the definition of $V_{\text{bias}}^{\text{MetaD}}$ in \eqref{eq:metad} produces a factorization of $\mathcal P_{\text{bias}}^{\text{MetaD}}$ involving a sum of rank-one functions:
\begin{equation}\label{coeff_tns_closed_form2}
    \mathcal P_{\text{bias}}^{\text{MetaD}} (i_1,\dots,i_D) = \sum_i h_i f^i_{i_1}f^i_{i_2}\cdots f^i_{i_D},
\end{equation}
where $f^i_{i_k} = \int g^i_k(x) \phi_{i_k}(x) \text{d}x$, which evaluates to
\begin{align}
    f^i_{i_k}=\begin{cases} h_i(\pi/L_k)^{1/2}\sigma_k,&~~~i_k=1,\\
    \exp\left[-\frac{(\pi\sigma_k\lfloor i_k/2\rfloor)^2}{2L^2}\right](2\pi/L_k)^{1/2}\sigma_k\\
    \quad\times\cos[\pi(x_{i,k}-a_k)\lfloor i_k/2\rfloor/L_k],&\mod(i_k,~2)=0,\\
    \exp\left[-\frac{(\pi\sigma_k\lfloor i_k/2\rfloor)^2}{2L^2}\right](2\pi/L_k)^{1/2}\sigma_k\\
    \quad\times\sin[\pi(x_{i,k}-a_k)\lfloor i_k/2\rfloor/L_k],&~~~\text{otherwise}.\end{cases}
\end{align}
The tensor diagram representation of Eq.~\eqref{coeff_tns_closed_form2} is given by Fig.~\ref{fig:Pbias_closedform}.

\subsection{TT-Sketch}
\label{sec:ttsketch_algo}

$\mathcal P_{\text{bias}}$ is a $D$-dimensional tensor that, in its plain form, suffers from the curse of dimensionality and thus cannot be stored or manipulated explicitly.
This is why a crucial aspect of our TT-MetaD pipeline is the decomposition into TT cores $G_1,\dots,G_D$ without ever forming the full tensor.
We achieve the compression of $\mathcal P_{\text{bias}}$ into a TT format via TT-Sketch~\cite{hur2023generative,chen2023combining,peng2024tensor}; we outline the procedure in Algorithm~\ref{alg:TT-Sketch}.
For $D$ CVs and $N$ Gaussians in the bias potential, the computational complexity of TT-Sketch is only $\mathcal O(DN)$.

The term ``sketch'' refers to multiplications of the large tensor by random ``test'' matrices from the left and right to produce small, tractable projections that capture the dominant structure.
Here we denote the left and right test  matrices (henceforth {\it sketch matrices}) by
$\{S_k(i_{1:k},\beta_{k})\}_{\beta_{k} = 1}^{R_k}$ and $\{T_k(i_{k+1:D},\alpha_{k})\}_{\alpha_{k}=1}^{R_k}$.
$R_k$ is the \emph{sketch size}, i.e., the number of random projections retained at each interface between adjacent dimensions, which controls the tradeoff between computational cost and approximation quality and is chosen independently of $D$; we discuss the choice of $R_k$ below.
Just as a random projection of a low-rank matrix retains its column space with high probability (e.g., as in randomized SVD~\cite{halko2011finding,huber2017randomized}), random projections of a low-rank tensor retain its TT structure.
At the cost of a minor loss of information, the sketch acts as a statistically reliable window into the tensor.

Each TT core $G_k$ is determined independently by solving a linear system of equations of the general form
\begin{equation}\label{eq:core_deter_eq}
    \sum_{\alpha_{k-1}=1}^{R_{k-1}}A_k(\beta_{k-1},\alpha_{k-1})G_k(\alpha_{k-1},i_k,\alpha_k)=B_k(\beta_{k-1},i_k,\alpha_k),
\end{equation}
where $A_k$ is a compressed view of $\mathcal{P}_{\text{bias}}$ obtained by contracting all dimensions except $k$ against the sketch matrices: $A_k$ contracts over all dimensions (including $k$) and serves as the coefficient matrix of the linear system, while $B_k$ leaves the physical index $i_k$ free and serves as the ``right-hand side,'' thereby encoding the full shape of $G_k$.
$A_k$ and $B_k$ are explicitly given by
\begin{align}
    \nonumber A_k(\beta_{k-1},\alpha_{k-1})=&\sum_{i_{1:k-1}}\sum_{i_{k:D}}S_{k-1}(i_{1:k-1},\beta_{k-1})\\
    &\times\mathcal{P}_{\text{bias}}(i_{1:k-1},i_{k:D})T_{k-1}(i_{k:D},\alpha_{k-1}), \label{A_k_def}\\
    \nonumber B_k(\beta_{k-1},i_k,\alpha_k)=&\sum_{i_{1:k-1}}\sum_{i_{k+1:D}}S_{k-1}(i_{1:k-1},\beta_{k-1})\\
    &\times\mathcal{P}_{\text{bias}}(i_{1:k-1},i_k,i_{k+1:D})T_k(i_{k+1:D},\alpha_k). \label{B_k_def}
\end{align}
Each equation (Eq.~\eqref{eq:core_deter_eq}) solves for a distinct core $G_k$ and is completely decoupled from all other equations; TT-Sketch is therefore fully parallelizable.
In each equation, we use the shorthand $i_{l:m}$ to denote the grouped multi-index $(i_l,i_{l+1},\dots,i_{m})$.
With this convention, $\mathcal{P}_{\text{bias}}(i_{1:k-1},i_{k:D})$ and $\mathcal{P}_{\text{bias}}(i_{1:k-1},i_k,i_{k+1:D})$ are the matrix and third-order tensor, respectively, obtained by reshaping the original \(D\)-dimensional tensor.
The cases $k=1$ and $k=D$ require special treatment, as indicated in Algorithm~\ref{alg:TT-Sketch}.
To visualize these equations, a four-dimensional example of this procedure is shown in Fig.~\ref{fig:sketch3} in the Supporting Information.

After solving the \emph{core-determining equations} given by Eq.~\eqref{eq:core_deter_eq}, we obtain a TT whose ranks are $R_1,\dots,R_{D-1}$ instead of the target ranks $r_1,\dots,r_{D-1}$.
The $R_k$ are independent of $D$, but must, in practice, be chosen to be considerably larger than the target ranks (typically a small integer multiple of $r_k$).
This results in an overdetermined system of equations; this ``oversampling'' ensures stability when solving the linear system.
To further reduce the ranks to the target $r_k$, we perform rank-$r_k$ singular value decompositions (SVD) on the matrices $A_k$ and then ``trim'' the oversketched TT by inserting the projector $W_k W_k^T$ between every two adjacent cores.
This trimming procedure is detailed in Algorithm~\ref{alg:TT-Sketch} and depicted in Fig.~\ref{fig:trim} in the Supporting Information.
In this work, we choose the target ranks $r_k$ such that the residual error of truncated SVD satisfies
\begin{equation}
    \sum_{\alpha_{k-1}=r_{k-1}+1}^{R_k}\left(\Sigma_k(\alpha_{k-1},\alpha_{k-1})\right)^2 < \epsilon  \sum_{\alpha_{k-1}=1}^{R_k}\left(\Sigma_k(\alpha_{k-1},\alpha_{k-1})\right)^2,
\end{equation}
where we set the threshold $\epsilon$ to $10^{-4}$ throughout this work.

The dominant computational cost in Algorithm \ref{alg:TT-Sketch} arises from forming $A_k$ and $B_k$ in Eq.~\eqref{eq:core_deter_eq}. A na{\"i}ve implementation based on direct tensor contractions incurs an exponential cost in $D$.
By exploiting the structure of $\mathcal{P}_{\text{bias}}$ (Eq.~\eqref{eq:coef}) and employing tensorized sketch matrices $S_k$ and $T_k$, these contractions can be computed efficiently through iterative procedures, with an overall complexity of $\mathcal O(DN)$. Details of this construction are provided in the next subsection.  

\begin{algorithm}[bt]
\caption{TT-Sketch}
\label{alg:TT-Sketch}
\begin{algorithmic}[1]
\Require Coefficient tensor $\mathcal P_{\text{bias}} = \mathcal{P}_{\text{bias}}^{\text{prev}} + \mathcal{P}_{\text{bias}}^{\text{MetaD}}$, stored implicitly as the sum of a rank-$r_k$ TT (the previous sketch $\mathcal{P}_{\text{bias}}^{\text{prev}}$, Eq.~\eqref{eq:coef}) and $N$ rank-one tensors (the current Gaussian list $\mathcal{P}_{\text{bias}}^{\text{MetaD}}$, Eq.~\eqref{coeff_tns_closed_form2}); target TT rank $r_k$;  tensorized sketch matrices $S_k\in \mathbb{R}^{n_1\cdots n_k\times R_k}$, $T_k\in \mathbb{R}^{n_{k+1}\cdots n_D\times R_k}$ for $k=1,\dots,D-1$
\Ensure Rank-$r_k$ TT approximation of $\mathcal P_{\text{bias}}$ with trimmed tensor cores $\bar{G}_1\in \mathbb{R}^{n_1\times r_1},\bar{G}_2\in \mathbb{R}^{r_1\times n_2\times r_2},\dots,\bar{G}_D\in \mathbb{R}^{r_{D-1}\times n_D}$
\State $G_1 := \text{fastcontract}(\mathcal{P}_{\text{bias}},T_1)$
\For{$k=2,\dots,D-1$}
\State  $A_k:= \text{fastcontract}(S_{k-1},\mathcal{P}_{\text{bias}},T_{k-1})$
\State $A_k^+\equiv\text{pseudoinverse}(A_k)$
\State  $B_k:= \text{fastcontract}(S_{k-1},\mathcal{P}_{\text{bias}},T_{k})$
\State $G_k(\alpha_{k-1},i_k,\alpha_k):=\sum_{\beta_{k-1}=1}^{R_{k-1}}A_k^+(\alpha_{k-1},\beta_{k-1})B_k(\beta_{k-1},i_k,\alpha_k)$.
\EndFor
\State $A_D:= \text{fastcontract}(S_{D-1},\mathcal{P}_{\text{bias}},T_{D-1})$.
\State $A_D^+\equiv\text{pseudoinverse}(A_D)$
\State $B_D:=\text{fastcontract}(S_{D-1},\mathcal{P}_{\text{bias}})$.
\State $G_D(\alpha_{D-1},i_k):=\sum_{\beta_{D-1}=1}^{R_{D-1}}A_D^+(\alpha_{D-1},\beta_{D-1})B_D(\beta_{D-1},i_D)$.
\State Rank-$r_k$ SVD $A_k \approx U_k \Sigma_k W_k^T$.
\Comment{Trimming}
 \State $\bar{G}_1(i_1,\gamma_1) := \sum_{\alpha_1=1}^{R_1}G_1(i_1,\alpha_1)W_2(\alpha_1,\gamma_1)$.
 \For{$k=2,\dots,D-1$}
 \State $\bar{G}_{k}(\gamma_{k-1},i_k,\gamma_{k}) := \sum_{\alpha_{k-1}=1}^{R_{k-1}}\sum_{\alpha_{k}=1}^{R_{k}}W_k(\alpha_{k-1},\gamma_{k-1}) G_{k}(\alpha_{k-1},i_k,\alpha_{k})W_{k+1}(\alpha_{k},\gamma_{k})$.
 \EndFor
 \State $\bar{G}_D(\gamma_{D-1},i_D) := \sum_{\alpha_{D-1}=1}^{R_{D-1}}W_D(\alpha_{D-1},\gamma_{D-1}) G_D(\alpha_{D-1},i_D)$.\\
 \Return $\bar{G}_1,\bar{G}_2,\dots,\bar{G}_D$
\end{algorithmic}
\end{algorithm}

\subsubsection{Fast contraction of the coefficient tensor and sketch matrices}
\label{sec:ttsketch}
\begin{figure}[bt]
    \centering
    \subfloat[Left sketch matrix]{{\includegraphics[width=0.4\textwidth]{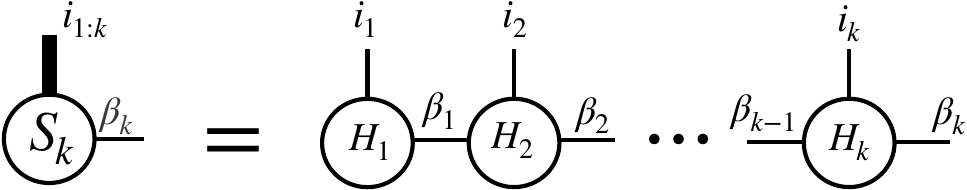}}}
    \qquad  \qquad  
    \subfloat[Right sketch matrix]{{\includegraphics[width=0.4\textwidth]{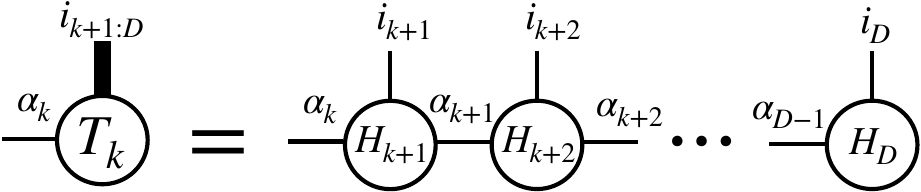}}}\\[20pt]
    \caption{Tensor diagrams illustrating the left and right sketch tensors used in TT-compression.
    Each sketch tensor contracts with a subset of the TT cores.
    The horizontal legs correspond to auxiliary indices $\beta_k$ and $\alpha_k$, while the thicker vertical legs represent composite indices formed by grouping the physical indices $i_1,\dots,i_k$ (left sketch) or $i_{k+1},\dots,i_D$ (right sketch).
    Although depicted schematically as ``matrices,'' the objects are technically higher-order tensors with multiple legs, and the term ``sketch matrix'' refers to their linear action on vectorized TT cores.
    Each tensor $H_k$ contains i.i.d. entries drawn from a standard normal distribution.}\label{fig:sketch_matrix_TT}    
\end{figure}

In this subsection, we detail the implementation of the subroutine ``fastcontract'' in Algorithm~\ref{alg:TT-Sketch}. We begin by discussing the choice of sketching matrices. Sketches na{\"i}vely constructed as full $D$-dimensional tensors prove highly impractical due to the exponential cost of contracting them with $\mathcal{P}_{\text{bias}}$.
In this work, we follow Refs.~\citenum{peng2024tensor,kressner2023streaming} and choose \emph{structured} TT sketches with prescribed ranks $\{R_k\}_{k=1}^{D-1}$; this way, contractions are performed separately for each dimension and, taken together, have a combined computational cost that is linear in $D$, as indicated by the linear chain structure of the sketch tensors in Fig.~\ref{fig:sketch_matrix_TT}, where each core $H_k$ contracts with only a single dimension at a time.
These ``sketch TTs'' have independent entries sampled from the standard normal distribution (i.e., Gaussian-distributed scalars with mean 0 and variance 1).
Such sketches have been shown to capture the dominant subspace of the original tensor $\mathcal{P}_{\text{bias}}$ with high probability~\cite{halko2011finding,huber2017randomized}. 

\begin{figure}[bt]
    \centering
     \subfloat[Tensor structure of $S_{k}\mathcal{P}_{\text{bias}}^{\text{MetaD}}T_{k}$ in $A_{k+1}$]{{\includegraphics[width=\textwidth]{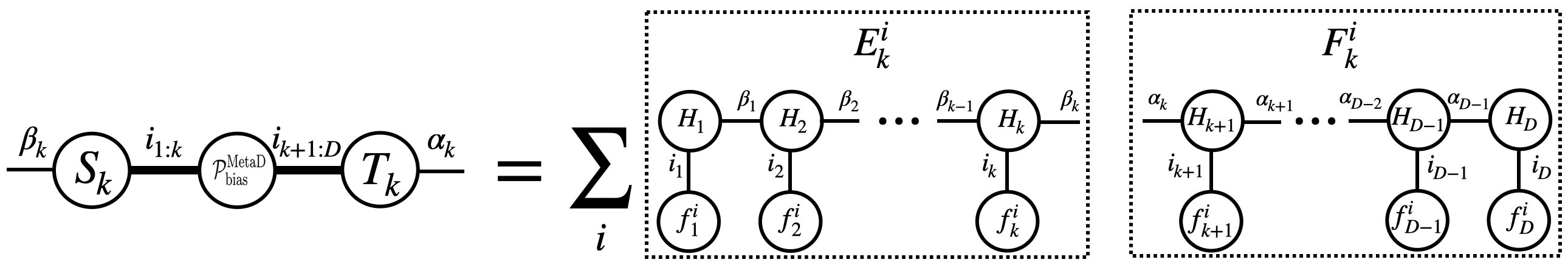}}}\\[20pt]
    \subfloat[Iterative computation of $E^i_k$]{{\includegraphics[width=0.32\textwidth]{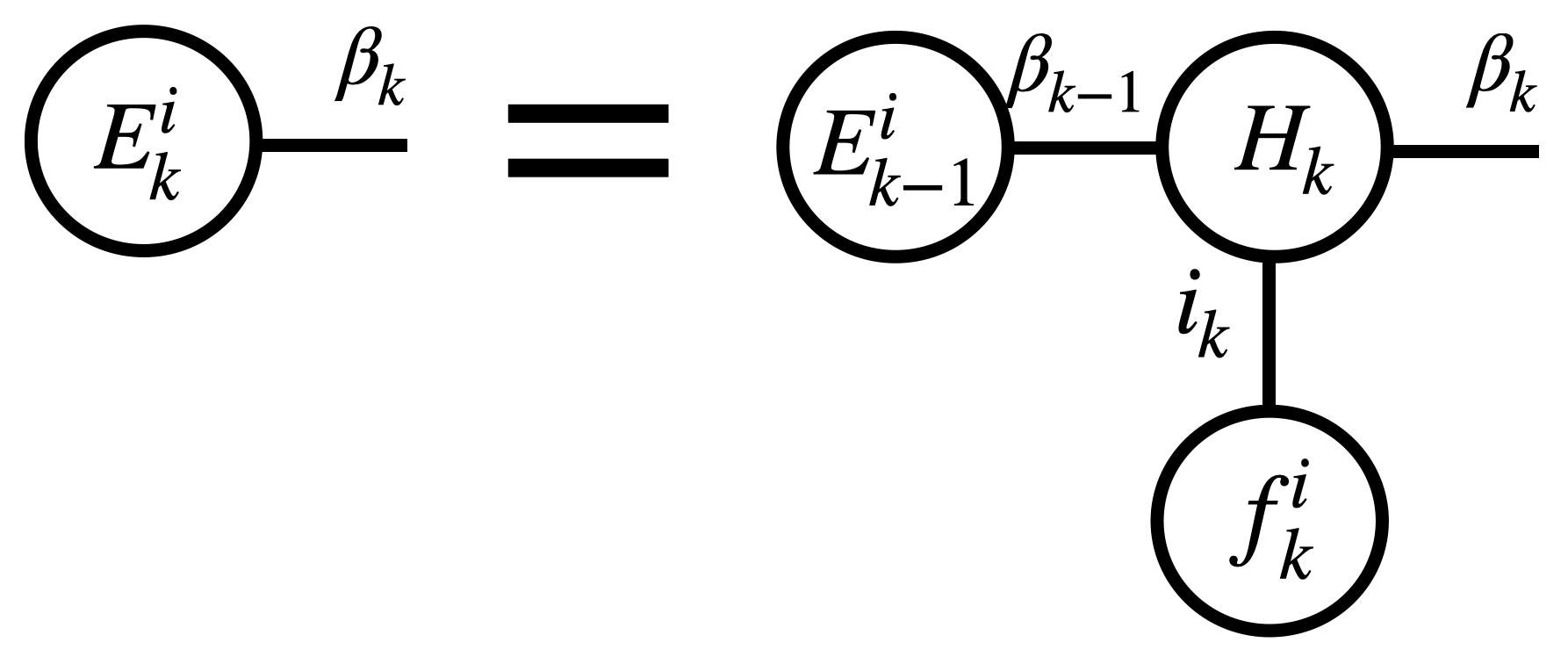}}}
    \qquad  \qquad  \qquad 
    \subfloat[Iterative computation of $F^i_k$]{{\includegraphics[width=0.32\textwidth]{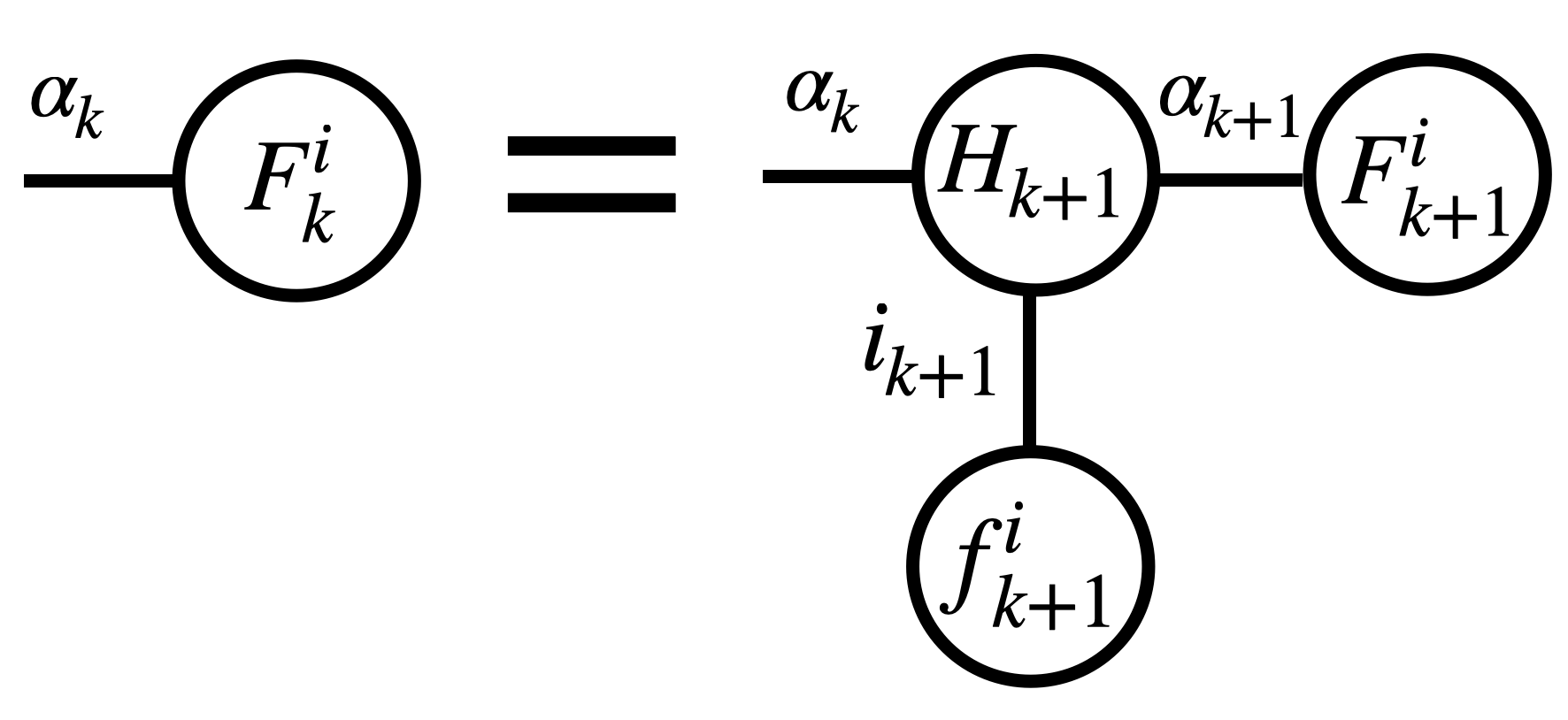}}}\\[20pt]
   \caption{Tensor diagrams illustrating the iterative computations of $S_{k}\mathcal{P}_{\text{bias}}^{\text{MetaD}}T_{k}$ when forming $A_{k+1}$ defined in Eq.~\eqref{A_k_def}.
By substituting the expressions of the sketch tensors $S_k, T_k$ from Fig.~\ref{fig:sketch_matrix_TT} and the bias tensor $\mathcal P_{\text{bias}}^{\text{MetaD}}$ from Fig.~\ref{fig:Pbias_closedform}, 
the matrix $S_{k}\mathcal{P}_{\text{bias}}^{\text{MetaD}}T_{k}$ can be written in the form $\sum_i h_i\, E_k^i ( F_k^i)^T$, as shown in (a), where the scalar weights $h_i$ are omitted in the tensor diagram for simplicity.
The column vectors $E_k^i$ and $F_k^i$ are computed recursively from $E_{k-1}^i$ and $F_{k+1}^i$, respectively, as illustrated in (b) and (c).
With this iterative procedure, for each fixed $i$, all vectors $E_k^i$ and $F_k^i$ for $k = 1, \dots, D$ can be obtained with $\mathcal{O}(D)$ computational cost.
Consequently, including all indices $i$, the overall computational and memory complexity scale as $\mathcal{O}(DN)$.
}\label{fig:Ak_MetaD}
\end{figure}

\begin{figure}[bt]
    \centering
     \subfloat[Tensor structure of $S_{k}\mathcal{P}_{\text{bias}}^{\text{prev}}T_{k}$ in $A_{k+1}$]{{\includegraphics[width=\textwidth]{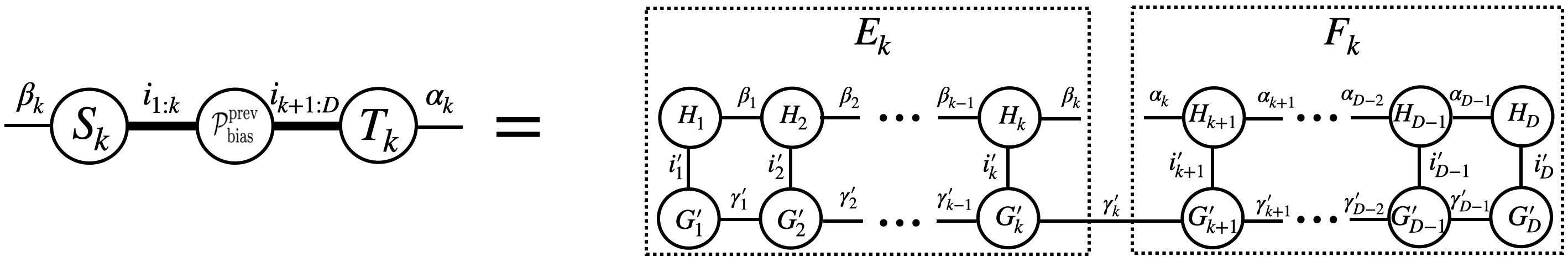}}}\\[20pt]
    \subfloat[Iterative computation of $E_k$]{{\includegraphics[width=0.32\textwidth]{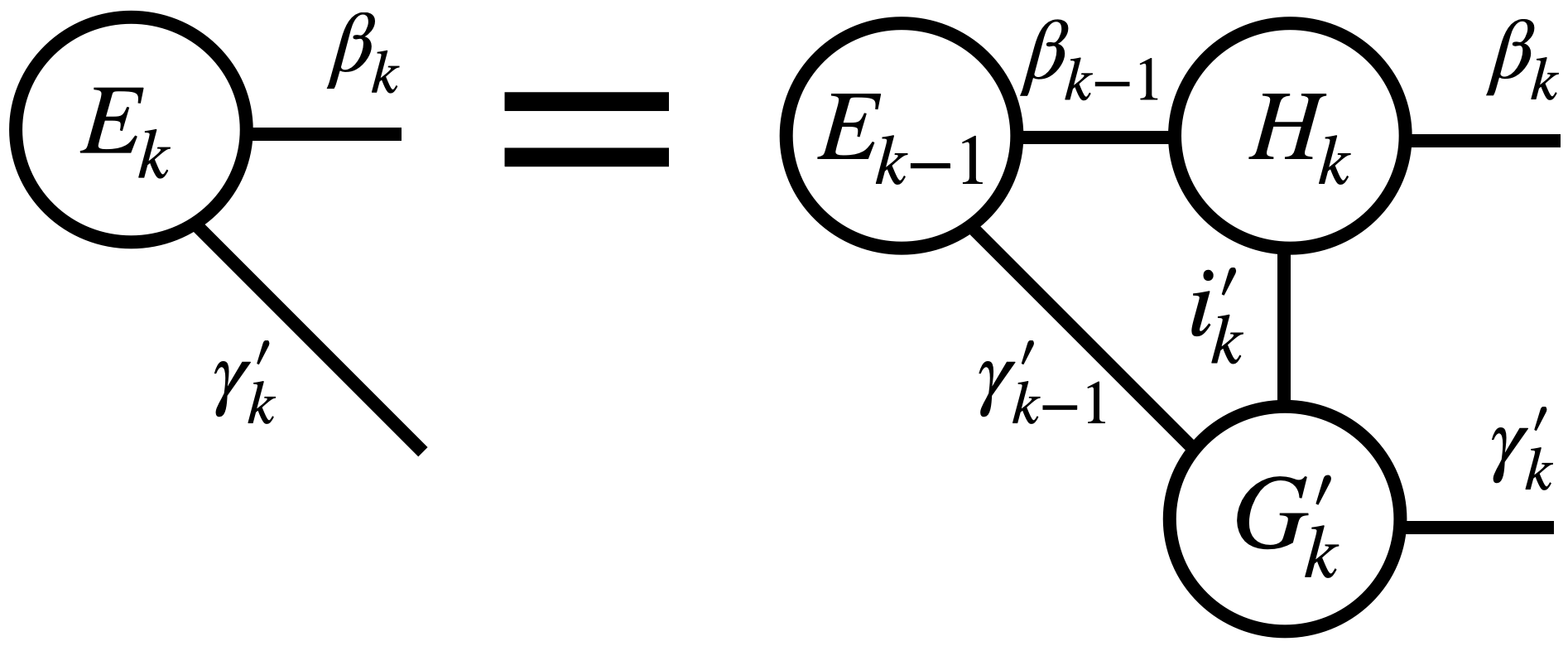}}}
    \qquad  \qquad  \qquad 
    \subfloat[Iterative computation of $F_k$]{{\includegraphics[width=0.32\textwidth]{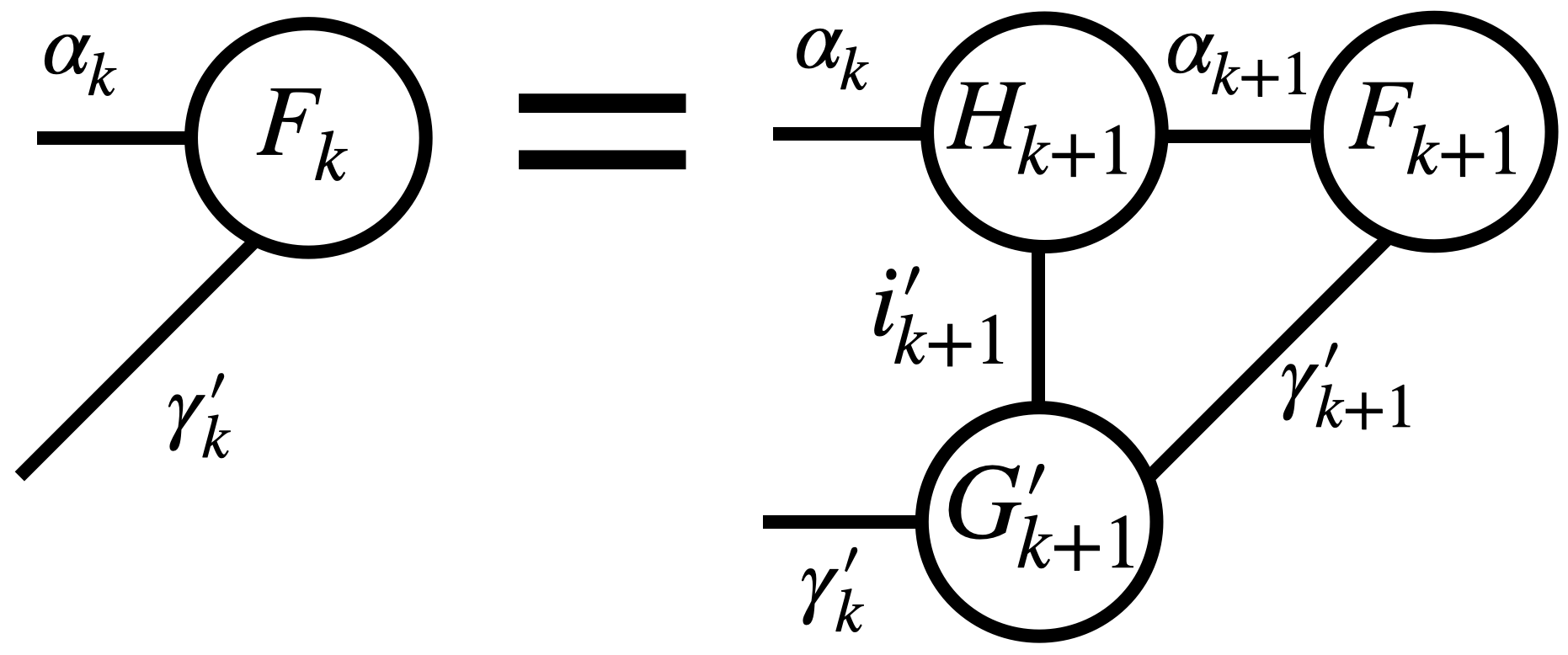}}}\\[20pt]
 \caption{Tensor diagrams illustrating the iterative computations of $S_{k}\mathcal{P}_{\text{bias}}^{\text{prev}}T_{k}$ when forming $A_{k+1}$ defined in Eq.~\eqref{A_k_def}.
By substituting the expressions of the sketch tensors $S_k, T_k$ from Fig.~\ref{fig:sketch_matrix_TT} and the bias tensor $\mathcal P_{\text{bias}}^{\text{prev}}$ in TT format like Fig.~\ref{fig:tt-diagram}(a), 
the matrix $S_{k}\mathcal{P}_{\text{bias}}^{\text{prev}}T_{k}$ can be written in the form $E_k ( F_k)^T$, as shown in (a).
The column vectors $E_k$ and $F_k$ are computed recursively from $E_{k-1}$ and $F_{k+1}$, respectively, as illustrated in (b) and (c).
With this iterative procedure, all vectors $E_k$ and $F_k$ for $k = 1, \dots, D$ can be obtained with $\mathcal{O}(D)$ computational cost.
}\label{fig:Ak_prev}
\end{figure}

 Next, we elaborate on how the structured sketch tensors interact with $\mathcal{P}_{\mathrm{bias}}$ to compute $
A_{k+1} = S_k \, \mathcal{P}_{\mathrm{bias}} \, T_k$ using the fastcontract procedure.
The computation of $B_k = S_k \, \mathcal{P}_{\mathrm{bias}} \, T_{k+1}$ follows an analogous treatment.
We begin by decomposing $A_{k+1}$ according to the structure of $\mathcal{P}_{\mathrm{bias}}$:
\begin{equation}
    A_{k+1}
    = S_k \, \mathcal{P}_{\mathrm{bias}} \, T_k
    = S_k \, \mathcal{P}_{\mathrm{bias}}^{\mathrm{MetaD}} \, T_k
    + S_k \, \mathcal{P}_{\mathrm{bias}}^{\mathrm{prev}} \, T_k.
    \end{equation}
Here, $\mathcal{P}_{\mathrm{bias}}^{\mathrm{MetaD}}$ is a sum of $N$ rank-one tensors, while $\mathcal{P}_{\mathrm{bias}}^{\mathrm{prev}}$ admits a TT representation.
By substituting the explicit structures of the sketch tensors and each component of $\mathcal{P}_{\mathrm{bias}}$, both $S_{k} \mathcal{P}_{\text{bias}}^{\text{MetaD}} T_k$ and $S_{k} \mathcal{P}_{\text{bias}}^{\text{prev}} T_k$ can be written as outer products of pairs of vectors, $\sum_i h_i E_k^i (F_k^i)^T$ and $E_k F_k^T$ respectively (Figs.~\ref{fig:Ak_MetaD}(a) and~\ref{fig:Ak_prev}(a)).
Here, $E_k^i$ and $E_k$ are \emph{left environments}, namely vectors that capture the cumulative effect of contracting the sketch and bias tensors over all dimensions to the \emph{left} of $k$, while $F_k^i$ and $F_k$ are the corresponding \emph{right environments}, capturing contractions over all dimensions to the \emph{right} of $k$.
The key insight is that these environments can be built up incrementally: $E_k^i$ depends only on $E_{k-1}^i$ and the local tensors at dimension $k$, and similarly $F_k^i$ depends only on $F_{k+1}^i$, as illustrated in Figs.~\ref{fig:Ak_MetaD}(b,c) and~\ref{fig:Ak_prev}(b,c).
This recursive structure means that all environments for $k=1,\dots,D$ can be obtained with $\mathcal O(D)$ cost per Gaussian labeled by index $i$.
Overall, the computational cost of forming all $A_k$ and $B_k$ is $\mathcal{O}(D N)$.
The complete procedures for forming all $A_k$ are summarized in Algorithm~\ref{alg:fast-contract}.

 \begin{algorithm}[bt]
\caption{Efficient computations of $A_{k}$ defined in Eq.\ \eqref{A_k_def}}
\label{alg:fast-contract}
\begin{algorithmic}[1]
\Require Coefficient tensor $\mathcal P_{\text{bias}} = \mathcal{P}_{\text{bias}}^{\text{prev}} + \mathcal{P}_{\text{bias}}^{\text{MetaD}}$, where $\mathcal{P}_{\text{bias}}^{\text{prev}}$ is a TT with cores $G'_1,G'_2,\cdots,G'_D$ and $\mathcal{P}_{\text{bias}}^{\text{MetaD}}$ defined as Eq.~\eqref{coeff_tns_closed_form2}); vectors $f^i_k \in \mathbb{R}^{n_k}$ (Eq.~\eqref{coeff_tns_closed_form2}) with entries $f^i_k(i_k) = \int g^i_k(x)\phi_{i_k}(x)\,\mathrm{d}x$, encoding the projection of the $i$th Gaussian onto the Fourier basis along dimension $k$, for $i = 1,\dots,N$ and $k = 1,\dots,D$; tensorized sketch matrices $S_k\in \mathbb{R}^{n_1\cdots n_k\times R_k}$ with tensor cores $H_1,\dots,H_k$, $T_k\in \mathbb{R}^{n_{k+1}\cdots n_D\times R_k}$ with tensor cores $H_{k+1},\dots H_D$, for $k=1,\cdots,D-1$
\Ensure $A_{k} \in \mathbb{R}^{R_{k-1}\times R_{k-1}}$ for $k= 2,\cdots,D$
\State $E_0 \equiv 1$; $E^i_0 \equiv 1$ for $i=1,\cdots, N$.
\For{$k=1,\dots,D-1$}
\Comment{Iterative computations of left environment}
\For{$i=1,\dots,N$}
\State $E^i_k(\beta_k) := \sum_{\beta_{k-1}=1}^{R_{k-1}}\sum_{i_k=1}^{n_k} E^i_{k-1}(\beta_{k-1})H_k(\beta_{k-1},i_k,\beta_{k})f^i_k(i_k)$.
\EndFor
\State $E_k(\beta_k,\gamma_k') := \sum_{\beta_{k-1}=1}^{R_{k-1}}\sum_{\gamma_{k-1}'=1}^{r_k}\sum_{i_k'=1}^{n_k} E_{k-1}(\beta_{k-1},\gamma_{k-1}')H_k(\beta_{k-1},i_k',\beta_{k})G'_k(\gamma_{k-1}',i'_k,\gamma_{k}')$.
\EndFor
\State $F_D \equiv 1$;  $F^i_D \equiv 1$ for $i=1,\cdots, N$.
\For{$k=D,\dots,2$}
\Comment{Iterative computations of right environment}
\For{$i=1,\dots,N$}
\State $F^i_{k-1}(\alpha_{k-1}) := \sum_{\alpha_{k}=1}^{R_{k}}\sum_{i_{k}=1}^{n_{k}} F_{k}(\alpha_{k})H_{k}(\alpha_{k-1},i_{k},\alpha_{k})f^i_{k}(i_{k})$.
\EndFor
\State $F_{k-1}(\alpha_{k-1},\gamma_{k-1}') := \sum_{\alpha_{k}=1}^{R_{k}}\sum_{\gamma_{k}'=1}^{r_{k}}\sum_{i_{k}'=1}^{n_{k}} F_{k}(\alpha_{k},\gamma_{k}')H_{k}(\alpha_{k-1},i_{k}',\alpha_{k})G'_{k}(\gamma_{k-1}',i'_{k},\gamma_{k}')$.
\EndFor
\For{$k=1,\dots,D-1$}
\Comment{Parallel computations of $A_k$}
\State $A_{k+1}(\beta_{k},\alpha_{k}) = \sum_{i=1}^N h_iE_{k}^i(\beta_{k})F_{k}^i(\alpha_{k}) + \sum_{\gamma'_{k}=1}^{r_{k}}E_{k}(\beta_{k},\gamma'_{k})F_{k}(\alpha_{k},\gamma'_{k})$.
\EndFor\\
\Return $A_2,A_3,\dots,A_D$
\end{algorithmic}
\end{algorithm}

\subsubsection{Kernel smoothing}
\label{sec:smoothing}

The TT-Sketch algorithm compresses a sum of Gaussian functions into a low-rank representation by solving a sequence of linear systems.
Because this compression operates on a finite, irregularly distributed set of Gaussian centers, the resulting TT can contain high-frequency oscillations that arise from fitting the local accumulation of kernels rather than the smooth underlying free energy landscape.
These artifacts manifest as large, spurious gradients in $V_{\text{bias}}$, which can destabilize the MD integrator and prevent the bias potential from converging to the negative of the free energy.
To suppress these artifacts, we apply a post-compression kernel smoothing step:
\begin{equation}
   \widetilde{v}_{\text{TT}}(\mathbf x)=\int_{-\infty}^\infty 
   v_{\text{TT}}(\mathbf x')\mathcal K(\mathbf x,\mathbf x')\mathrm d \mathbf x',
   \label{eq:smoothing}
\end{equation}
where $v_{\text{TT}}$ is the TT-approximated bias potential and $\mathcal K(\mathbf x,\mathbf x')\equiv\prod_{k=1}^DK(x_k,x_k')$ is a separable kernel composed of single-dimensional Gaussian kernels $K(x_k,x_k')$ with bandwidth vector $\boldsymbol\rho$.
Because the kernel is separable and the bias is represented in a Fourier basis, the convolution in Eq.~\eqref{eq:smoothing} is computed analytically: $\widetilde{v}_{\text{TT}}(\mathbf x)$ takes the same functional TT form as in Fig.~\ref{fig:tt-diagram}(b), but with each univariate basis function $\phi_i(x_k)$ replaced by its smoothed counterpart
\begin{equation}
    \widetilde{\phi}_i(x_k)=\begin{cases}(2L_k)^{-1/2},&~~~i=1,\\
    \exp\left[-\frac{(\pi \rho_k\lfloor i/2\rfloor)^2}{2L^2}\right](L_k)^{-1/2}\\
    \quad\times\cos\left[\pi(x_k-a_k)\lfloor i/2\rfloor/L_k\right],&\mod(i,~2)=0,\\
    \exp\left[-\frac{(\pi \rho_k\lfloor i/2\rfloor)^2}{2L^2}\right](L_k)^{-1/2}\\
    \quad\times\sin\left[\pi(x_k-a_k)\lfloor i/2\rfloor/L_k\right],&~~~\text{otherwise}.\end{cases}
\end{equation}
The Gaussian factor in $\widetilde{\phi}_i$ progressively damps high-frequency Fourier modes: modes with wavenumber $\lfloor i/2 \rfloor / L_k$ larger than $\sim1/(\pi\rho_k)$ are suppressed exponentially.
Thus, $\boldsymbol\rho$ directly controls the spatial resolution of the bias potential, and its choice involves a tradeoff: too small a bandwidth leaves high-frequency noise in $V_{\text{bias}}$, while too large a bandwidth smooths over genuine free energy barriers and reduces the ability of the bias potential to promote transitions.

When evaluating $V_{\text{bias}}$ and its gradient during molecular dynamics steps, i.e., step~\ref{step:force} of Algorithm~\ref{alg:full_algorithm}, we use $\widetilde{\phi}_i(x)$ in place of $\phi_i(x)$.
However, during TT-Sketch, Eq.~\eqref{coeff_tns_closed_form2} still uses the unmodified basis functions $\phi_i(x)$.
The coefficient TT $\mathcal P_{\text{bias}}$ is unaffected by the kernel smoothing and therefore remains unmodified until the next TT-Sketch procedure.

In this work, we use the constant $\rho=0.05$~rad for each CV for all systems with $D>2$.
For alanine dipeptide ($D=2$), standard grid-based metadynamics serves as the primary reference, and we set $\rho=0$ to enable a direct comparison with the TT approximation itself.
A brief discussion of qualitative and quantitative effects of the choice of bandwidth on the quality of sampling is provided in Section~\ref{sec:6d8d}.

\subsubsection{Reweighting}
\label{sec:reweighting}

Samples collected throughout the TT-Metadynamics-biased simulations must be reweighted to obtain unbiased statistics.
In standard metadynamics calculations, the standard approach is to update a quantity $c(t)$, which, if subtracted from the time-dependent bias potential $V_{\text{bias}}(\mathbf x,t)$, yields a time-independent estimator of the free energy, that is, $\mathcal F(\mathbf x)\equiv-(V_{\text{bias}}(\mathbf x,t)-c(t))$ \cite{tiwary2015time}.
Histograms and other statistical quantities from the unbiased distribution can be easily computed by assigning importance weights $w_i$ to samples $i$ as $w_i\propto\exp\left(-\beta\mathcal F(\mathbf x^{(i)})\right)$.
The approximation of $c(t)$ involves nontrivial multidimensional integrals, whose computation is made practical by grid storage of the bias potential, rendering this approach only viable in low-dimensional settings (a few CVs).
In the case of TT-Metadynamics, frequent evaluation of $c(t)$ is therefore computationally prohibitive.
For the higher-dimensional systems studied in this work (beyond alanine dipeptide), we instead adopt a simpler reweighting strategy in which each sample $\mathbf x^{(i)}$, collected at time $t$, is assigned a weight proportional to the instantaneous bias potential,
\begin{equation}
\label{eq:reweighting}
    w_i\propto\exp\left(\beta V_{\text{bias}}(\mathbf x^{(i)},t)\right).
\end{equation}

This simple reweighting scheme is exact only in the limit that $V_{\text{bias}}$ has fully converged to $-\mathcal{F}(\mathbf x)/\gamma + C'$ (the well-tempered limit~\cite{dama2014well}), such that the time dependence of the bias can be ignored.
Before convergence, two sources of error are present.
First, because the bias is still evolving, samples collected at different times carry systematically different weights, and the simple scheme does not correct for this non-stationarity; this is precisely what $c(t)$ is designed to handle in the Tiwary--Parrinello approach.
Second, even after convergence, the simple weights can have high variance when the bias is large since the exponential amplifies small differences in $V_{\text{bias}}$.
This effect can, in practice, be mitigated by discarding an initial portion of the trajectory (as we do for AIB$_9$).
Moreover, we report only low-dimensional projections, which average over much of the high-dimensional variation.
This approach avoids the explicit computation of $c(t)$ and was found to be sufficient for accurate estimation of low-dimensional free energy projections in the systems considered.

\subsection{Simulation protocol and systems studied}
\label{sec:simsetup}

Listed in order of increasing complexity and dimensionality of $V_{\text{bias}}$, the molecules used to test TT-Metadynamics in this work were alanine dipeptide (i.e., $N$-acetyl-alanyl-$N'$-methylamide), trialanine, ditryptophan, and a peptide of nine $\alpha$-aminoisobutyric acids (AIB$_9$).
Alanine dipeptide was mainly included for easy visualization of the bias potential (in 2D) and validation of TT-Metadynamics.
Trialanine and ditryptophan involved a dimensionality of $V_{\text{bias}}$ large enough (6D and 8D, respectively) for the practical benefits of TT-Metadynamics over its traditional variant to become apparent.
Finally, AIB$_9$ was chosen because it is known to exhibit strong metastability, with several local minima separating two stable helical conformations \cite{buchenberg2015hierarchical,sittel2017principal,biswas2018metadynamics,mehdi2022accelerating,wang2022data,strahan2023inexact}; a dimension of $V_{\text{bias}}$ as high as 14 was required for our simulations to sample the free energies along selected dihedral angles accurately.

Simulations were performed using the Langevin integrator in GROMACS 2024.2~\cite{abraham2015gromacs}, with the timestep and temperature set to 2 fs and 300 K, respectively.
All systems except AIB$_9$ were modeled in vacuum with the AMBER99SB-ILDN force field~\cite{hornak2006comparison}.
In the case of AIB$_9$, we used the same starting structure and force field (CHARMM36m~\cite{huang2017charmm36m}) as in Ref.~\citenum{mehdi2022accelerating}, without the water molecules.
All simulations were preceded by an initial steepest descent energy minimization.

TT-Sketch is performed every $\tau$ steps; we use $\tau=5\times10^5$ for the alanine dipeptide calculations and $\tau=5\times10^6$ for all other calculations.
The Gaussian list defined in Eq.~\eqref{eq:metad} is reset after each sketching step to keep the active list small.
After the first $\tau$ steps, the sum of Gaussian functions defined in Eq.~\eqref{eq:metad} is fit to a TT, which becomes the new bias potential.
The bias potential is then computed by adding evaluations of this TT function with sums over the new Gaussian list.
In subsequent sketching steps, both the Gaussian list and the prior TT are inputs into the TT-Sketch algorithm to form the next TT.
A potential concern with this approach is that errors can accumulate. However, metadynamics is self-correcting:  
Gaussians are deposited in regions where the bias is deficient.  
The procedure is summarized in Algorithm~\ref{alg:full_algorithm}.

The TT-Metadynamics code was developed as part of the PLUMED package, version 2.10b \cite{plumed}; as such, PLUMED was used to patch GROMACS simulations and for CV analysis.
Tensor-related computations (e.g., contractions, SVD) were performed with the ITensor library~\cite{itensor}.
All simulations involved the use of multiple parallel walkers \cite{raiteri2005efficient}; in all cases, the number of walkers was 10.
The simulation times reported are total simulation times, that is, when we say that we ran a simulation that lasted a time $t_{\text{sim}}$, we mean that each walker ran a total time $t_{\text{sim}}/10$.
Moreover, each walker adds a Gaussian function every $\omega$ timesteps, but only $\tau/10$ timesteps within a single walker elapse between consecutive TT-Sketch procedures.
Every time a walker deposits a Gaussian function or TT-Sketch occurs, the bias potential shown in Algorithm~\ref{alg:full_algorithm} is communicated between all walkers to ensure that each feels the same bias potential at all times.

Several user-specified parameters were carefully tuned for the systems considered in this work.
Metadynamics-related parameters, namely the bias factor $\gamma$, the common kernel bandwidth $\sigma$ used for all dihedral angles, and the initial Gaussian height $h_0$ (see Eq.~\eqref{eq:height}) are listed in Table~\ref{timescales} in the Supporting Information.
The Gaussian deposition rate $\omega$ is kept at 500 steps throughout.
For all calculations, the number of basis functions $n_k$ along all dimensions and the sketch tensor ranks $R_k$ used for TT-Sketch were kept fixed at 31 (15 cosines/15 sines) and 60, respectively.

To validate our method for all systems, 1D potentials of mean force (PMFs) obtained from the TT-Metadynamics simulations were compared with similar curves obtained from temperature replica exchange molecular dynamics (REMD) simulations.
In all cases, exchange moves were proposed every 4~ps.
The temperatures for each system were selected using the algorithm from Ref.~\citenum{patriksson2008temperature}, with the target acceptance ratio set to 0.5.
All temperatures are tabulated in Table~\ref{temps} in the Supporting Information.
In the case of alanine dipeptide, each replica was run for a total time 200~ns, while for all other systems, each was run for a total time $2~\upmu$s.
Samples from all replicas were combined and used to obtain reweighted histograms with the help of the PyMBAR software \cite{shirts2008statistically}.
The reported root mean square displacements (RMSD) for PMFs exclude points where the REMD PMF exceeds 13~$\beta^{-1}$ above the minimum value to focus on regions that contribute significantly to molecular populations at 300 K.

\begin{figure}[bt]
\centering
\begin{tikzpicture}
\node at (0,0) {\includegraphics[width=0.9\textwidth]{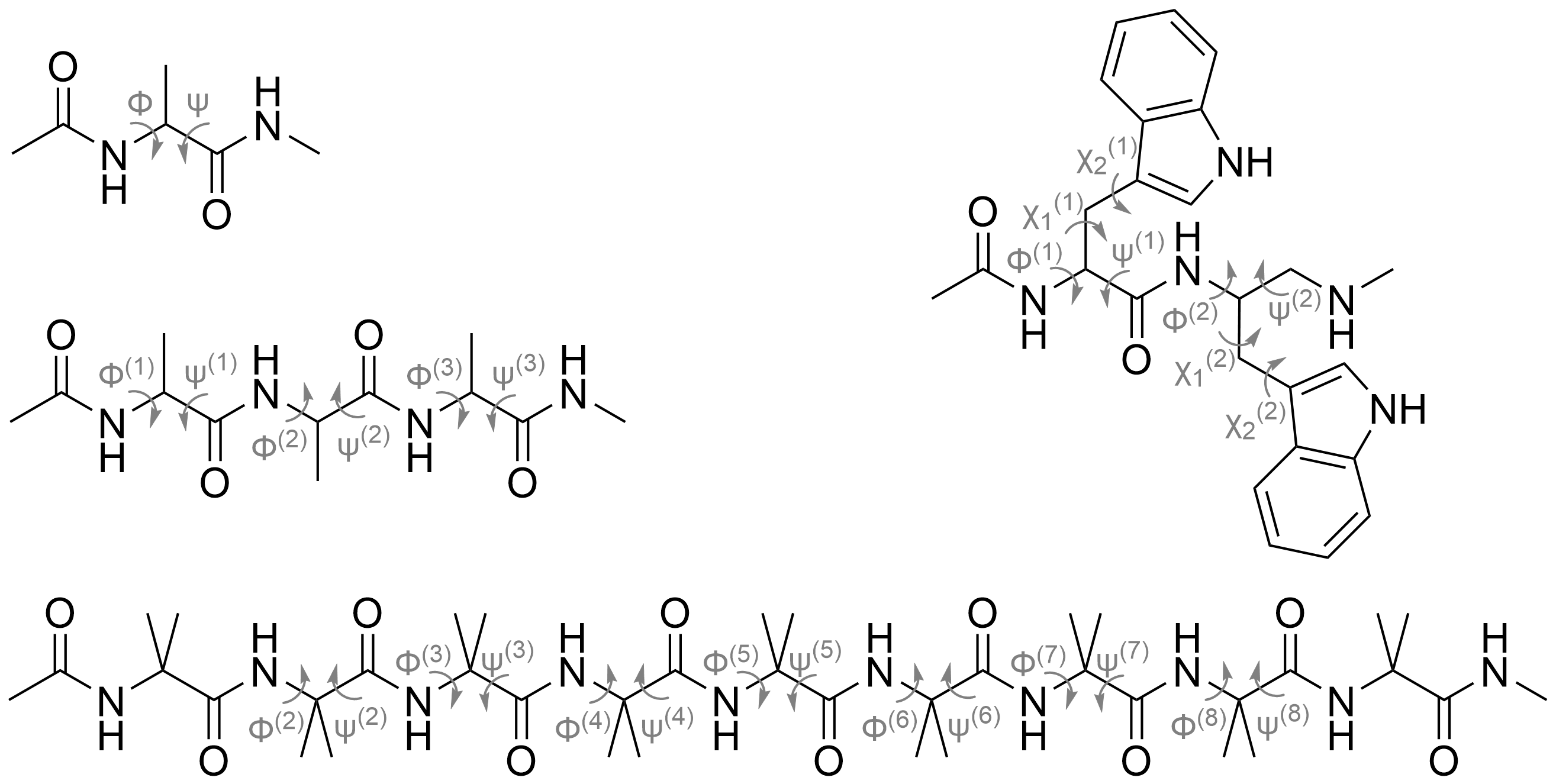}};
\node at (-7.7,3) {(a)};
\node at (-7.7,0.4) {(b)};
\node at (1.1,3) {(c)};
\node at (-7.7,-2.2) {(d)};
\end{tikzpicture}
\caption{2D structures of alanine dipeptide (a), trialanine (b), ditryptophan (c), and AIB$_9$ (d) with annotated dihedral angles.
}
\label{fig:structures}
\end{figure}

The CVs that we used are the dihedral angles labeled in gray in Fig.\ \ref{fig:structures}.
For alanine dipeptide, trialanine, and ditryptophan, there are 2, 6, and 8 CVs, respectively.
For AIB$_9$, we performed two separate simulations: one with 10 CVs, which excluded dihedral angles from residues 1, 2, 8, and 9, and one with 14 CVs, which excluded dihedral angles from residues 1 and 9.

\section{Results and discussion}
\label{sec:results}

\subsection{Overview of test systems}

We studied four systems to probe different aspects of TT-Metadynamics.  
The 22-atom alanine dipeptide ($D=2$) serves primarily as a validation case: because grid-based metadynamics with Tiwary--Parrinello reweighting is fully tractable at this dimensionality, it provides a baseline for isolating the accuracy of the TT approximation itself from other sources of error.
Trialanine ($D=6$) and ditryptophan ($D=8$) represent the regime where grid storage becomes infeasible, and the practical advantages of TT-Metadynamics over storing the full Gaussian sum start to become apparent; these systems allow us to assess the convergence rate, computational cost, and the effect of the kernel smoothing bandwidth as a function of dimension.
Ditryptophan is a particularly stringent test because one of its dihedral barriers exceeds $15~\beta^{-1}$, making transitions rare and the bias potential difficult to converge.
Finally, AIB$_9$ ($D=10$ and $D=14$) represents a qualitatively harder problem, as it is a peptide with several local free-energy minima separating two stable helical conformations.
For this benchmark system, dimensionalities large enough to prohibit any comparison with standard metadynamics are required to achieve accurate free energy estimates.
Together, these systems allow us to characterize both the accuracy and the scaling behavior of TT-Metadynamics across a range of conditions relevant to biomolecular simulation.

\subsection{Alanine dipeptide}

\begin{figure}
\centering
\begin{tikzpicture}
\node at (0,0) {\includegraphics[width=0.45\textwidth]{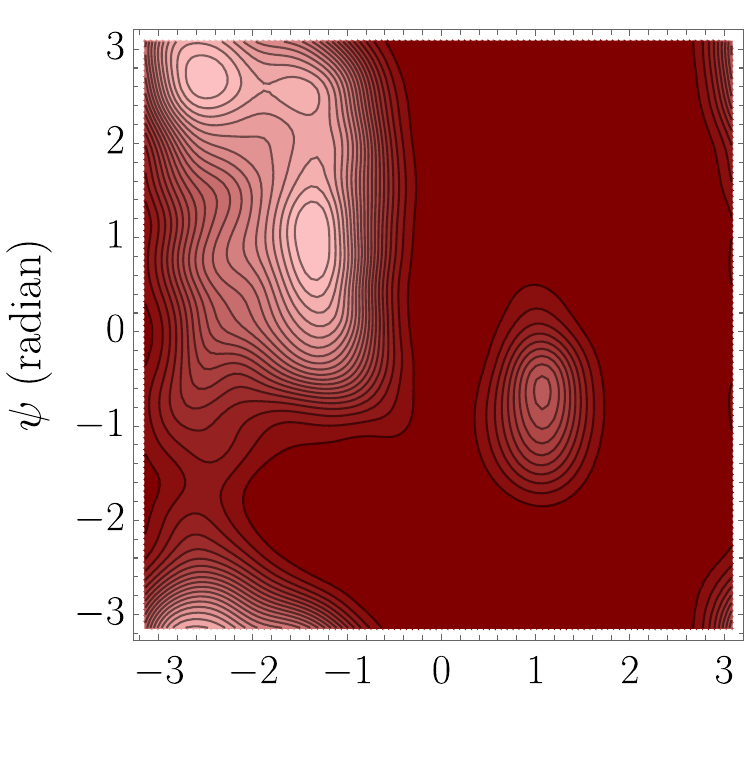}};
\node at (6.8,0) {\includegraphics[width=0.45\textwidth]{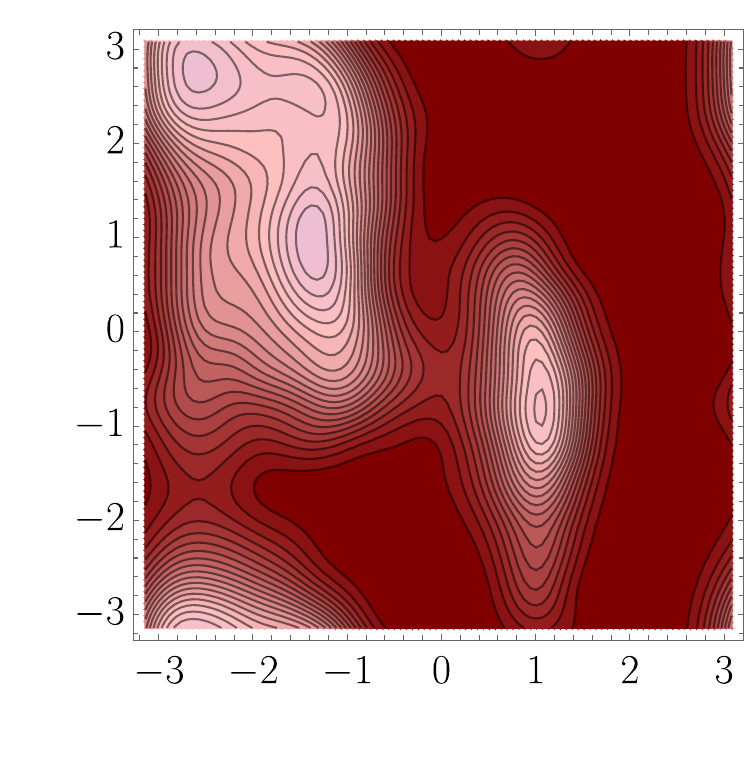}};
\node at (0,-6.8) {\includegraphics[width=0.45\textwidth]{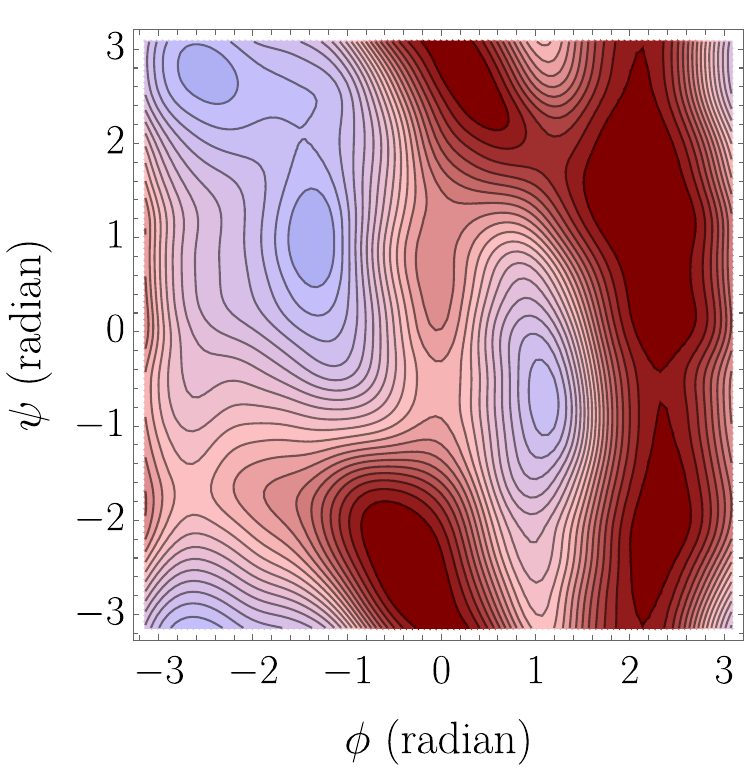}};
\node at (6.8,-6.8) {\includegraphics[width=0.45\textwidth]{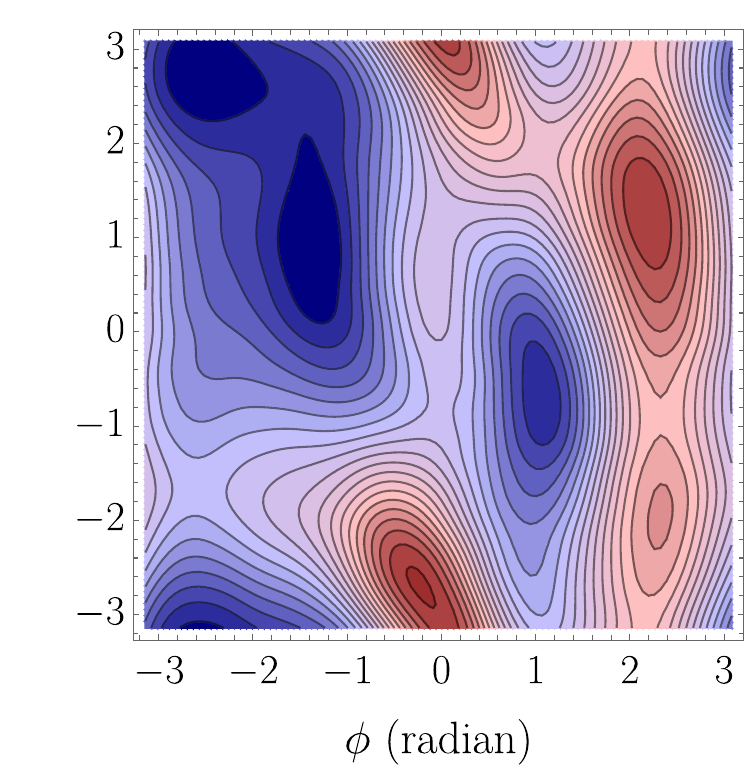}};
\node at (12.2,-2) {\includegraphics[width=0.18\textwidth]{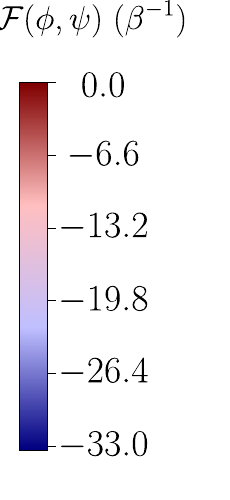}};
\end{tikzpicture}
\caption{Evolution of free energy of alanine dipeptide during a 50 ns TT-Metadynamics simulation.
2D free energies are approximated as the negative of the instantaneous bias potential, and the examples shown immediately follow sketching (performed every 1 ns).
Top left: 1 ns; top right: 2 ns; bottom left: 10 ns; bottom right: 50 ns.}
\label{fig:alanine}
\end{figure}

Alanine dipeptide has free-energy minima at $(\phi,\psi)\approx (-2.5,+2.7)$, $(\phi,\psi)\approx (-1.4,+1.0)$, and $(\phi,\psi)\approx (+1.1,-0.7)$ with a barrier of approximately $10~k_{\text{B}}T$ between the latter two minima at 300 K.
TTs require a dimensionality of at least two, and we use TT-Metadynamics to obtain the bias potential as a function of $\phi$ and $\psi$.
The free energy can be estimated as the negative of the instantaneous bias potential.
The evolution of this free energy during a 50~ns TT-Metadynamics simulation is shown in Fig.~\ref{fig:alanine}.

\begin{figure}
\centering
\begin{tikzpicture}
\node at (0,0) {\includegraphics[width=0.45\textwidth]{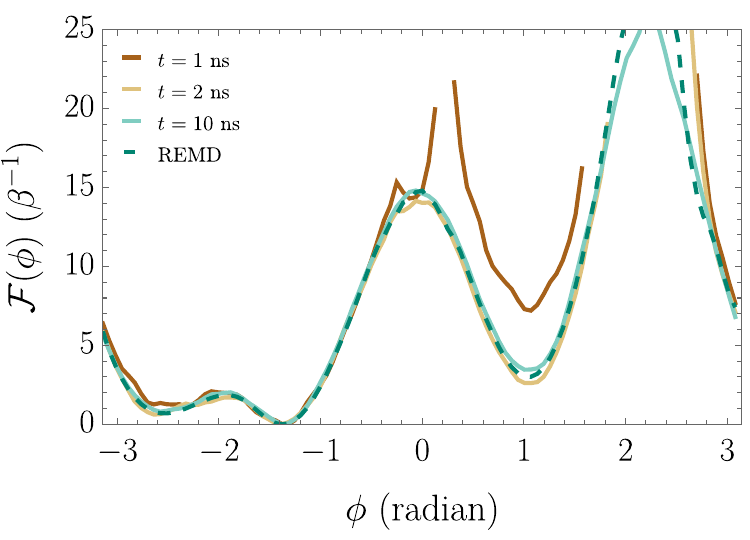}};
\node at (8.2,0) {\includegraphics[width=0.48\textwidth]{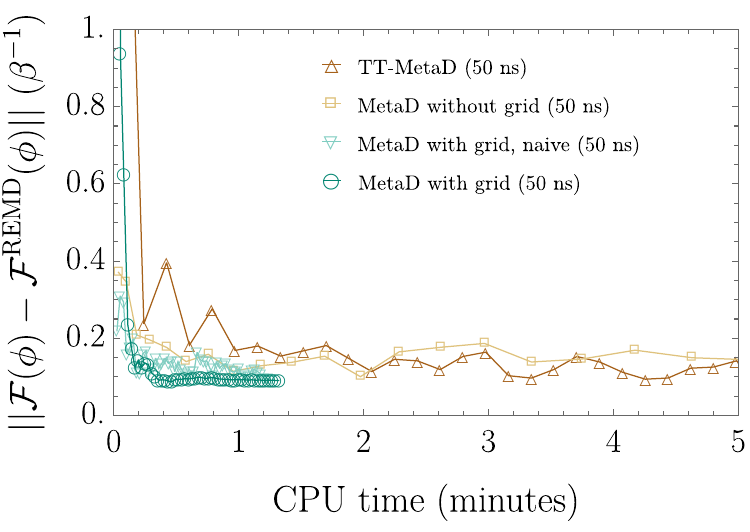}};
\node at (0,-5.8) {\includegraphics[width=0.45\textwidth]{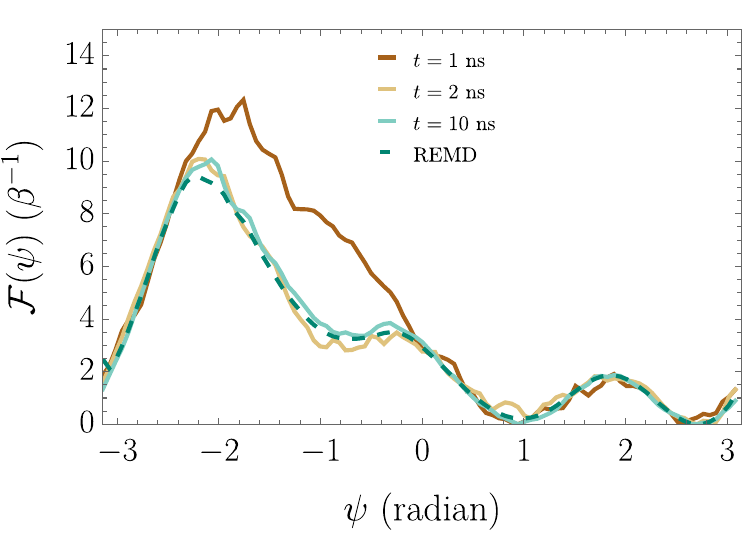}};
\node at (8.2,-5.8) {\includegraphics[width=0.48\textwidth]{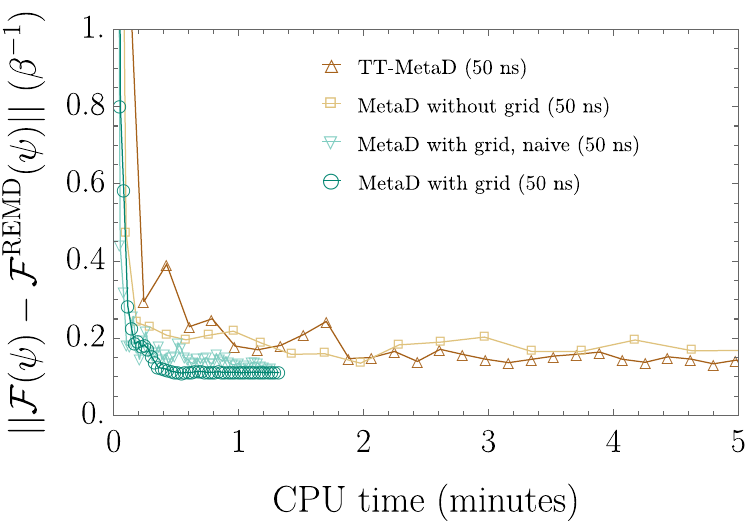}};
\end{tikzpicture}
\caption{1D free energy convergence for alanine dipeptide across metadynamics variants.
Left: evolution of free energies as functions of $\phi$ (top) and $\psi$ (bottom) from a TT-Metadynamics run.
Curves shown immediately follow sketching (performed every 1~ns).
For reference, the single dashed curve is the final free energy resulting from a replica exchange run.
Right: root mean square displacements (RMSD) between 1D free energy profiles from all four metadynamics versions ((a), (b), (c), and TT-Metadynamics) and those from replica exchange (see text).
The RMSDs are plotted as functions of the amount of CPU time (per simulation walker).
The primary reason RMSDs plateau around 0.1~$\beta^{-1}$ is an accumulation of errors in the estimated free energies within high-energy regions.
The Tiwary--Parrinello reweighting scheme \cite{tiwary2015time} was used for samples collected from version (a) of metadynamics.
The simple reweighting scheme was used for TT-Metadynamics and versions (b) and (c) of metadynamics.
}
\label{fig:alanine_1d}
\end{figure}

1D free energy profiles as a function of $\phi$ and $\psi$ were generated from TT-Metadynamics samples using the ``simple'' reweighting scheme discussed in Section \ref{sec:reweighting}.
We assess the effectiveness of the TT-Metadynamics bias by comparing these 1D curves with ones obtained from several regular metadynamics simulations (and a REMD reference, as described in the previous section).
Three versions of standard metadynamics were considered: (a) grid storage with Tiwary--Parrinello reweighting \cite{tiwary2015time}; (b) grid storage with simple reweighting; and (c) full kernel storage (no grid) with simple reweighting.
The relative accuracy and efficiency of all approaches are shown in Fig.~\ref{fig:alanine_1d}.
When $D$ is low, as it is here, grid lookup is fast, making bias potential computations using grid storage cheaper than those using full kernel storage or TT-Metadynamics.
Consequently, version (a) provides both the best accuracy and efficiency, and version (b) is second best on both counts; the latter converges more slowly because it introduces errors not present in the Tiwary--Parrinello scheme \cite{tiwary2015time}.
Version (c) and TT-Metadynamics achieve comparable accuracies  to methods (a) and (b) within 50~ns runs.
For this system, there is no clear advantage in using TTs over storing all kernels in memory.
These results validate that the TT approximation introduces negligible additional error relative to grid storage at low dimensionality, justifying its use as a drop-in replacement in higher-dimensional settings where grid storage becomes infeasible.

\subsection{Trialanine and ditryptophan}
\label{sec:6d8d}

\begin{figure}
\centering
\begin{tikzpicture}
\node at (0,0) {\includegraphics[width=0.45\textwidth]{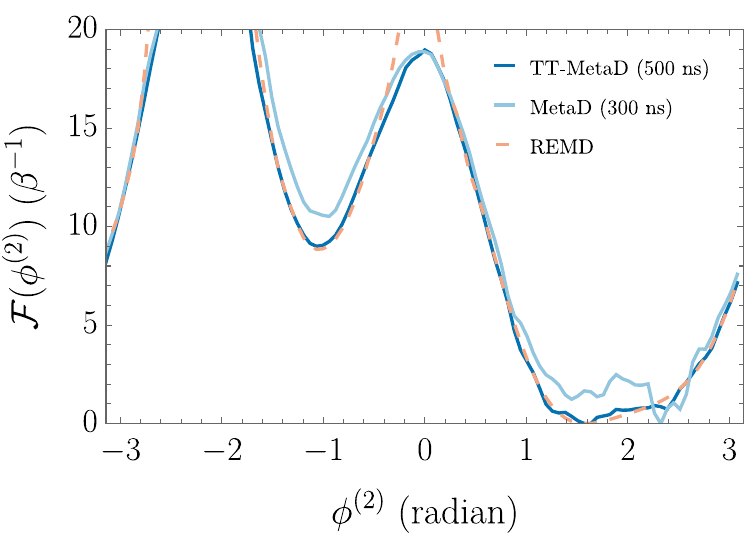}};
\node at (8.2,0) {\includegraphics[width=0.48\textwidth]{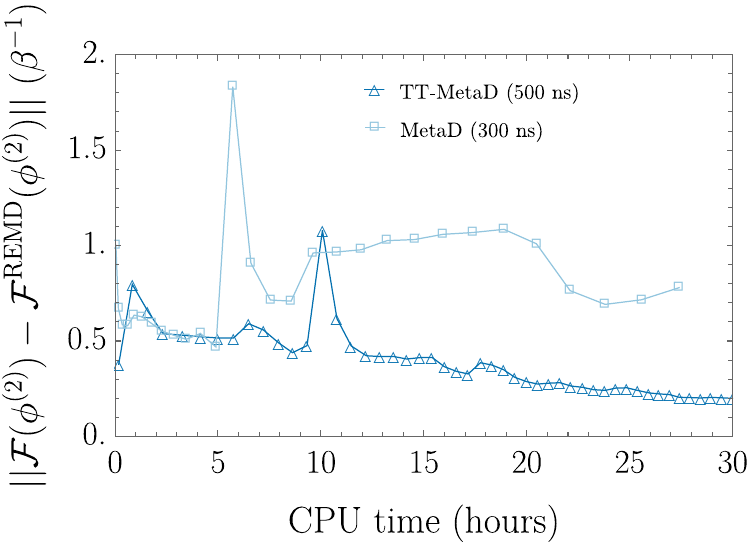}};
\node at (0,-5.8) {\includegraphics[width=0.45\textwidth]{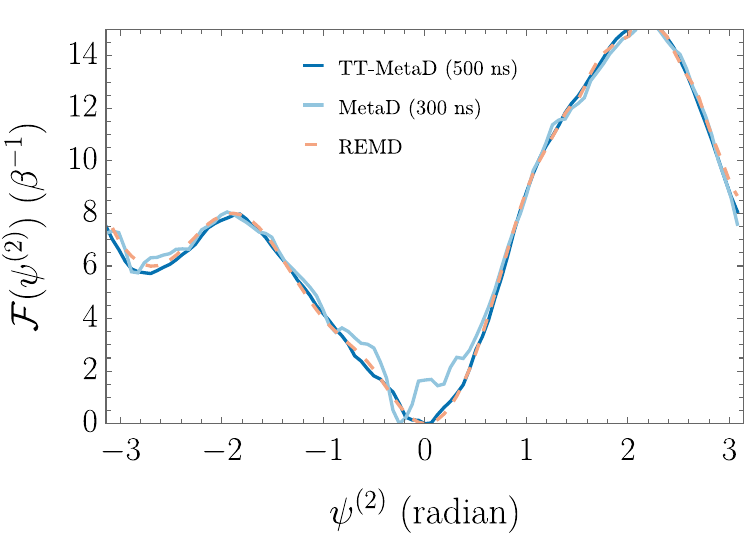}};
\node at (8.2,-5.8) {\includegraphics[width=0.48\textwidth]{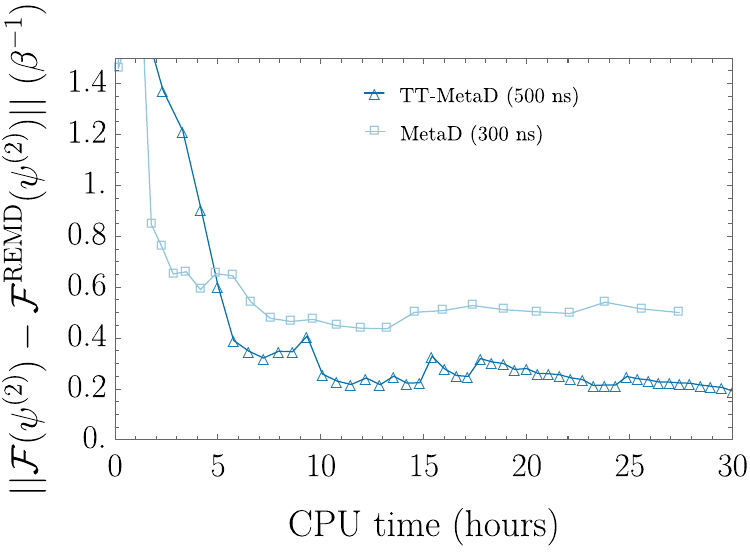}};
\end{tikzpicture}
\caption{1D free energy convergence for ditryptophan. 
Left: 1D free energy profiles as a function of the $\phi$ (top) and $\psi$ (bottom) of the second residue from a 300 ns metadynamics run and a 500 ns TT-Metadynamics run compared with reference replica exchange simulations.
Right: Root mean square deviations (RMSDs) of the indicated 1D free energy profiles from replica exchange results.
The data points shown are the results of free energies obtained at simulation snapshots separated by 10~ns.
The spike visible in the RMSD curves corresponds to the moment when the tall free energy barrier (${\sim}15\,k_{\rm B}T$) separating the two main basins is crossed for the first time: prior to crossing, the RMSD decreases steadily as the estimate within the initially explored basin improves; upon crossing, the RMSD transiently increases as the newly discovered basin is poorly sampled, before rapidly decreasing again as sampling accumulates.
}
\label{fig:ditryptophan}
\end{figure}

For both trialanine ($D=6$) and ditryptophan ($D=8$), grid storage is no longer feasible given our computational resources, so we can only compare version (c) and TT-Metadynamics.
As seen in Fig.~\ref{fig:ditryptophan} and  Figs.~\ref{fig:trialaninephi1}--\ref{fig:ditryptophanchi22} in the Supporting Information, TT-Metadynamics initially converges more slowly but outperforms standard metadynamics beyond roughly 150~ns for both systems.
As expected, the advantage of TT-Metadynamics grows with dimension (i.e., is more noticeable for ditryptophan than trialanine).
In Fig.~\ref{fig:ditryptophan}, both the dark blue (TT-Metadynamics) and light blue (standard metadynamics) curves show sharp spikes.
These correspond to the times when the simulations first crossed the barrier of $>15~k_{\text{B}}T$ at 0 radians and discovered the free energetic basin at around 1~radian.
Even though the spike occurs sooner in the standard metadynamics case, beyond these spikes, free energy curves from TT-Metadynamics consistently remain more accurate than ones from standard metadynamics.

For both trialanine and diptryptophan, accuracies of standard metadynamics-based free energies appear to plateau.
In addition to severely hampering the speed of the simulation, the boundless accumulation of Gaussian functions in memory likely causes a significant buildup of numerical precision errors.

\begin{figure}
\centering
\begin{tikzpicture}
\node at (0,0) {\includegraphics[width=0.45\textwidth]{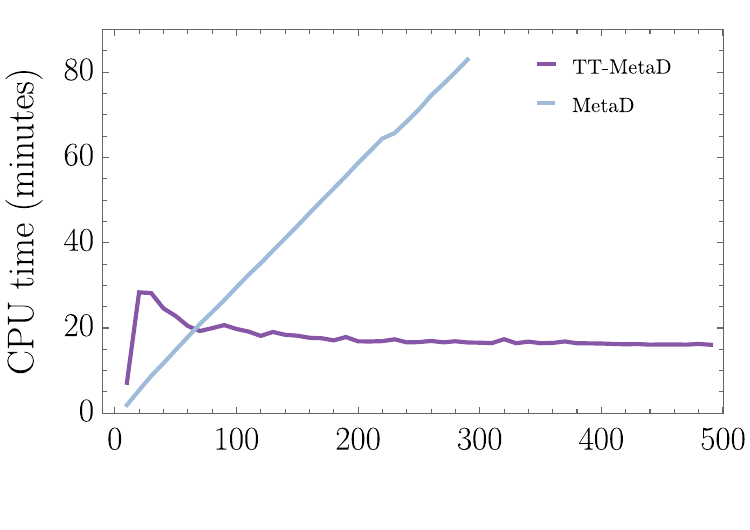}};
\node at (8.6,0) {\includegraphics[width=0.45\textwidth]{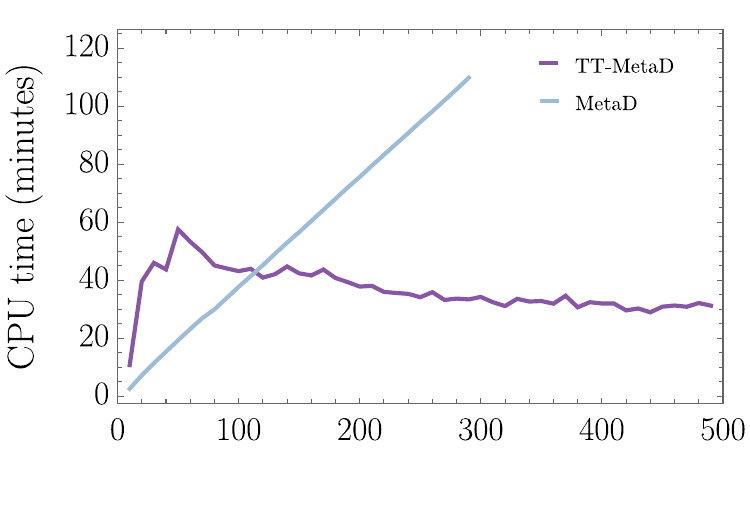}};
\node at (0,-4.5) {\includegraphics[width=0.44\textwidth]{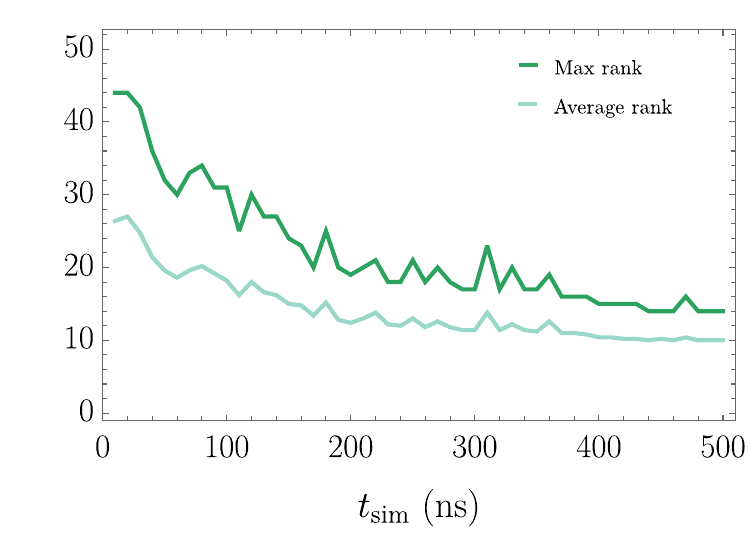}};
\node at (8.7,-4.5) {\includegraphics[width=0.43\textwidth]{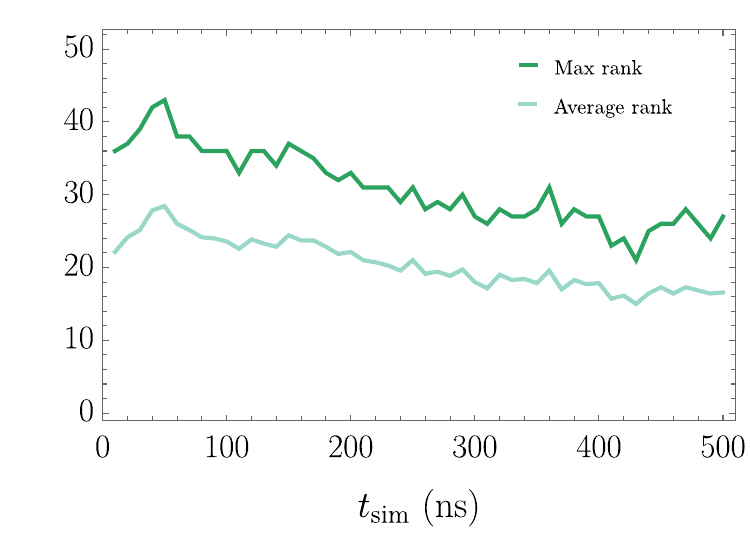}};
\end{tikzpicture}
\caption{Top: evolution of the computational time spent within every 10~ns-long simulation time window over the course of simulations of trialanine (left) and ditryptophan (right).
The MetaD times are for simulations with full Gaussian storage.
The TT-MetaD times include both the TT-Sketch step and MD integration with the TT-parametrized bias potential.
Bottom: evolution of the TT rank (max, average) over the course of simulations of trialanine (left) and ditryptophan (right).
Ranks shown are collected every 10 ns.}
\label{fig:timeevolution}
\end{figure}

A subtle feature of  the free energy curves obtained from the standard metadynamics simulations is that the computational cost per 10~ns window gradually increases.  This behavior can be observed in the spacing between the symbols in Fig.\  \ref{fig:ditryptophan}, and we plot it more clearly in the top panels of Fig.~\ref{fig:timeevolution}.
This behavior is expected, as the cost of evaluating the bias potential increases as the number of Gaussian functions increases.
By contrast, the computational cost per 10 ns window shrinks over the course of the TT-Metadynamics simulations.  
This behavior is due to the way the TT rank evolves.
As mentioned in Section~\ref{sec:ttsketch_algo}, the ranks are adaptively determined to satisfy a fixed relative SVD truncation tolerance during each sketching step and thus reflect the intrinsic complexity of the accumulated bias.
The rank initially increases as the complexity of the bias potential grows; once all minima are explored, the rank decreases, and the computational cost per window drops.
This gradual drop in rank can be used as a heuristic for sampling convergence.
This behavior is illustrated in the bottom panels of Fig.~\ref{fig:timeevolution}.

To assess the sensitivity of TT-Metadynamics to the kernel smoothing bandwidth $\boldsymbol\rho$, we ran two additional 500~ns ditryptophan simulations with $\rho=0$ (no smoothing) and $\rho=0.15$~rad (more aggressive smoothing than the default $\rho=0.05$~rad); RMSDs comparing potentials of mean force (PMF) from TT-MetaD with respect to those from the same REMD reference as before are provided in Figs.~\ref{fig:ditryptophanphi1_rho}--\ref{fig:ditryptophanchi22_rho} in the Supporting Information.
Without smoothing, free energy profiles for the majority of CVs slowly diverge from the REMD reference beyond 200~ns of simulation, indicating that the unsmoothed TT bias actively prevents convergence.
With moderate smoothing ($\rho=0.05$~rad), the TT-MetaD curves steadily approach the REMD reference.
The largest bandwidth considered ($\rho=0.15$~rad) yields similarly accurate results for most CVs, but performs worse for $\phi^{(2)}$, which separates two minima by a barrier of approximately $15~\beta^{-1}$; with the broader kernel, substantially more simulation time elapses before this barrier is crossed for the first time.
This behavior is consistent with the idea that too much smoothing prevents the bias potential from reflecting the features of the free-energy landscape.
We also note that larger $\boldsymbol\rho$ tends to produce modestly higher TT ranks over the course of the simulation, reflecting the slower convergence of the bias potential and correspondingly higher computational cost per sketching step.

In practice, a reference distribution is rarely available, so the bandwidth must be chosen based on simulation data alone.
We suggest the following heuristic for systems where at least one CV with a tall barrier can be identified in advance (e.g., $\phi^{(2)}$ for ditryptophan).
One should first run short simulations with a few values of $\boldsymbol\rho$ spanning a range from zero to a value comparable to the Gaussian deposition bandwidth $\boldsymbol\sigma$.
One should monitor the evolution of the free energy along the chosen CV and note the time of the first crossing of a tall barrier (or, equivalently, the spike in the corresponding RMSD curve, as seen in Fig.~\ref{fig:ditryptophan} for ditryptophan).
We expect this time to increase when $\boldsymbol\rho$ is too large and the kernel is too broad to capture the fine features and shape of the barrier.
We recommend setting $\boldsymbol\rho$ to the largest value that does not significantly delay the first barrier crossing to balance capturing fine features of the free energy with the suppression of high-frequency artifacts that arise at $\rho = 0$.

\subsection{Helical peptide AIB$_9$}

We tested TT-Metadynamics's ability to sample higher dimensional PMFs using AIB$_9$. This system is representative of many biomolecular and materials systems in that it has many local free-energy minima separating its main metastable states (left- and right-handed helices) \cite{buchenberg2015hierarchical,sittel2017principal,biswas2018metadynamics,mehdi2022accelerating,wang2022data,strahan2023inexact}.
We ran two TT-Metadynamics calculations:  one with a 10D bias potential that was a function of the $\phi$ and $\psi$ dihedral angles of residues 3 through 7 and another with a 14D bias potential that was a function of the $\phi$ and $\psi$ dihedral angles of residues 2 through 8.
These dimensions prohibit comparison with standard metadynamics.

\begin{figure}
\centering
\begin{tikzpicture}
\node at (0,0) {\includegraphics[width=0.39\textwidth]{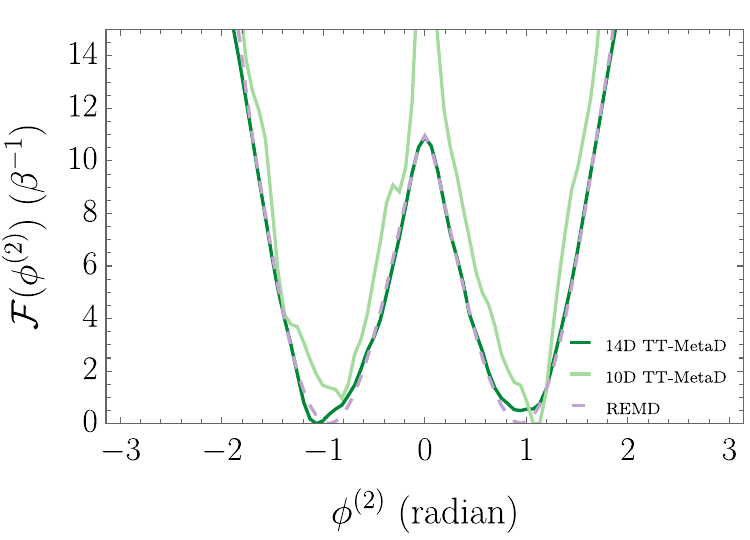}};
\node at (8,0) {\includegraphics[width=0.41\textwidth]{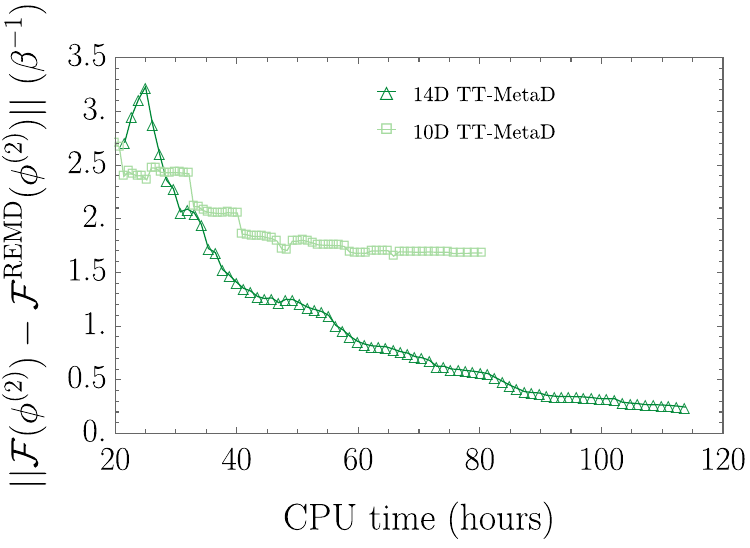}};
\node at (0,-4.9) {\includegraphics[width=0.39\textwidth]{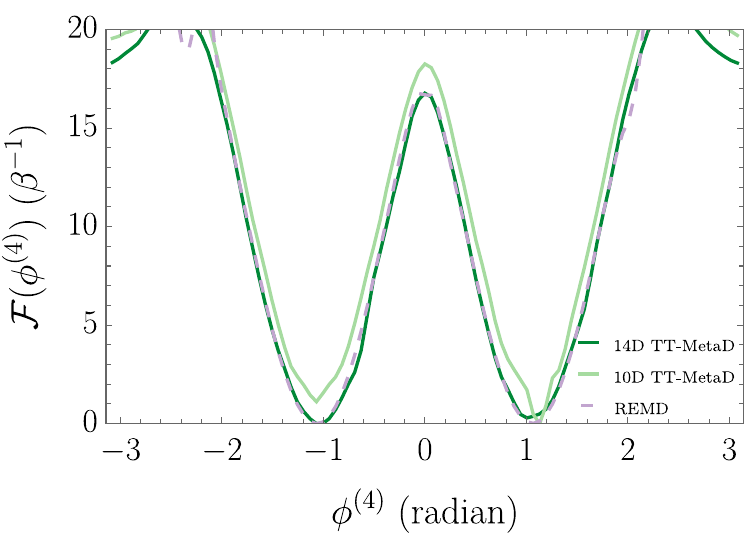}};
\node at (8,-4.9) {\includegraphics[width=0.41\textwidth]{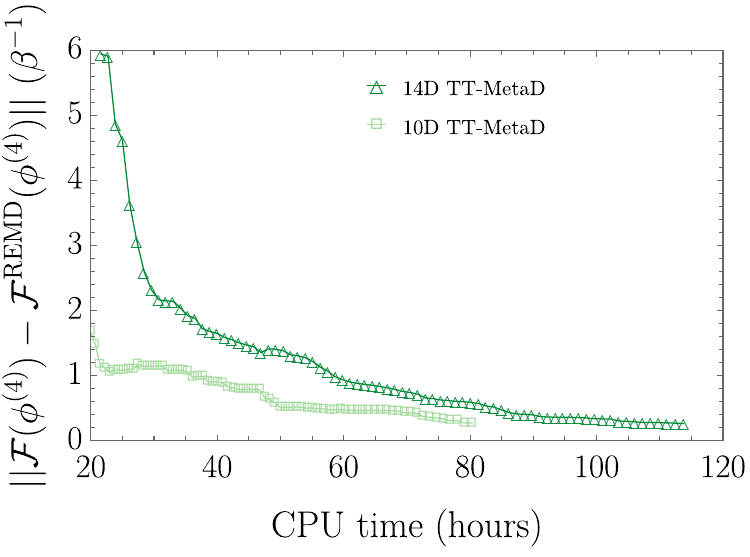}};
\node at (0,-9.8) {\includegraphics[width=0.39\textwidth]{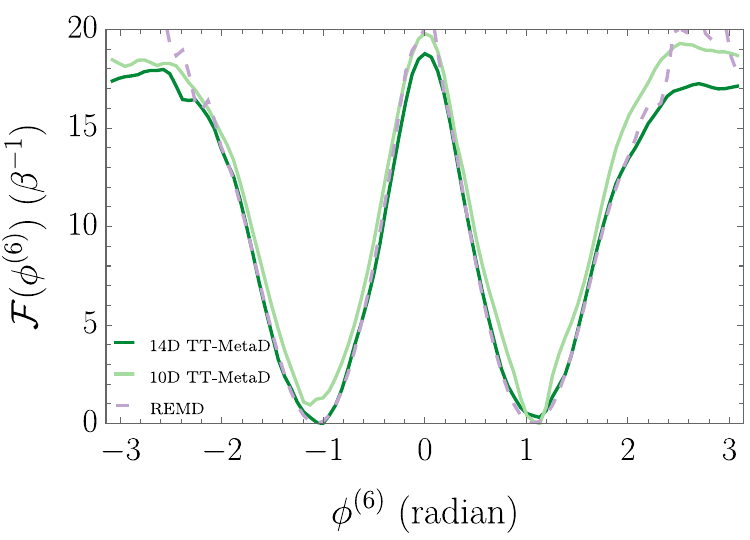}};
\node at (8,-9.8) {\includegraphics[width=0.41\textwidth]{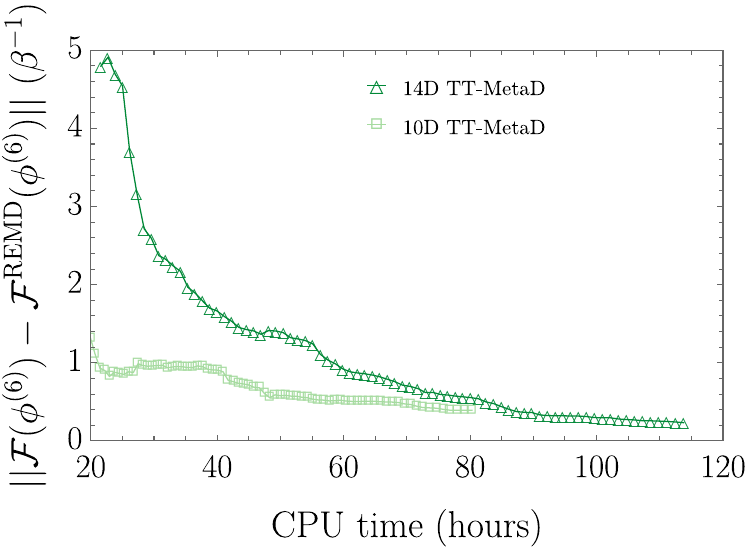}};
\node at (0,-14.7) {\includegraphics[width=0.39\textwidth]{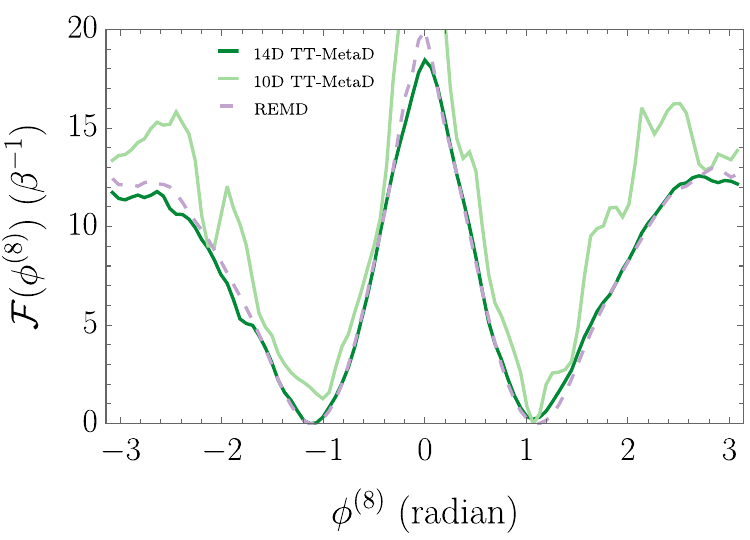}};
\node at (8,-14.7) {\includegraphics[width=0.41\textwidth]{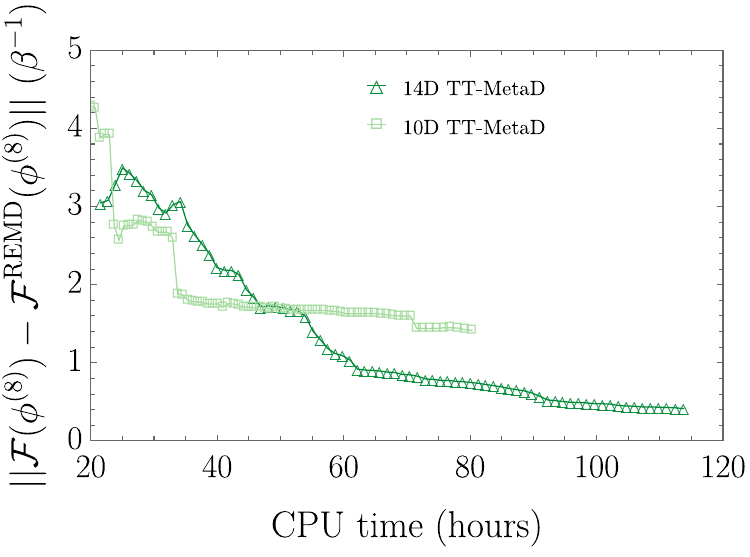}};
\end{tikzpicture}
\caption{Left: 1D free energy convergence for AIB$_9$.
Results shown are for selected dihedral angles ($\phi^{(2)}$, $\phi^{(4)}$, $\phi^{(6)}$, and $\phi^{(8)}$) after 1~$\upmu$s simulations.
The first 200~ns of samples are omitted from potential of mean force (PMF) estimates for both the 10D and the 14D cases.
Right: Root mean square deviations (RMSDs) of the indicated 1D free energy profiles from replica exchange results.
The data points shown are the results of free energies obtained at simulation snapshots separated by 10~ns.
}
\label{fig:aib9}
\end{figure}

\begin{figure}
\centering
\begin{tikzpicture}
\node at (0,0) {\includegraphics[width=0.42\textwidth]{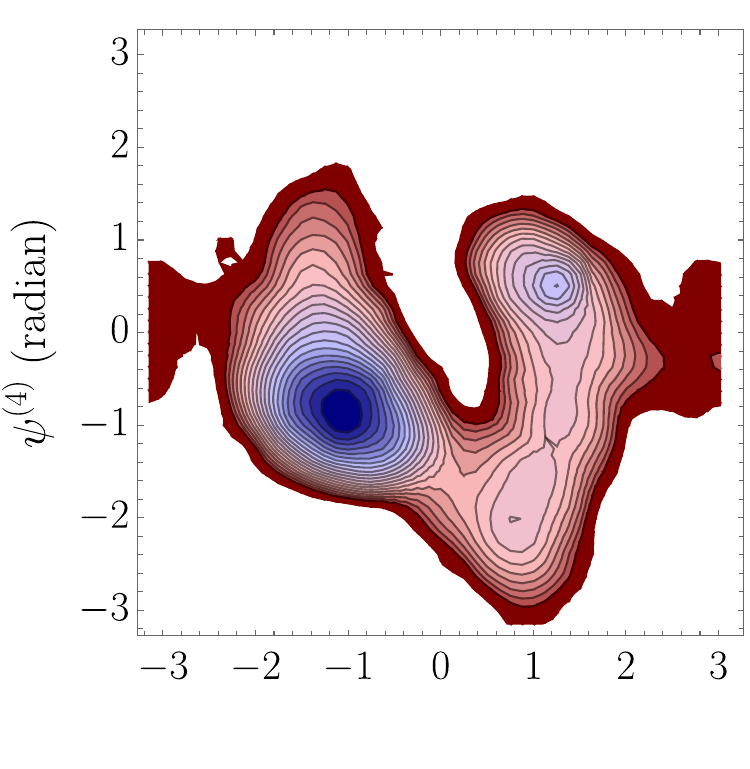}};
\node at (7.3,0) {\includegraphics[width=0.42\textwidth]{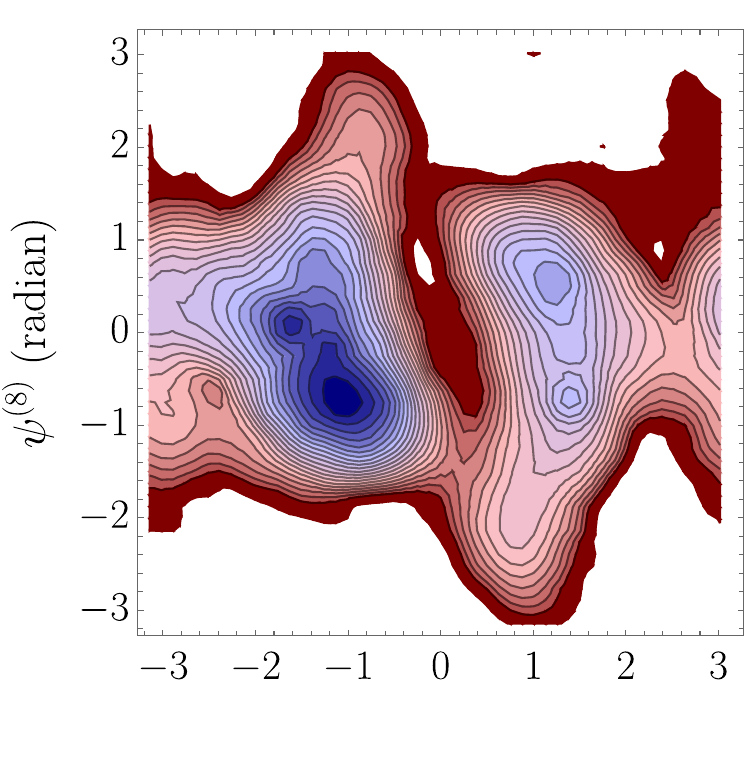}};
\node at (0,-6.6) {\includegraphics[width=0.42\textwidth]{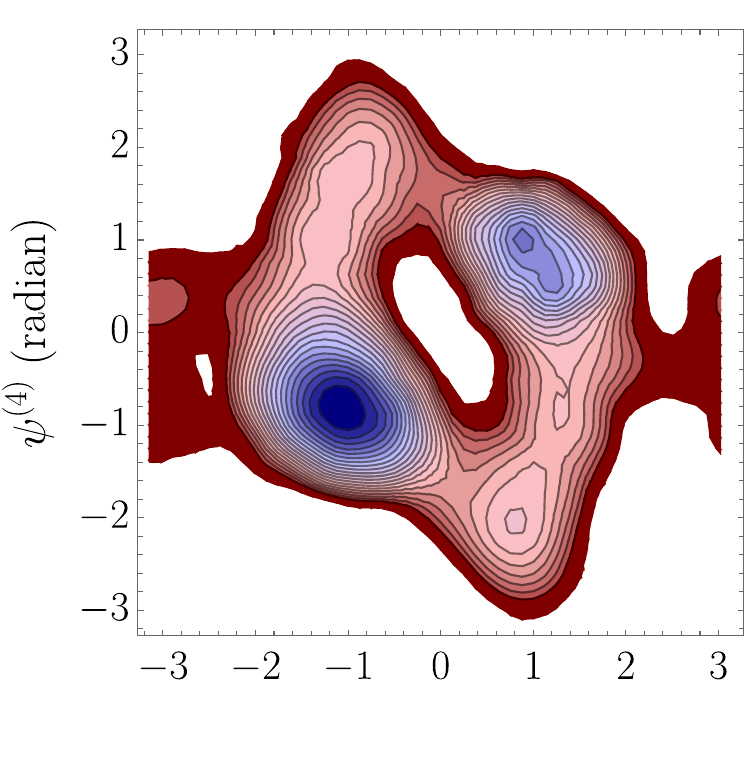}};
\node at (7.3,-6.6) {\includegraphics[width=0.42\textwidth]{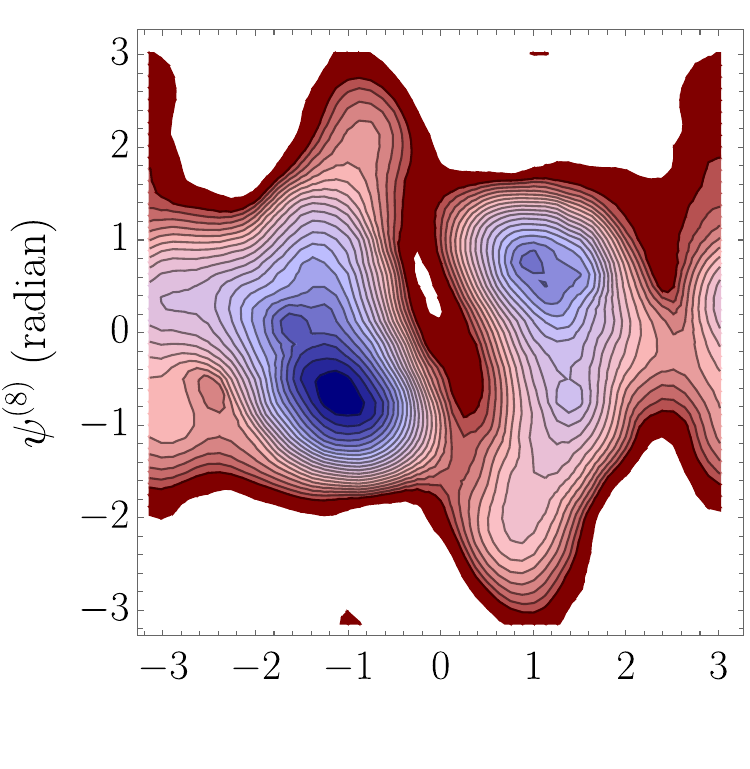}};
\node at (0,-13.2) {\includegraphics[width=0.42\textwidth]{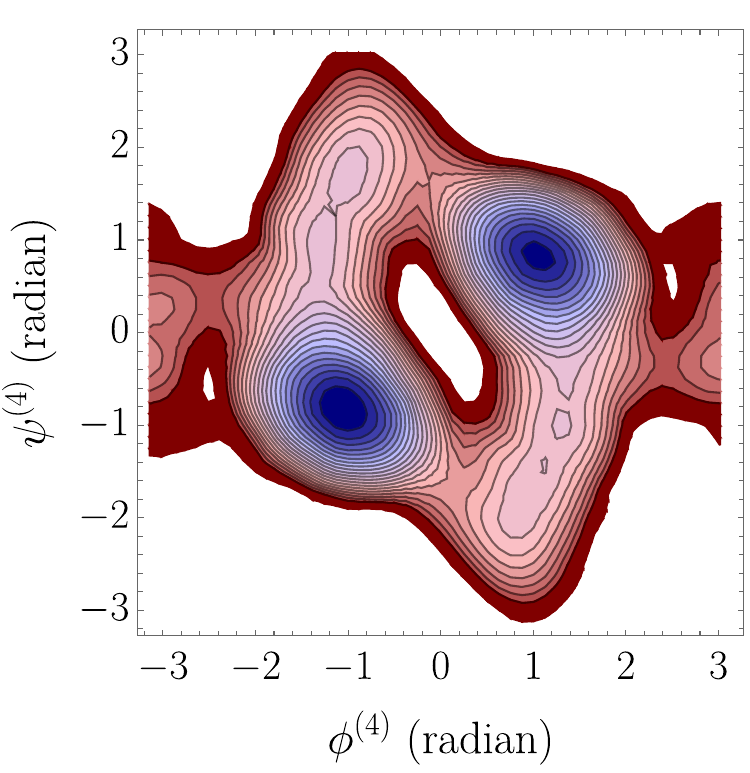}};
\node at (7.3,-13.2) {\includegraphics[width=0.42\textwidth]{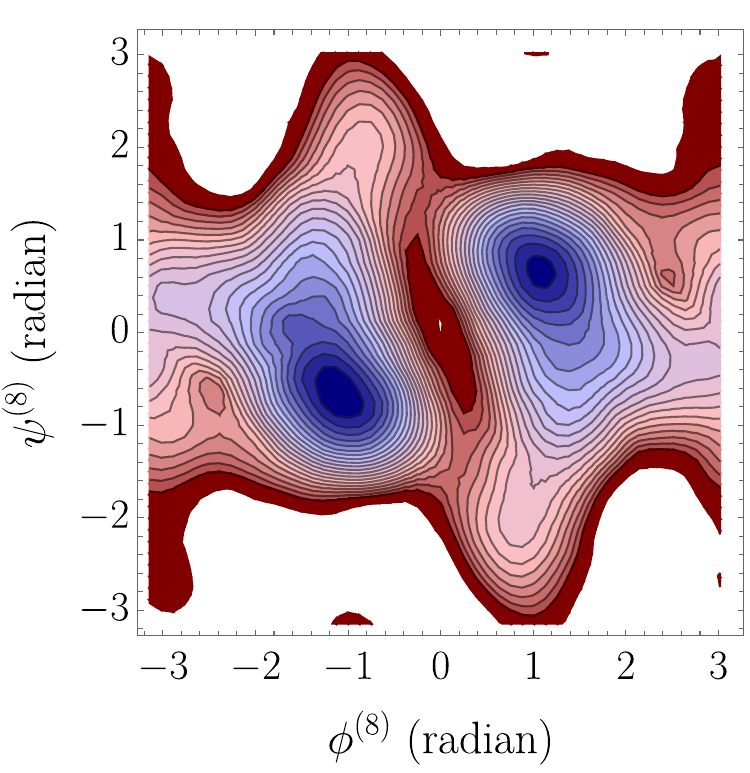}};
\node at (12.5,-6) {\includegraphics[width=0.18\textwidth]{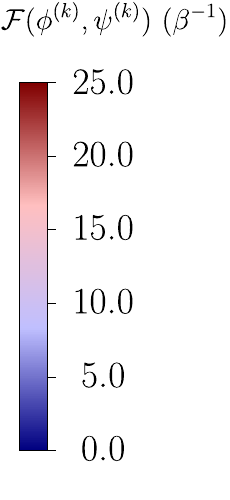}};
\node at (2.2,2.5) {\Large100~ns};
\node at (2.2,-4.1) {\Large300~ns};
\node at (2.2,-10.7) {\Large1~$\upmu$s};
\end{tikzpicture}
\caption{2D free energy convergence for AIB$_9$. Results shown are for selected pairs of dihedral angles  $(\phi^{(k)},\psi^{(k)})$ for $k=4,8$ at various snapshots of the 1~$\upmu$s-long 14D-biased simulation involving AIB$_9$, showcasing the gradual progress of the conformational space exploration. Top: 100~ns; middle: 300~ns; bottom: 1~$\upmu$s.
No samples were omitted for these potential of mean force (PMF) estimates.}
\label{fig:aib9_2d}
\end{figure}

1D PMFs as functions of selected dihedral angles indicate that the 14D bias potential outperforms the 10D bias potential within 1 $\upmu$s considered (Figs.~\ref{fig:aib9} and \ref{fig:aib9_2d}).
The ranks at the end of the simulations are lower than those at the halfway point, mirroring the behavior in Fig.~\ref{fig:timeevolution}.
Interestingly, although the 14D simulation was more costly due to the linear scaling of the TT evaluations with dimension, the TT ranks are consistently smaller in the 14D case than in the 10D case.
For example, as depicted in Fig.~\ref{fig:aib9evolution} in the Supporting Information, at the halfway points (ends) of the simulations, the maximum ranks of the TTs for the simulations with the 10D and 14D bias potentials are 26 and 18 (23 and 16), respectively.
One possible explanation is that the simulation with the 14D bias potential, unlike that with the 10D bias potential, is able to anneal artifacts and, in turn, decrease the complexity of the bias potential.

\section{Conclusions}
\label{sec:conclusion}

We have shown that TT-Metadynamics provides a scalable adaptive biasing scheme that outperforms standard metadynamics \cite{laio2002escaping} for bias potentials in 6D and beyond.
Simulations with compressed representations of the bias potential consistently outperformed those with storage of the full sum over kernels. For 2D bias potentials, grid representations are preferred because they require minimal overhead.  However, as the dimension of the bias potential increases, grid storage quickly becomes infeasible, and TTs are effective alternatives.
Key elements of our implementation are the use of an efficient \emph{sketching} procedure and \emph{smoothing} of the TT bias potential with a Gaussian kernel.
We find that the TT rank initially increases as modes are discovered, and then decreases as artifacts are annealed; its stabilization can be used as a heuristic for convergence.
Owing to the overhead associated with the TT, we expect TT-Metadynamics to be more competitive for systems with computationally costly force evaluations (e.g., systems larger than those considered here and/or simulations based on  machine-learned interatomic potentials and/or quantum chemical calculations).

A practical question that arises when biasing along many CVs is how to interpret the resulting high-dimensional free energy.  In the present work, we visualized the results through 1D and 2D projections on dihedral angle variables.
Although the choice of directions onto which to project may not be obvious in less well-characterized systems, it need not be made in advance: because TT-Metadynamics produces a converged, reweighted ensemble of configurations, standard unsupervised dimensionality reduction methods, such as PCA or diffusion maps, can be applied to the collected samples \emph{after} the simulation to identify the directions of greatest variance or slowest relaxation.
These, in turn, can be compared with physically motivated CVs.  We thus advocate a two-step strategy: sample in a high-dimensional CV space to ensure convergence and then reduce the dimension based on the data for interpretation.

One can imagine various extensions to our approach.
It would be worth considering other tensor decompositions \cite{ehrlacher2022adaptive}  and approaches that combine TT-Metadynamics with other techniques.
An example of the latter would be to seed the bias potential for TT-Metadynamics with that obtained either from bias-exchange metadynamics \cite{piana2007bias,marinelli2009kinetic} or from parallel-bias metadynamics~\cite{pfaendtner2015efficient}.
It may also be possible to use TTs to estimate other quantities, such as the coefficient $c(t)$ of the Tiwary--Parrinello approach \cite{tiwary2015time}.
Another way to incorporate robust reweighting would be to develop a TT variant of on-the-fly probability enhanced sampling (OPES), which differs from metadynamics in that the bias potential is built from probability densities obtained from \emph{weighted} kernel density estimation \cite{invernizzi2020rethinking}.  
Overall, TT-Metadynamics offers a practical and scalable framework for high-dimensional enhanced sampling.

\begin{acknowledgement}

We thank Miles Stoudenmire, Michael Lindsey, and Jonathan Weare for insightful discussions, as well as Pratyush Tiwary and Shams Mehdi for providing starting structures and simulation files for AIB$_9$.
This work was supported primarily by National Institutes of Health award R35 GM136381 and National Science Foundation (NSF) award DMS-2111563. This project was also supported in part by the Margot and Tom Pritzker Science Foundation.
Y.K. was partially funded by NSF DMS-2339439, DOE DE-SC0022232, and a Sloan Research Fellowship.  
The authors acknowledge the University of Chicago Research Computing Center for computational resources and
the National Institute for Theory and Mathematics in Biology (NITMB), supported by grants from the NSF (DMS-2235451) and Simons Foundation (MPS-NITMB-00005320). 

\end{acknowledgement}

\begin{suppinfo}

The following figures and tables are available free of charge.
\begin{itemize}
  \item Table~\ref{timescales}: Metadynamics-related parameters for all systems studied
  \item Table~\ref{temps}: Temperatures used for all replica exchange molecular dynamics (REMD) simulations
  \item Fig.~\ref{fig:sketch3}: Tensor diagram depiction of core-determining equations during TT-Sketch
  \item Fig.~\ref{fig:trim}: Tensor diagram depiction of the trimming step of TT-Sketch
  \item Figs.~\ref{fig:trialaninephi1}--\ref{fig:ditryptophanchi22}: 1D free energy convergence and root mean square deviations (RMSD) for trialanine and ditryptophan
  \item Figs.~\ref{fig:aib9phi1}--\ref{fig:aib9psi9}: 1D free energy convergence and root mean square deviations (RMSD) for AIB$_9$
  \item Figs.~\ref{fig:aib9_2d_1}--\ref{fig:aib9_2d_9}: 2D free energy convergence of $(\phi^{(k)},\phi^{(k)})$ for $k=1,\dots,9$ for AIB$_9$
  \item Figs.~\ref{fig:ditryptophanphi1_rho}--\ref{fig:ditryptophanchi22_rho}: 1D free energy convergence and root mean square deviations (RMSD) for ditryptophan as a function of the kernel smoothing bandwidth $\rho$
  \item Fig.~\ref{fig:aib9evolution}: Evolution of the TT ranks during simulations of AIB$_9$ using both 10D and 14D bias potentials
\end{itemize}

\end{suppinfo}

%\bibliography{biblio}
\providecommand{\latin}[1]{#1}
\makeatletter
\providecommand{\doi}
  {\begingroup\let\do\@makeother\dospecials
  \catcode`\{=1 \catcode`\}=2 \doi@aux}
\providecommand{\doi@aux}[1]{\endgroup\texttt{#1}}
\makeatother
\providecommand*\mcitethebibliography{\thebibliography}
\csname @ifundefined\endcsname{endmcitethebibliography}  {\let\endmcitethebibliography\endthebibliography}{}

\pagebreak

\setcounter{figure}{0}  % 1=section, 2=subsection, 3=subsubsection
\setcounter{table}{0}  

\renewcommand{\thesection}{S\arabic{section}}
\renewcommand{\thefigure}{S\arabic{figure}}
\renewcommand{\theequation}{\thesection\arabic{equation}}
\renewcommand{\thetable}{S\arabic{table}}

\begin{table}[htb]
    \begin{tabular}{llllll}
        \hline
        System & $t_{\text{sim}}$ (ns) & $\gamma$ & $\sigma$ (radians) & $h_0$ (kJ/mol) & $\rho$ (radians) \\
        \hline\hline
        Alanine dipeptide & 50 & 8 & 0.25 & 1 & 0 \\
        Trialanine & 500 & 8 & 0.3 & 1 & 0.05 \\
        Ditryptophan & 500 & 8 & 0.35 & 1 & 0.05 \\
        AIB$_9$ (10D) & $10^3$ & 6 & 0.4 & 1 & 0.05 \\
        AIB$_9$ (14D) & $10^3$ & 6 & 0.4 & 0.6 & 0.05 \\
        \hline
    \end{tabular}
    \caption{Simulation times and TT-Metadynamics parameters for all systems studied.}
    \label{timescales}
\end{table}

\begin{table}[htb]
    \begin{tabular}{llll}
        \hline
        Alanine dipeptide & Trialanine & Ditryptophan & AIB$_9$ \\
        \hline\hline
        300 & 300 & 300 & 300 \\
        357 & 341 & 334 & 323 \\
        424 & 387 & 372 & 348 \\
        503 & 439 & 414 & 373 \\
        596 & 498 & 460 & 401 \\
        697 & 564 & 511 & 430 \\
        798 & 639 & 567 & 462 \\
        899 & 723 & 629 & 497 \\
        1000 & 818 & 698 & 534 \\
        & 919 & 774 & 574 \\
        & 1000 & 858 & 617 \\
        & & 900 & 663 \\
        & & & 712 \\
        & & & 764 \\
        & & & 820 \\
        & & & 880 \\
        & & & 944 \\
        & & & 1000 \\
        \hline
    \end{tabular}
    \caption{Temperatures (in K) used for temperature replica exchange molecular dynamics (REMD) simulations.}
    \label{temps}
\end{table}

\begin{figure}[bt]
    \centering
    \subfloat[$G_1$-determining equation]{{\includegraphics[width=0.3\textwidth]{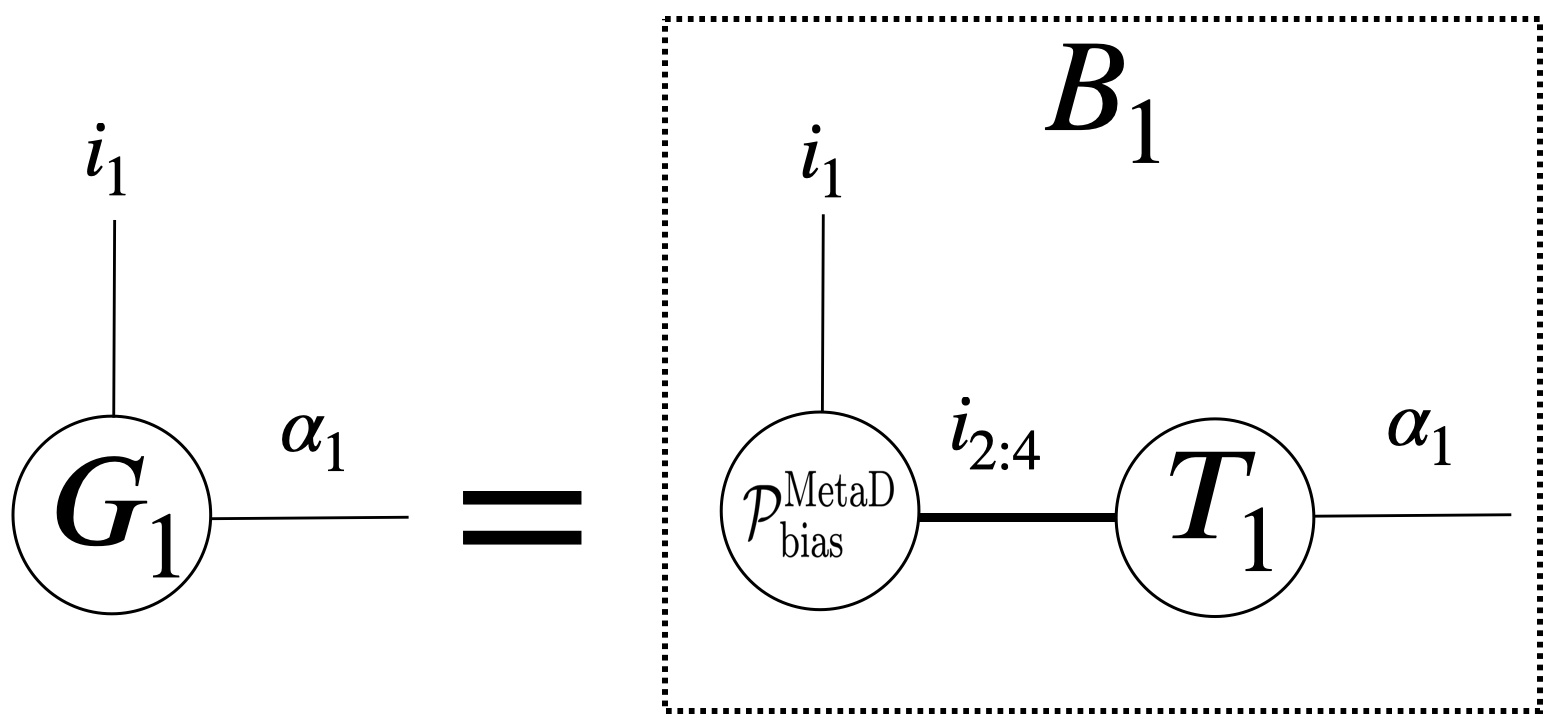}}}
    \qquad  \qquad  
    \subfloat[$G_2$-determining equation]{{\includegraphics[width=0.45\textwidth]{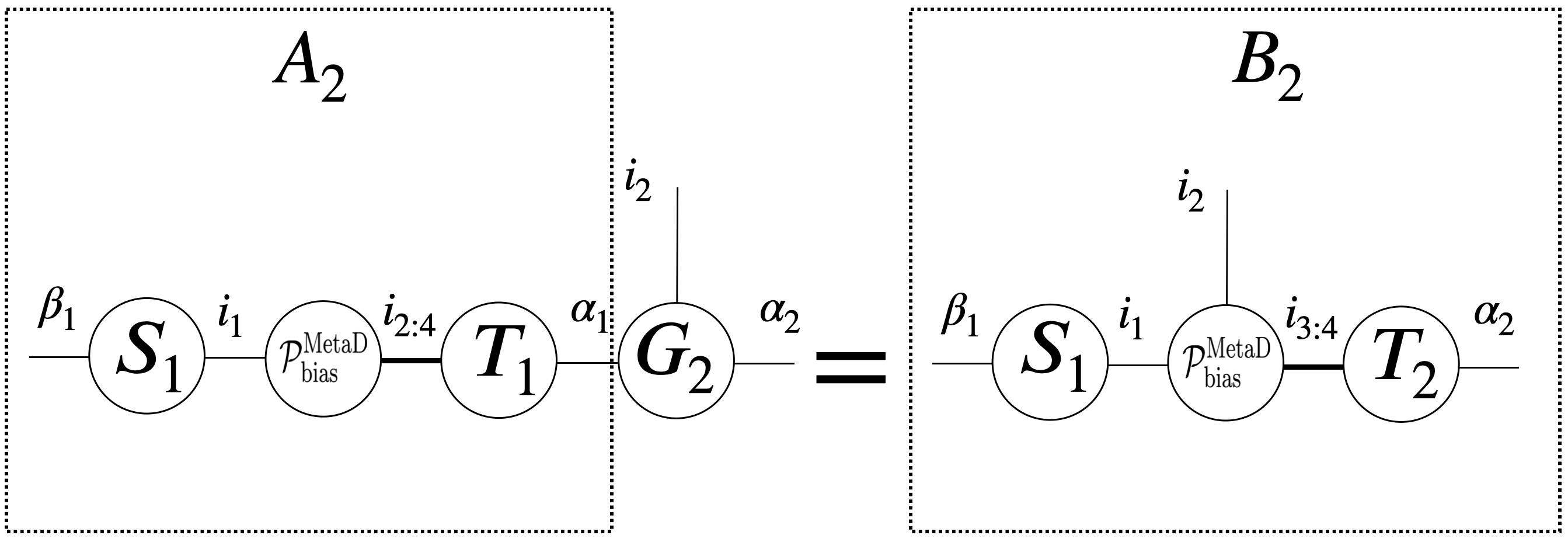}}}\\[20pt]
    \subfloat[$G_3$-determining equation]{{\includegraphics[width=0.45\textwidth]{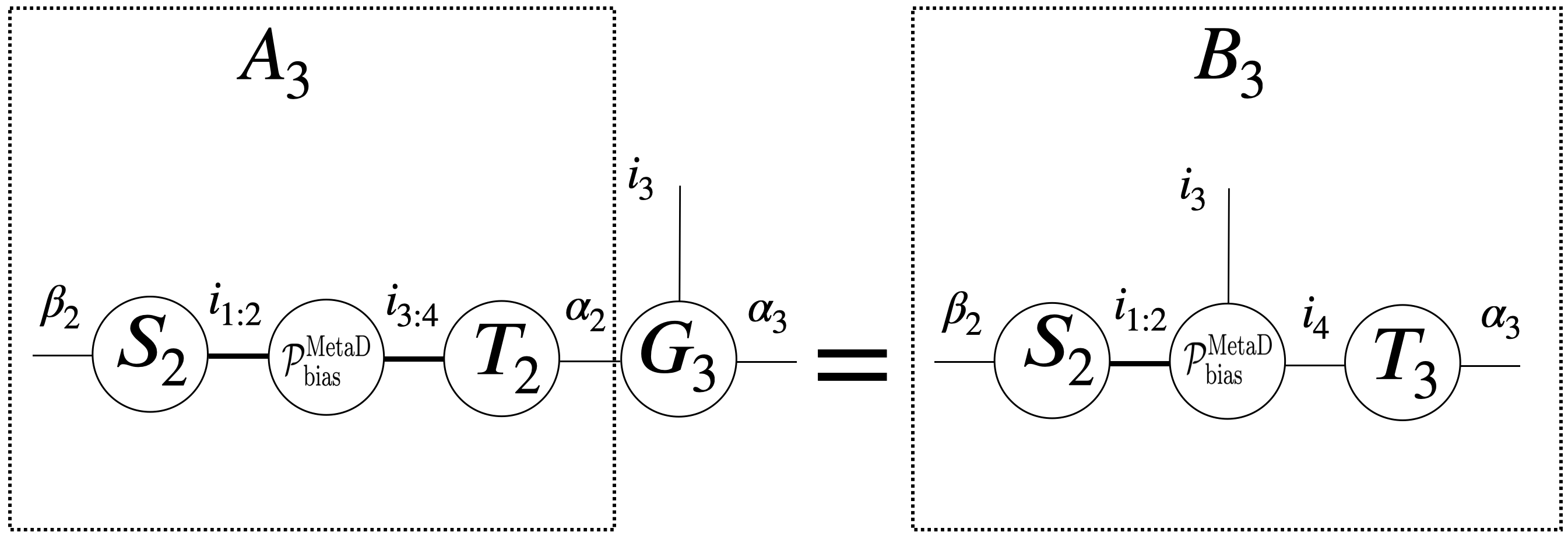}}}
    \qquad  \qquad 
    \subfloat[$G_4$-determining equation]{{\includegraphics[width=0.35\textwidth]{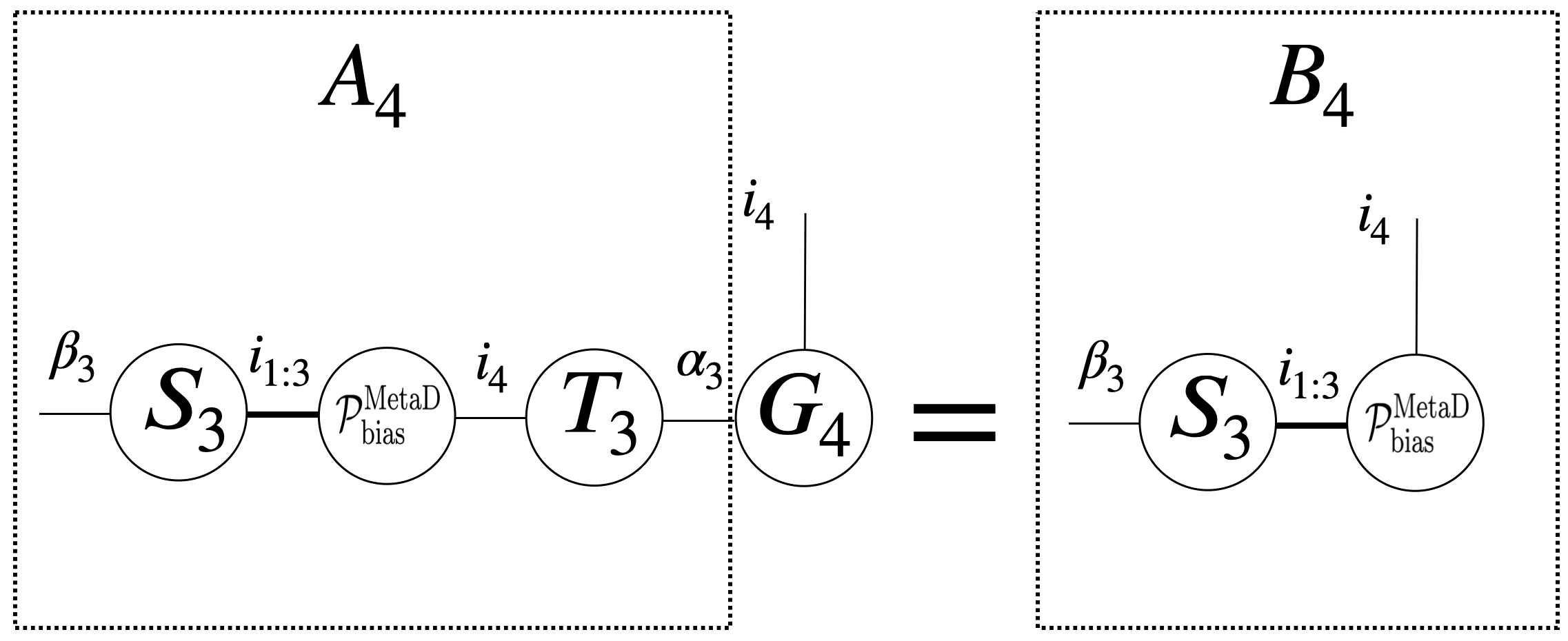}}}
    \caption{Tensor diagrams illustrating the core-determining equations for $D=4$. The four equations can be solved in parallel using least squares optimization, except the first core $G_1$ can be directly assigned as shown in (a). The total cost for solving these equations is linear in $D$ with suitable choices of sketch matrices $S_k$ and $T_k$.}\label{fig:sketch3}    
\end{figure}

\begin{figure}[ht!]
\centering
\includegraphics[width=\textwidth]{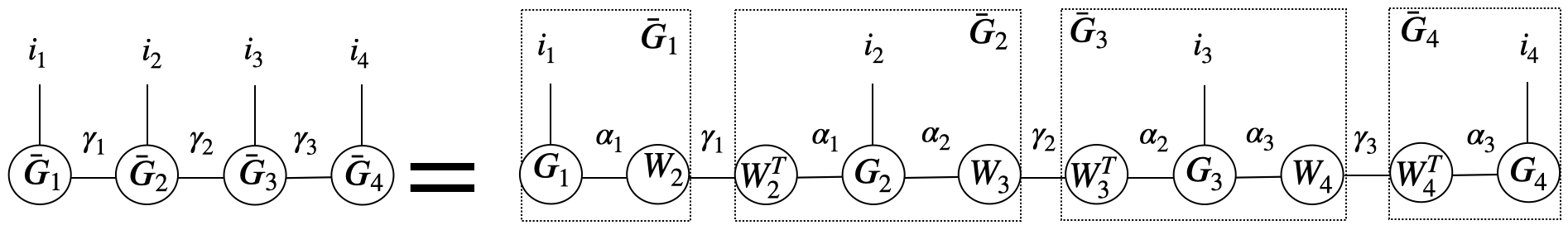}
\caption{Tensor diagram illustrating the trimming step for $D = 4$. $\bar{G}_1,\dots, \bar{G}_D$ form the final TT estimation for the coefficient $\mathcal{P}_{\text{bias}}$ in the ansatz given by Eq.~\eqref{functional_TT} for bias potential.}
\label{fig:trim}
\end{figure}

\clearpage

\begin{figure}
\centering
\begin{tikzpicture}
\node at (0,0) {\includegraphics[width=0.45\textwidth]{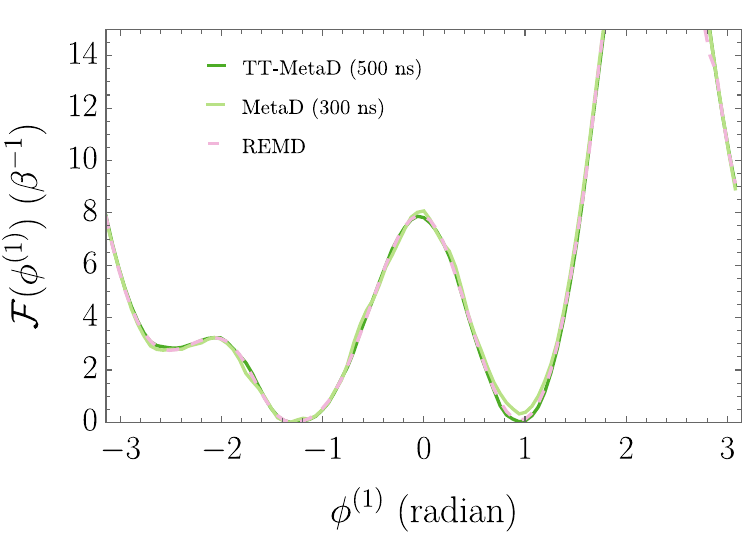}};
\node at (8.2,0) {\includegraphics[width=0.48\textwidth]{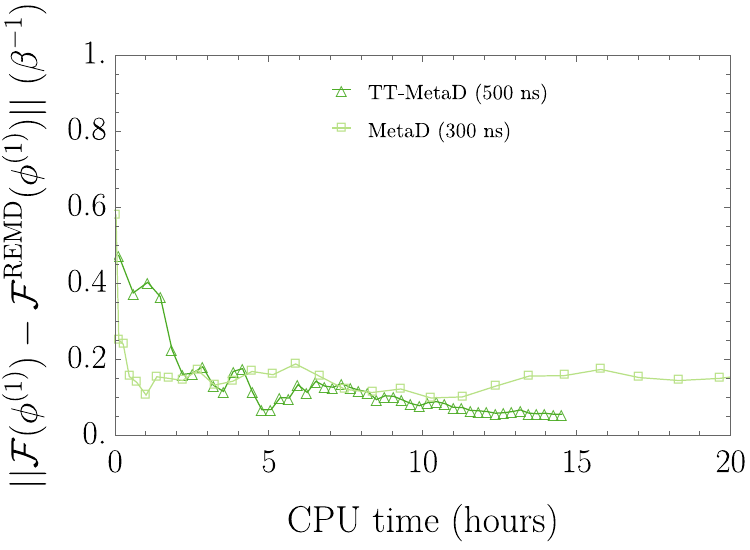}};
\end{tikzpicture}
\caption{1D free energy convergence and root mean square deviations (RMSD) of $\phi^{(1)}$ for trialanine.}
\label{fig:trialaninephi1}
\end{figure}

\begin{figure}
\centering
\begin{tikzpicture}
\node at (0,0) {\includegraphics[width=0.45\textwidth]{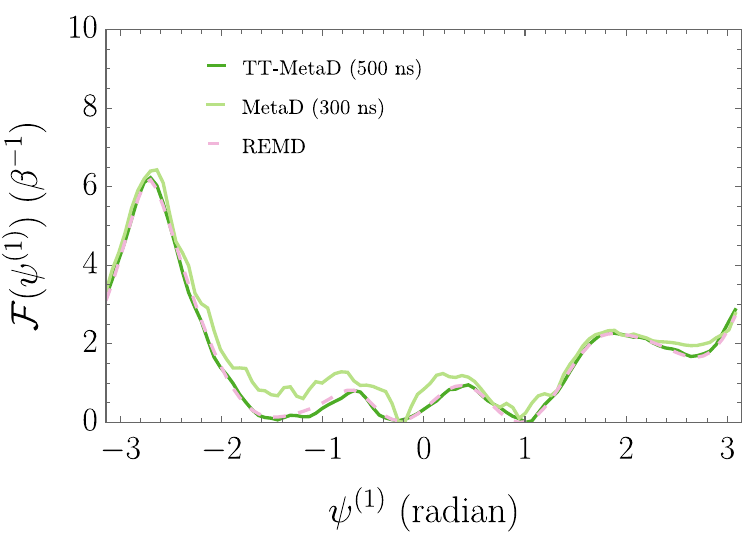}};
\node at (8.2,0) {\includegraphics[width=0.48\textwidth]{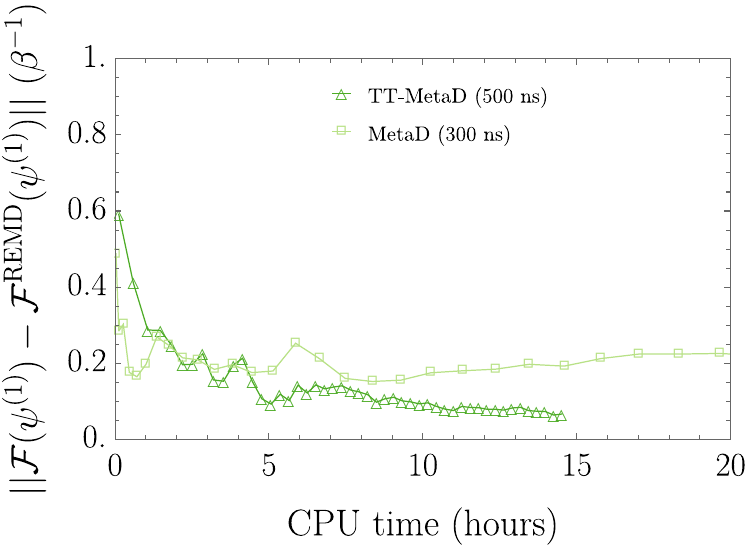}};
\end{tikzpicture}
\caption{1D free energy convergence and root mean square deviations (RMSD) of $\psi^{(1)}$ for trialanine.}
\label{fig:trialaninepsi1}
\end{figure}

\begin{figure}
\centering
\begin{tikzpicture}
\node at (0,0) {\includegraphics[width=0.45\textwidth]{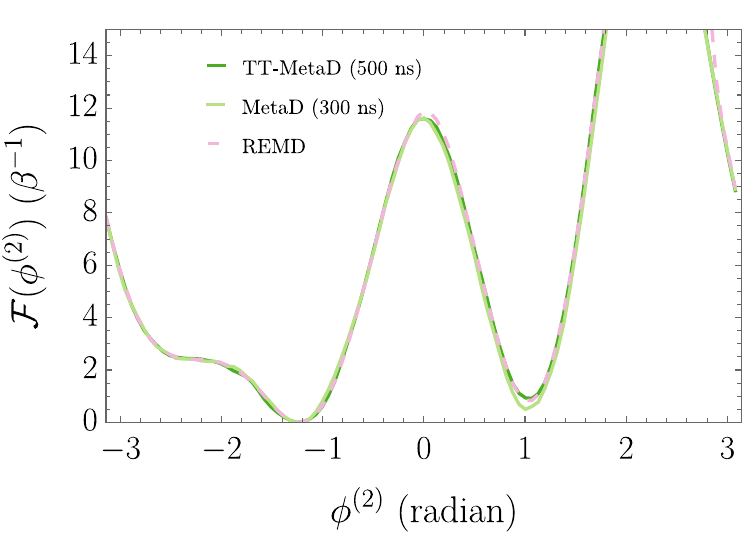}};
\node at (8.2,0) {\includegraphics[width=0.48\textwidth]{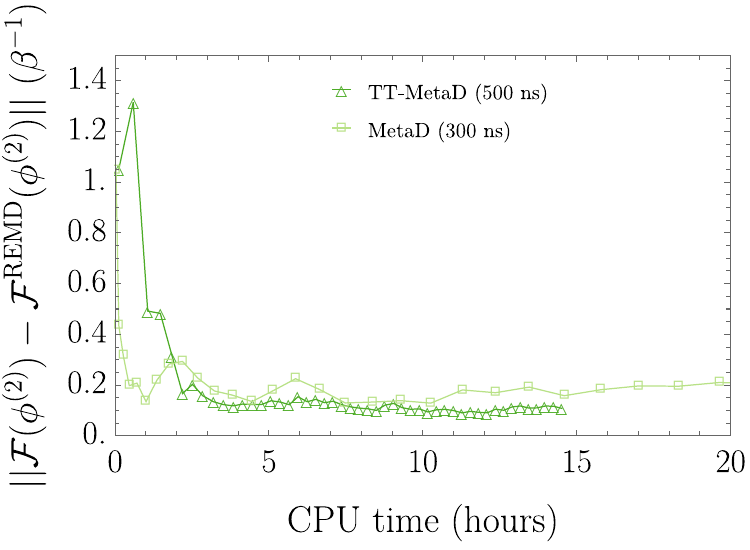}};
\end{tikzpicture}
\caption{1D free energy convergence and root mean square deviations (RMSD) of $\phi^{(2)}$ for trialanine.}
\label{fig:trialaninephi2}
\end{figure}

\begin{figure}
\centering
\begin{tikzpicture}
\node at (0,0) {\includegraphics[width=0.45\textwidth]{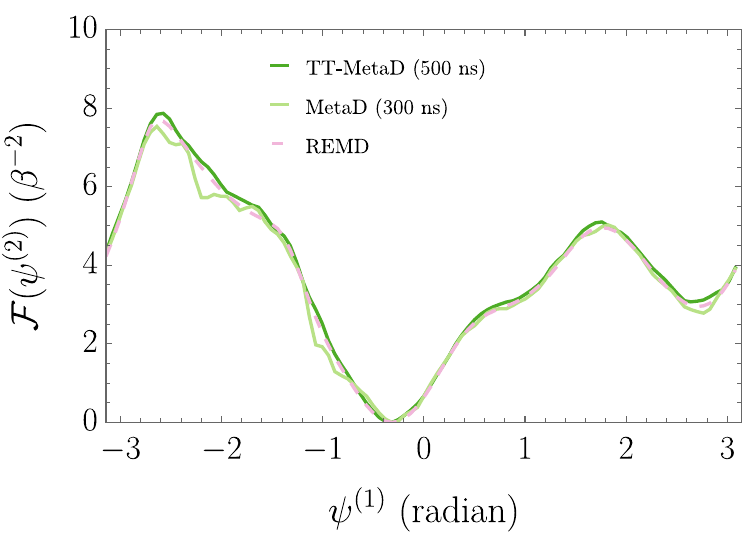}};
\node at (8.2,0) {\includegraphics[width=0.48\textwidth]{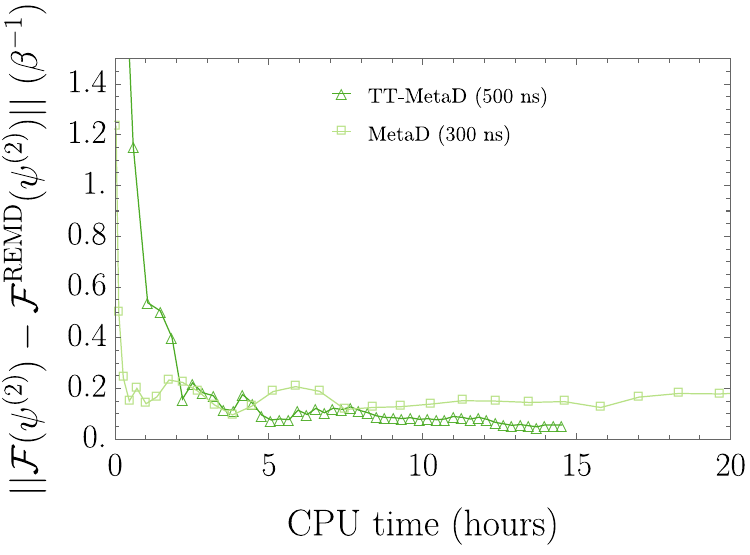}};
\end{tikzpicture}
\caption{1D free energy convergence and root mean square deviations (RMSD) of $\psi^{(2)}$ for trialanine.}
\label{fig:trialaninepsi2}
\end{figure}

\begin{figure}
\centering
\begin{tikzpicture}
\node at (0,0) {\includegraphics[width=0.45\textwidth]{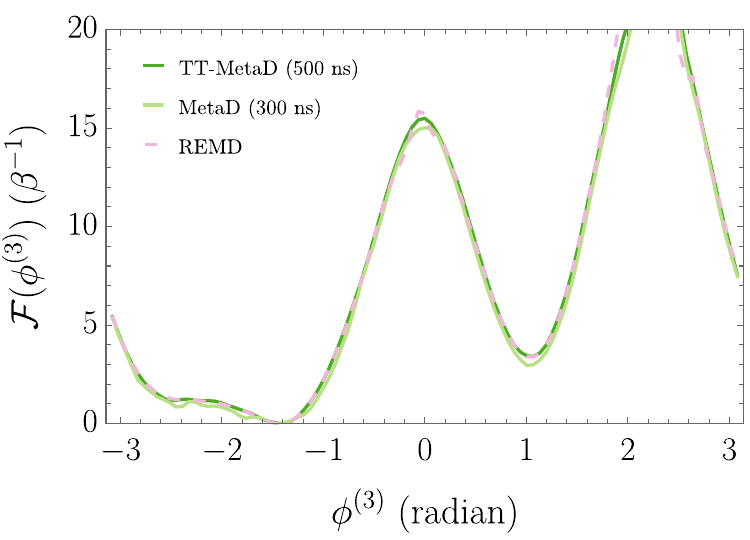}};
\node at (8.2,0) {\includegraphics[width=0.48\textwidth]{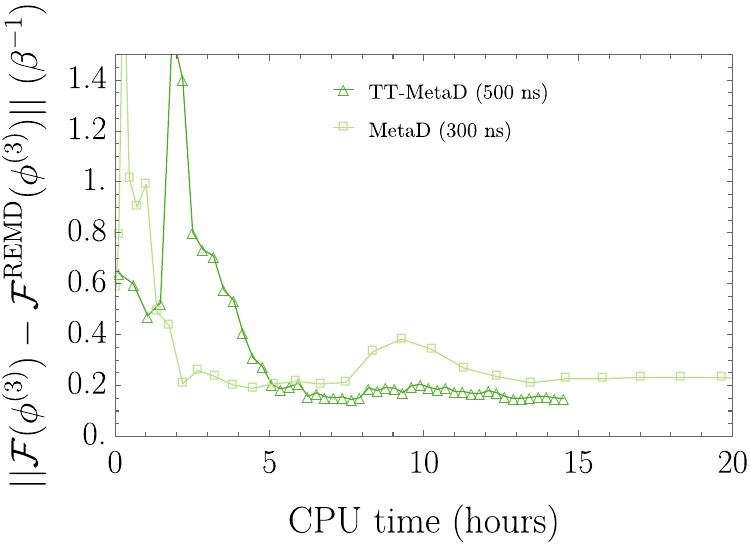}};
\end{tikzpicture}
\caption{1D free energy convergence and root mean square deviations (RMSD) of $\phi^{(3)}$ for trialanine.}
\label{fig:trialaninephi3}
\end{figure}

\begin{figure}
\centering
\begin{tikzpicture}
\node at (0,0) {\includegraphics[width=0.45\textwidth]{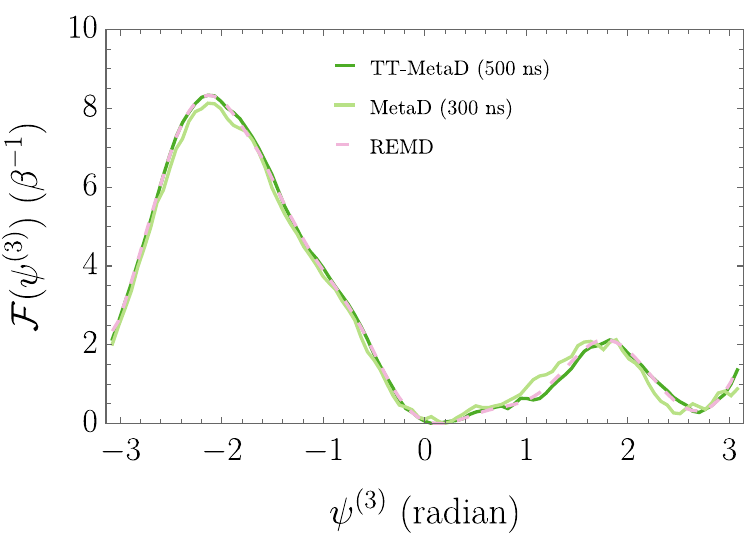}};
\node at (8.2,0) {\includegraphics[width=0.48\textwidth]{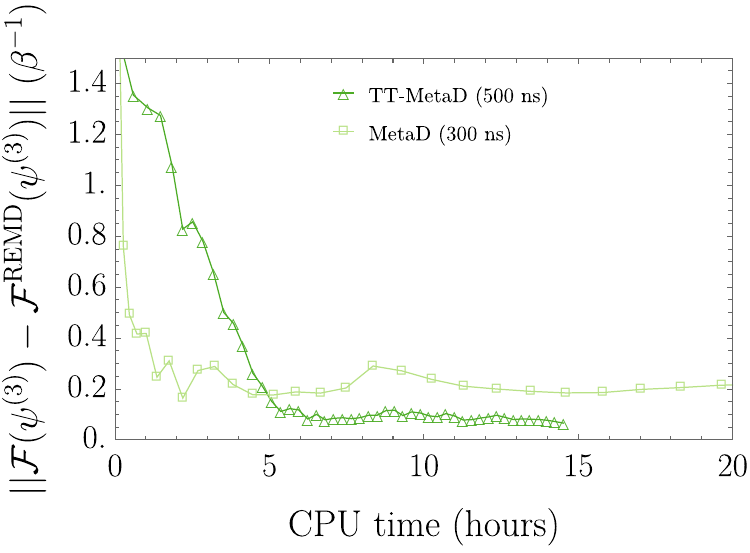}};
\end{tikzpicture}
\caption{1D free energy convergence and root mean square deviations (RMSD) of $\psi^{(3)}$ for trialanine.}
\label{fig:trialaninepsi3}
\end{figure}

\begin{figure}
\centering
\begin{tikzpicture}
\node at (0,0) {\includegraphics[width=0.45\textwidth]{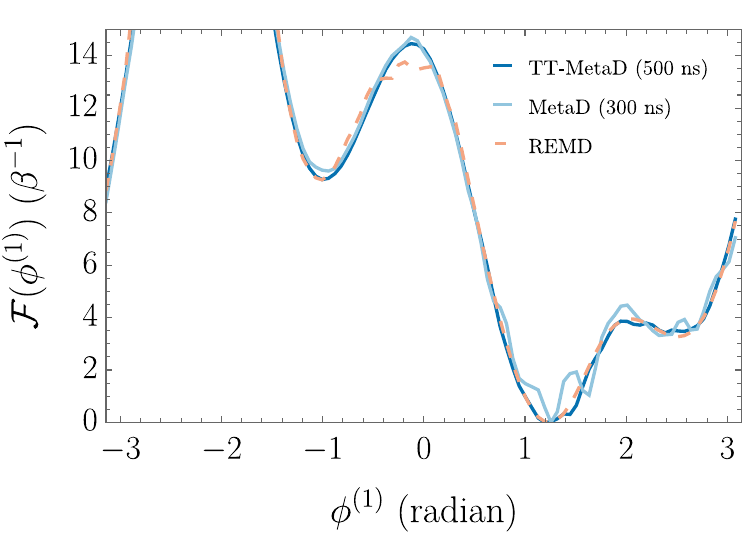}};
\node at (8.2,0) {\includegraphics[width=0.48\textwidth]{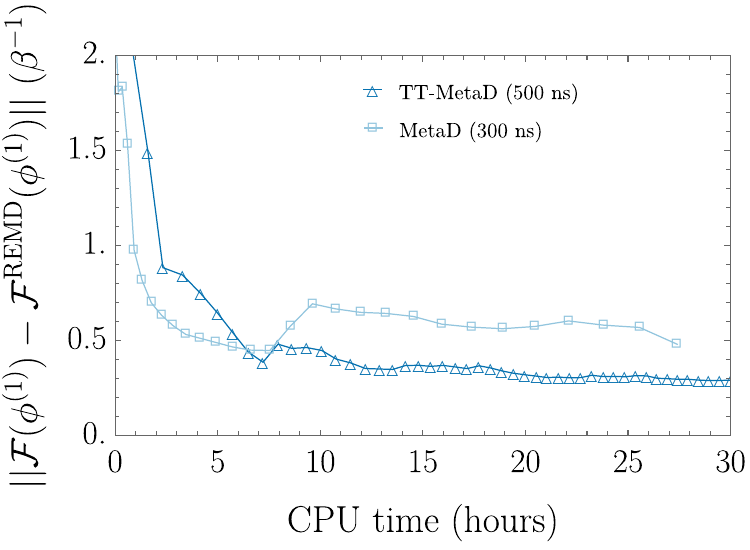}};
\end{tikzpicture}
\caption{1D free energy convergence and root mean square deviations (RMSD) of $\phi^{(1)}$ for ditrpyptophan.}
\label{fig:ditryptophanphi1}
\end{figure}

\begin{figure}
\centering
\begin{tikzpicture}
\node at (0,0) {\includegraphics[width=0.45\textwidth]{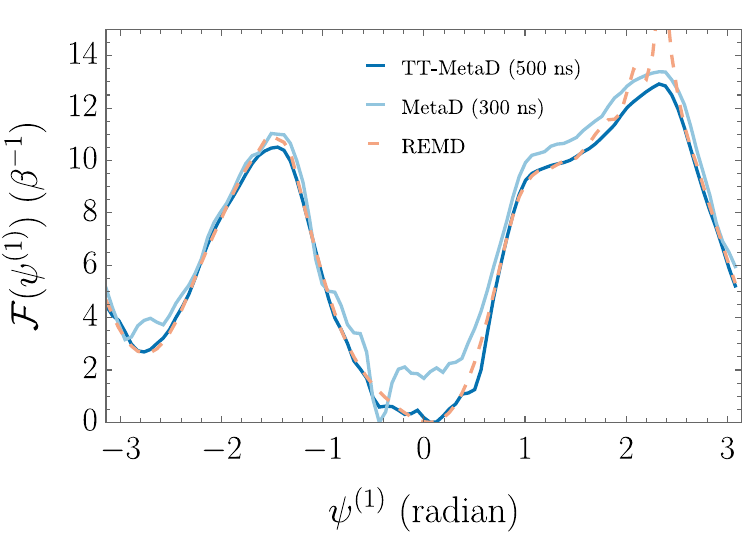}};
\node at (8.2,0) {\includegraphics[width=0.48\textwidth]{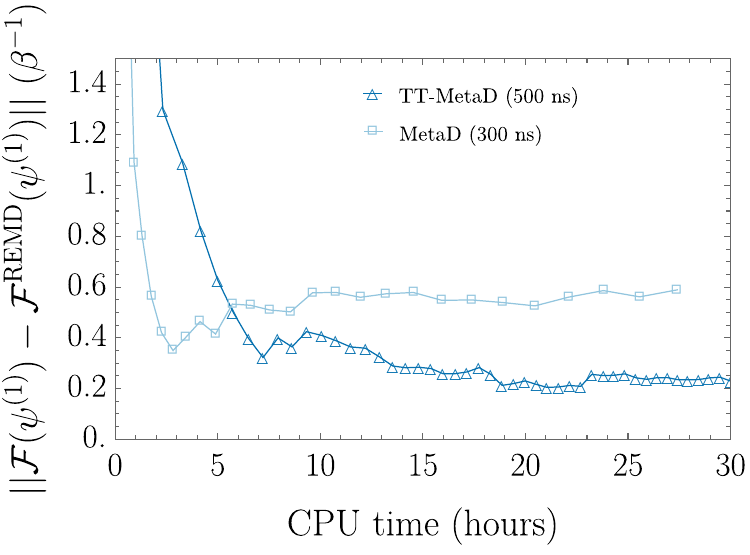}};
\end{tikzpicture}
\caption{1D free energy convergence and root mean square deviations (RMSD) of $\psi^{(1)}$ for ditrpyptophan.}
\label{fig:ditryptophanpsi1}
\end{figure}

\begin{figure}
\centering
\begin{tikzpicture}
\node at (0,0) {\includegraphics[width=0.45\textwidth]{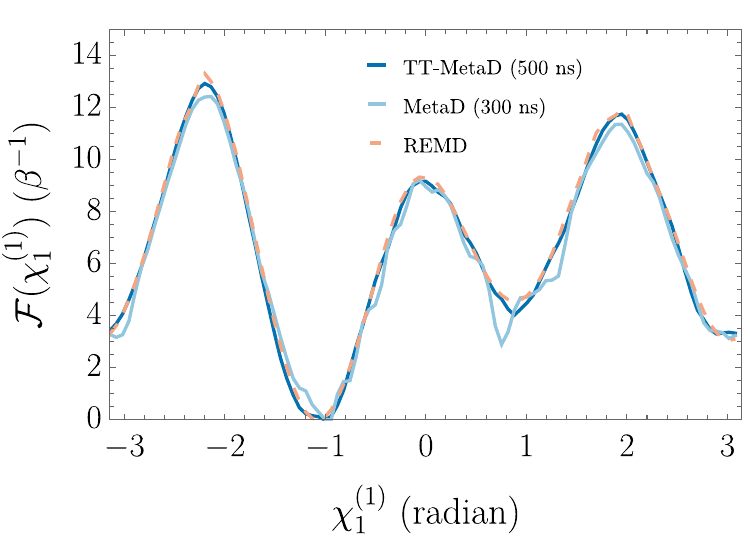}};
\node at (8.2,0) {\includegraphics[width=0.48\textwidth]{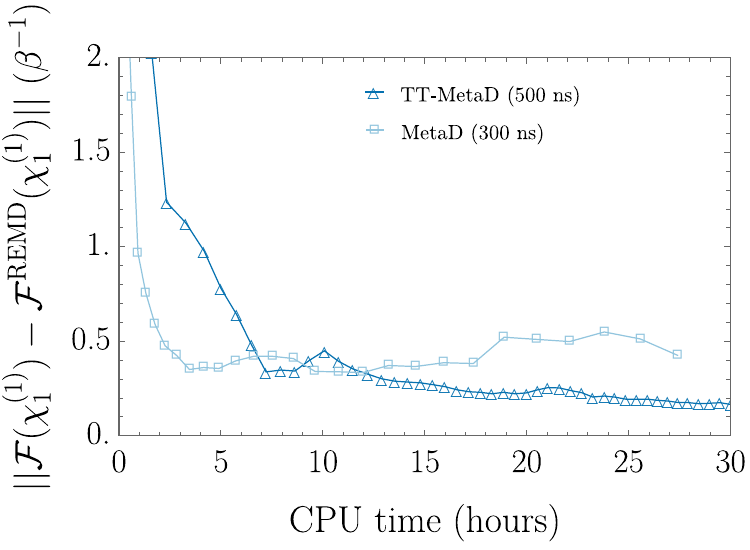}};
\end{tikzpicture}
\caption{1D free energy convergence and root mean square deviations (RMSD) of $\chi_1^{(1)}$ for ditrpyptophan.}
\label{fig:ditryptophanchi11}
\end{figure}

\begin{figure}
\centering
\begin{tikzpicture}
\node at (0,0) {\includegraphics[width=0.45\textwidth]{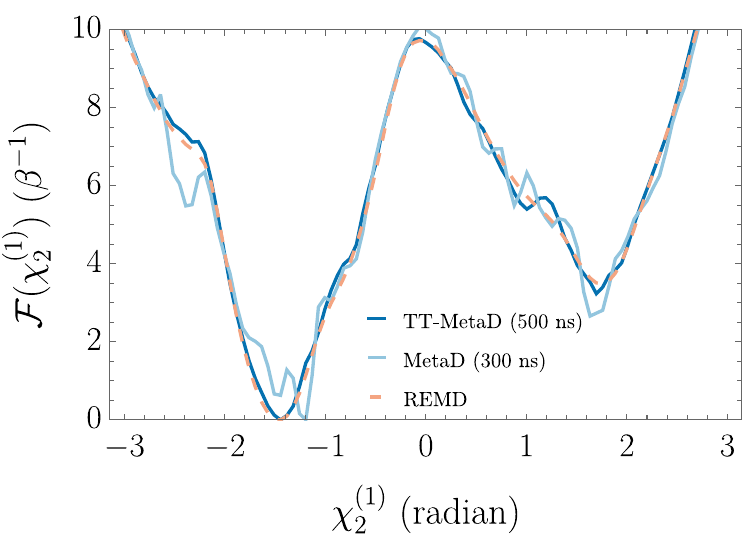}};
\node at (8.2,0) {\includegraphics[width=0.48\textwidth]{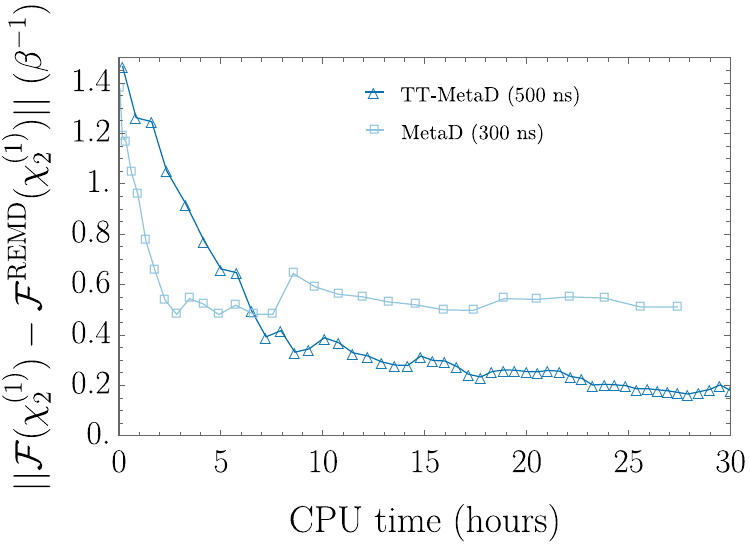}};
\end{tikzpicture}
\caption{1D free energy convergence and root mean square deviations (RMSD) of $\chi_2^{(1)}$ for ditrpyptophan.}
\label{fig:ditryptophanchi12}
\end{figure}

\begin{figure}
\centering
\begin{tikzpicture}
\node at (0,0) {\includegraphics[width=0.45\textwidth]{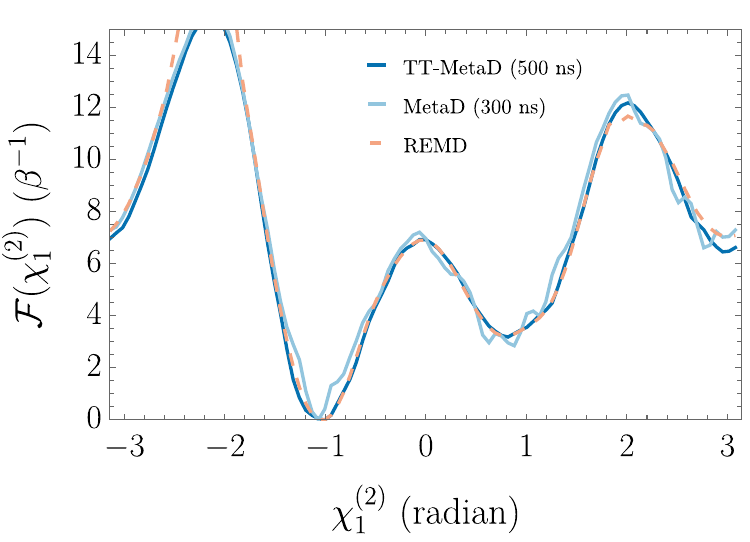}};
\node at (8.2,0) {\includegraphics[width=0.48\textwidth]{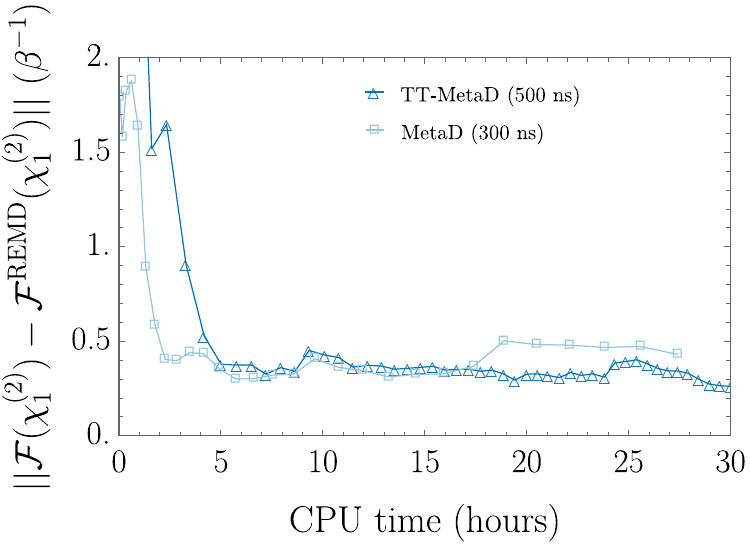}};
\end{tikzpicture}
\caption{1D free energy convergence and root mean square deviations (RMSD) of $\chi_1^{(2)}$ for ditrpyptophan.}
\label{fig:ditryptophanchi21}
\end{figure}

\begin{figure}
\centering
\begin{tikzpicture}
\node at (0,0) {\includegraphics[width=0.45\textwidth]{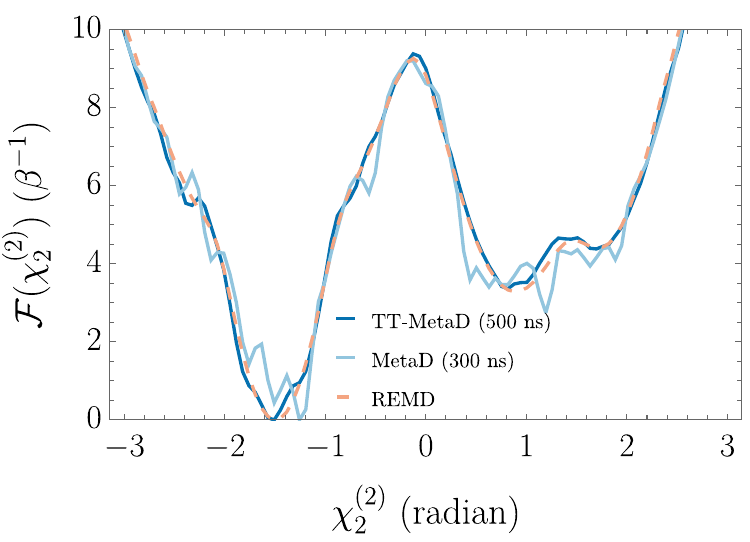}};
\node at (8.2,0) {\includegraphics[width=0.48\textwidth]{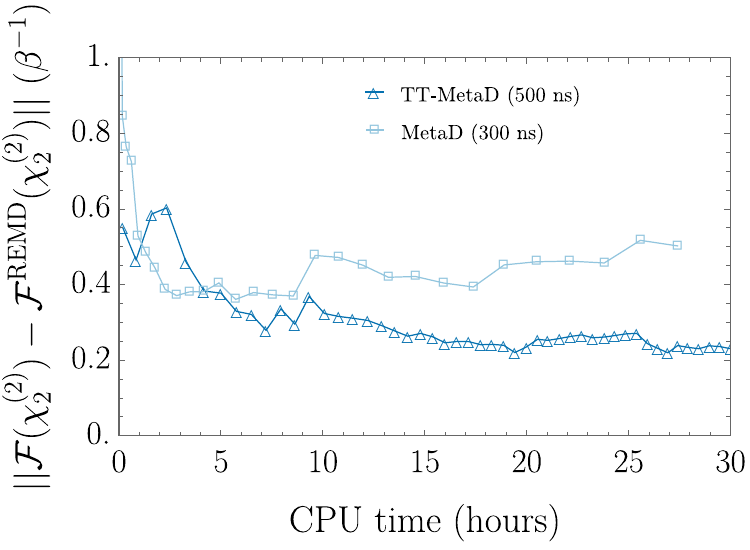}};
\end{tikzpicture}
\caption{1D free energy convergence and root mean square deviations (RMSD) of $\chi_2^{(2)}$ for ditrpyptophan.}
\label{fig:ditryptophanchi22}
\end{figure}

\begin{figure}
\centering
\begin{tikzpicture}
\node at (0,0) {\includegraphics[width=0.45\textwidth]{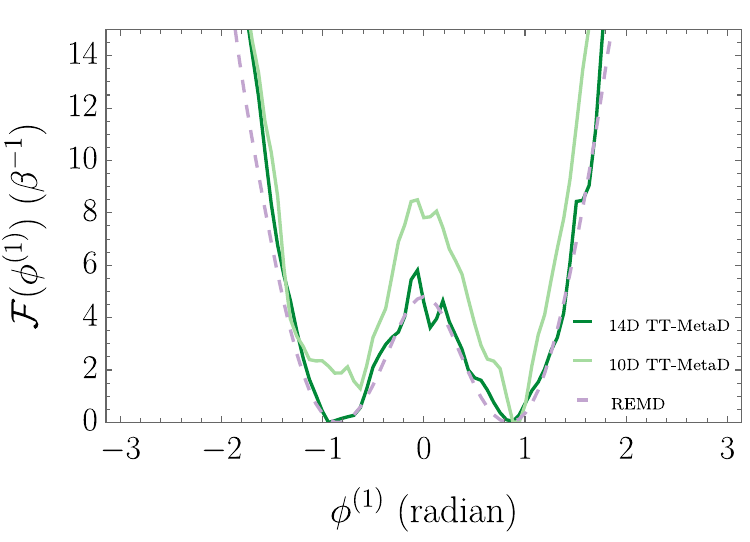}};
\node at (8.2,0) {\includegraphics[width=0.48\textwidth]{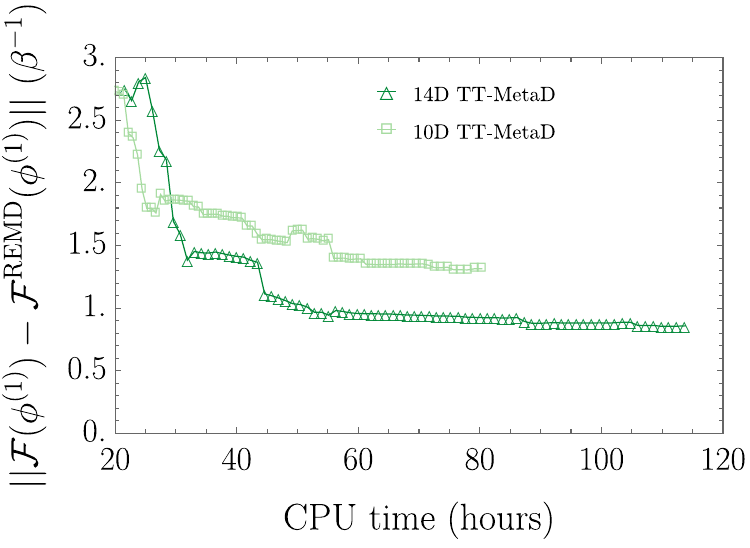}};
\end{tikzpicture}
\caption{1D free energy convergence and root mean square deviations (RMSD) of $\phi^{(1)}$ for AIB$_9$.}
\label{fig:aib9phi1}
\end{figure}

\begin{figure}
\centering
\begin{tikzpicture}
\node at (0,0) {\includegraphics[width=0.45\textwidth]{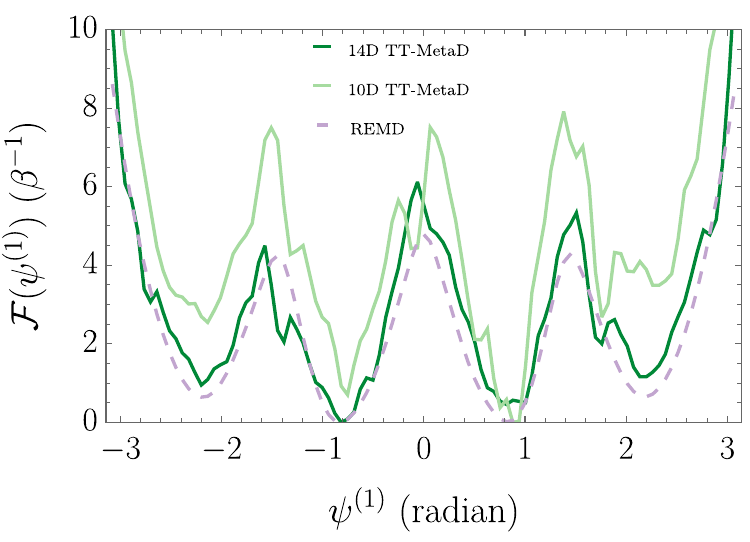}};
\node at (8.2,0) {\includegraphics[width=0.48\textwidth]{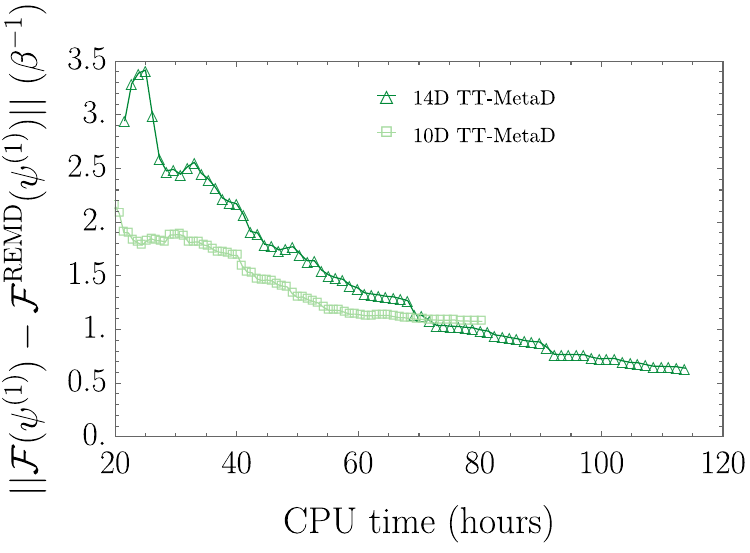}};
\end{tikzpicture}
\caption{1D free energy convergence and root mean square deviations (RMSD) of $\psi^{(1)}$ for AIB$_9$.}
\label{fig:aib9psi1}
\end{figure}

\begin{figure}
\centering
\begin{tikzpicture}
\node at (0,0) {\includegraphics[width=0.45\textwidth]{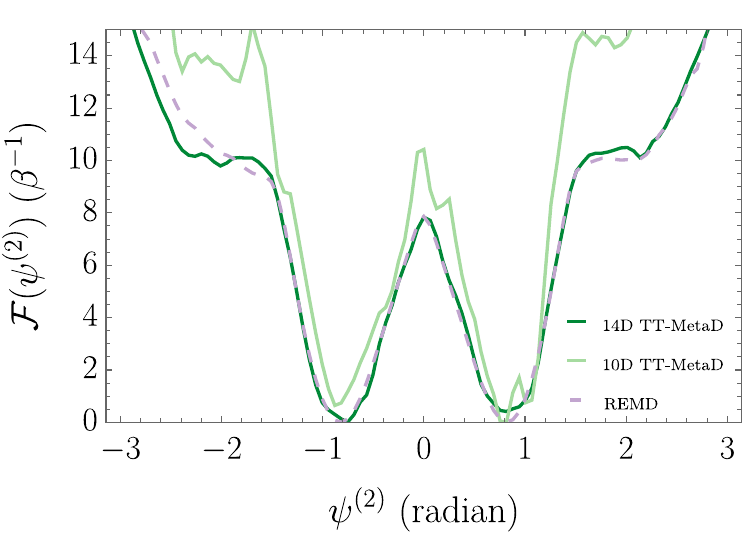}};
\node at (8.2,0) {\includegraphics[width=0.48\textwidth]{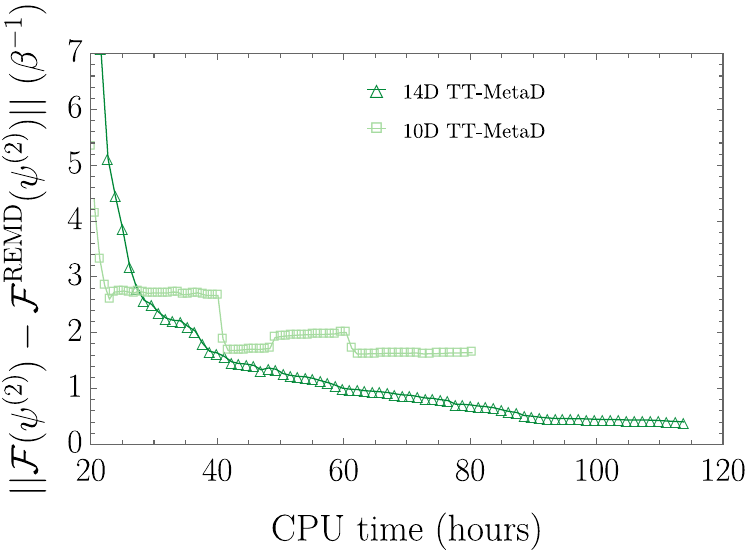}};
\end{tikzpicture}
\caption{1D free energy convergence and root mean square deviations (RMSD) of $\psi^{(2)}$ for AIB$_9$.}
\label{fig:aib9psi2}
\end{figure}

\begin{figure}
\centering
\begin{tikzpicture}
\node at (0,0) {\includegraphics[width=0.45\textwidth]{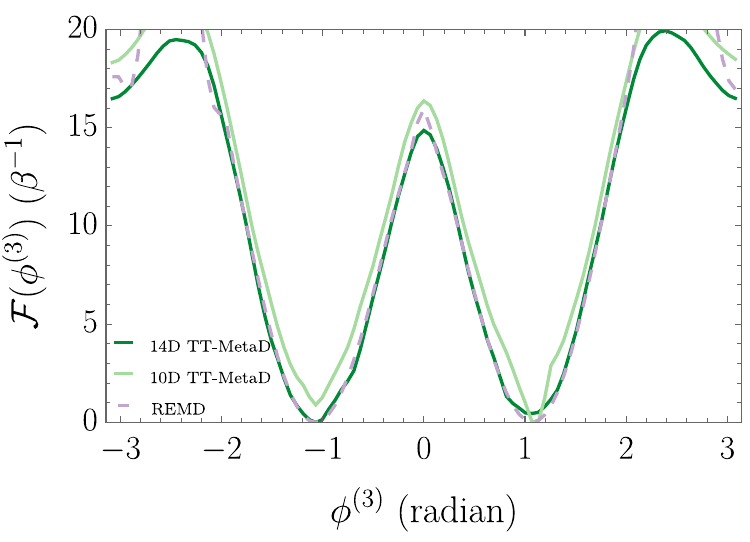}};
\node at (8.2,0) {\includegraphics[width=0.48\textwidth]{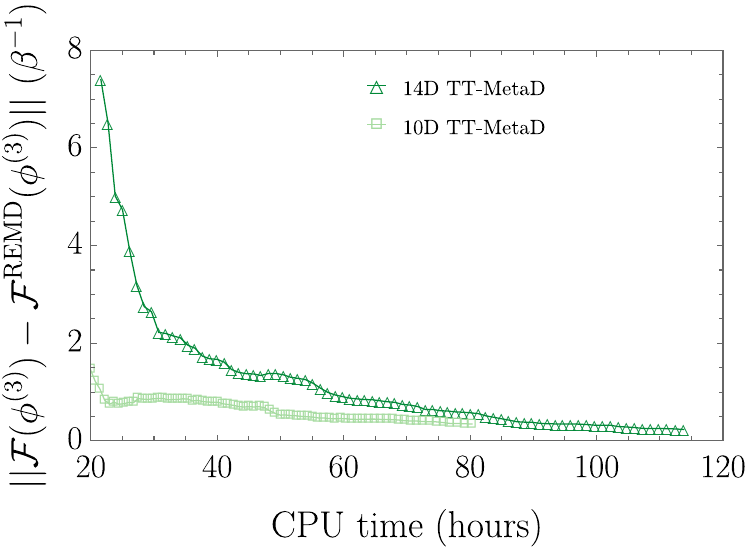}};
\end{tikzpicture}
\caption{1D free energy convergence and root mean square deviations (RMSD) of $\phi^{(3)}$ for AIB$_9$.}
\label{fig:aib9phi3}
\end{figure}

\begin{figure}
\centering
\begin{tikzpicture}
\node at (0,0) {\includegraphics[width=0.45\textwidth]{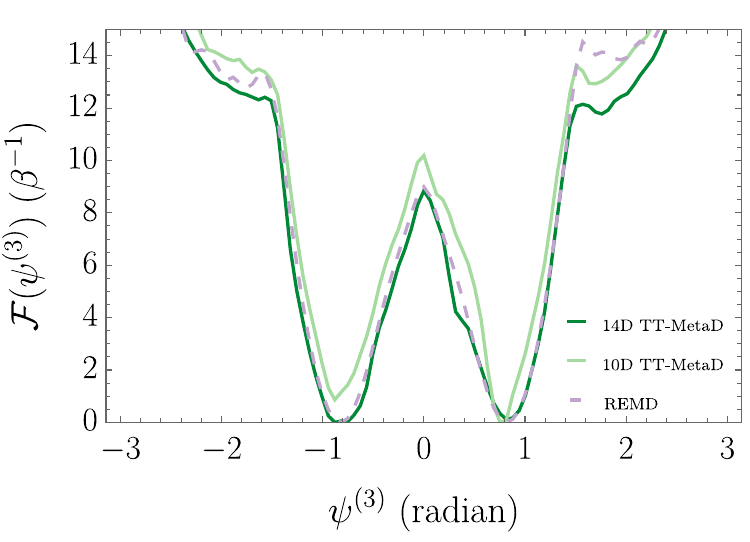}};
\node at (8.2,0) {\includegraphics[width=0.48\textwidth]{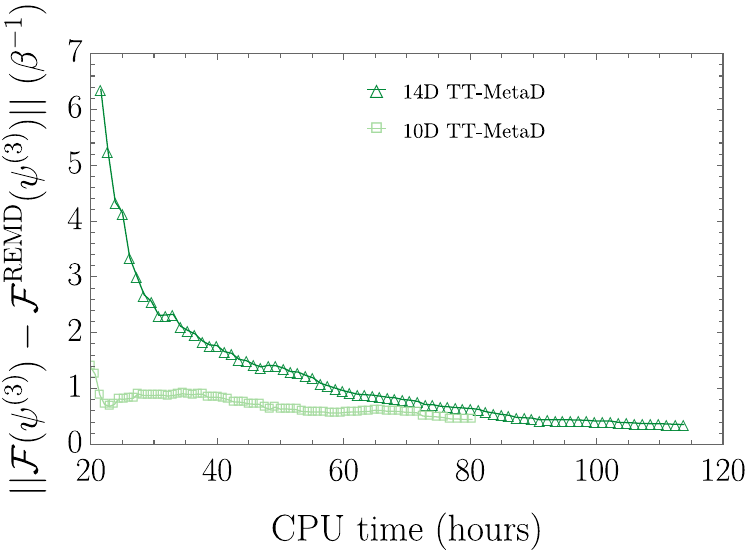}};
\end{tikzpicture}
\caption{1D free energy convergence and root mean square deviations (RMSD) of $\psi^{(3)}$ for AIB$_9$.}
\label{fig:aib9psi3}
\end{figure}

\begin{figure}
\centering
\begin{tikzpicture}
\node at (0,0) {\includegraphics[width=0.45\textwidth]{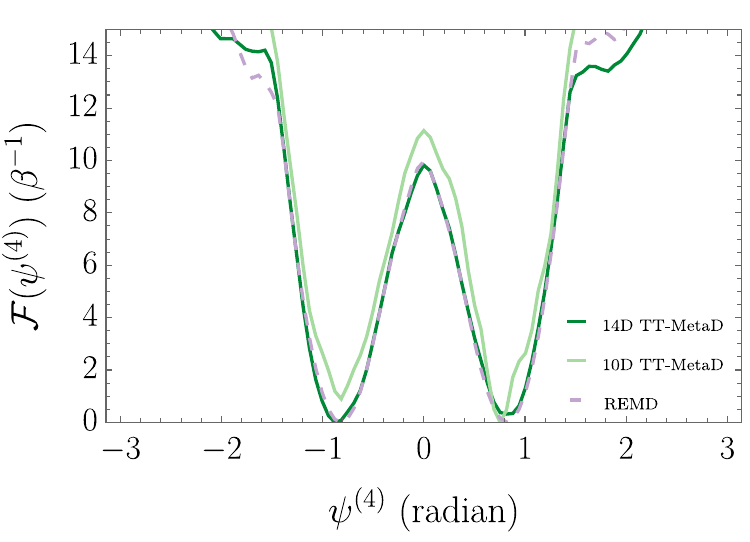}};
\node at (8.2,0) {\includegraphics[width=0.48\textwidth]{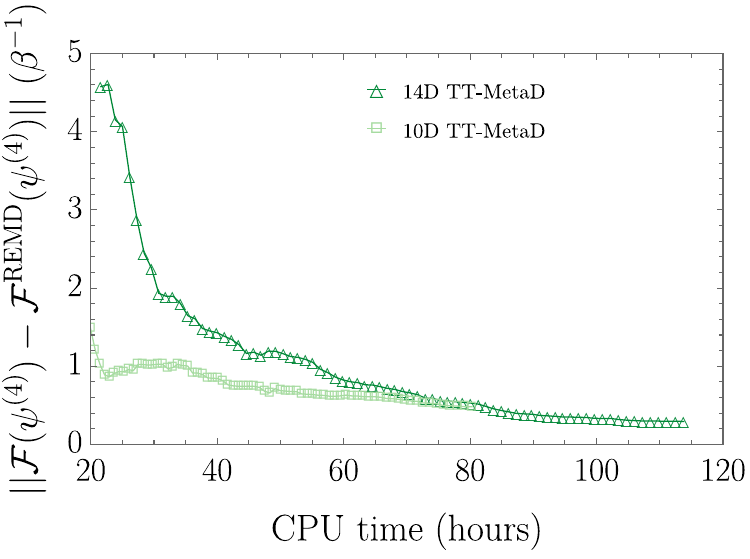}};
\end{tikzpicture}
\caption{1D free energy convergence and root mean square deviations (RMSD) of $\psi^{(4)}$ for AIB$_9$.}
\label{fig:aib9psi4}
\end{figure}

\begin{figure}
\centering
\begin{tikzpicture}
\node at (0,0) {\includegraphics[width=0.45\textwidth]{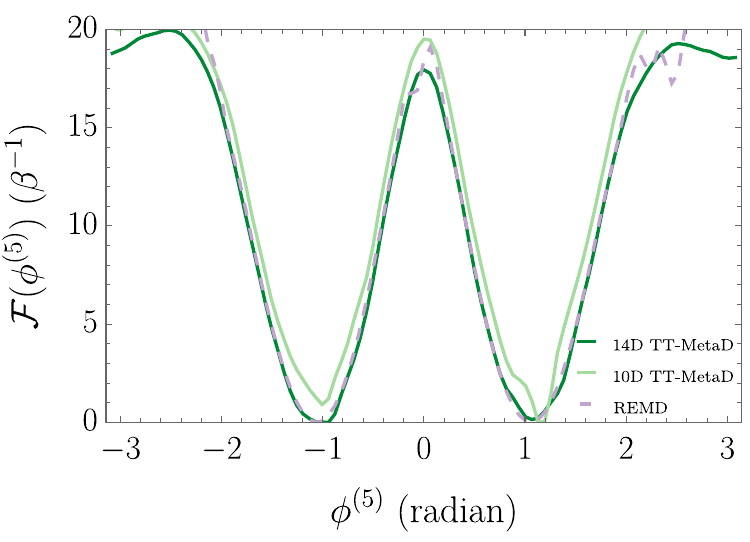}};
\node at (8.2,0) {\includegraphics[width=0.48\textwidth]{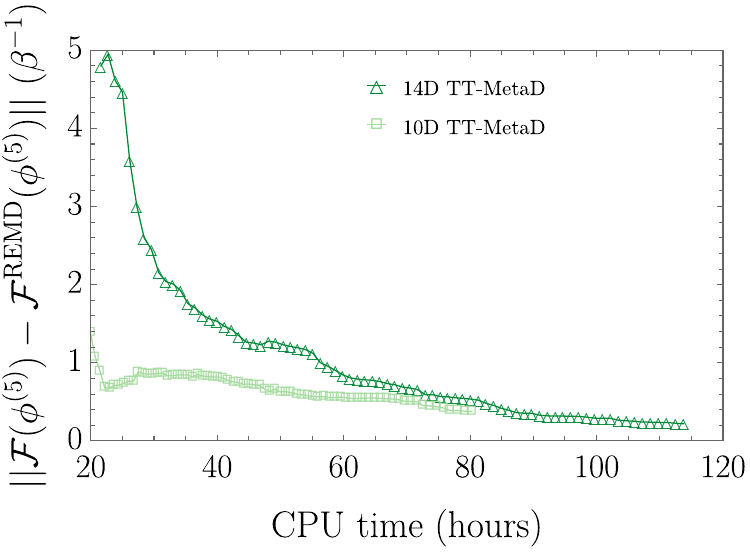}};
\end{tikzpicture}
\caption{1D free energy convergence and root mean square deviations (RMSD) of $\phi^{(5)}$ for AIB$_9$.}
\label{fig:aib9phi5}
\end{figure}

\begin{figure}
\centering
\begin{tikzpicture}
\node at (0,0) {\includegraphics[width=0.45\textwidth]{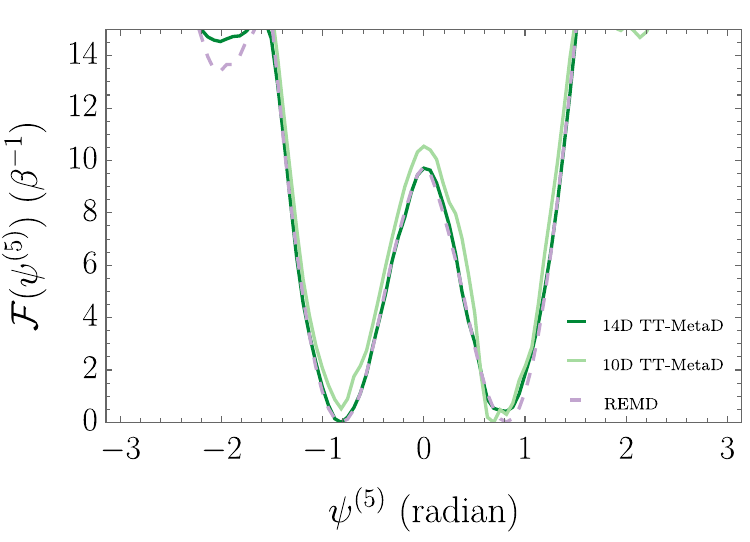}};
\node at (8.2,0) {\includegraphics[width=0.48\textwidth]{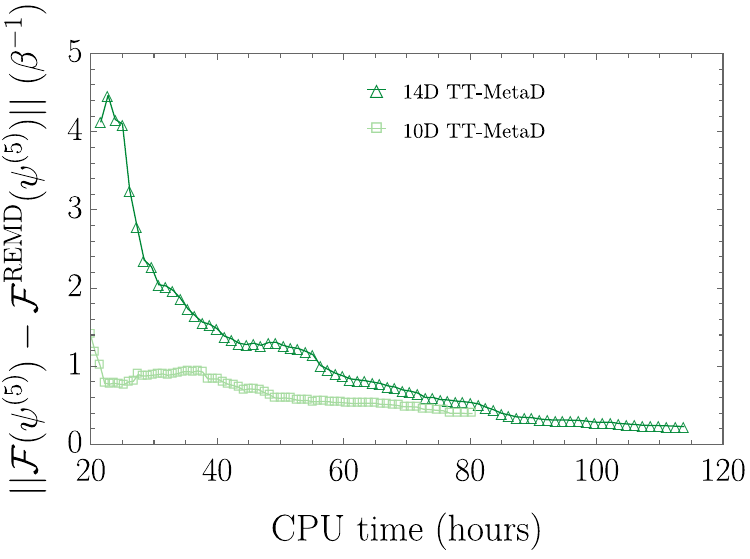}};
\end{tikzpicture}
\caption{1D free energy convergence and root mean square deviations (RMSD) of $\psi^{(5)}$ for AIB$_9$.}
\label{fig:aib9psi5}
\end{figure}

\begin{figure}
\centering
\begin{tikzpicture}
\node at (0,0) {\includegraphics[width=0.45\textwidth]{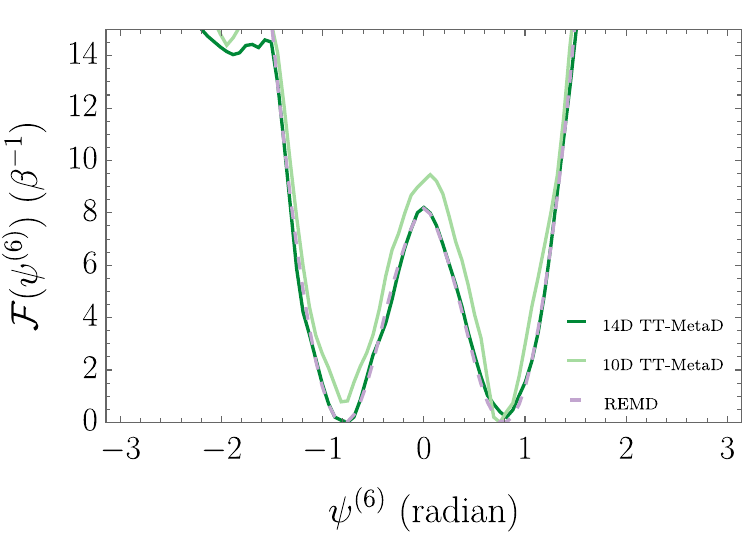}};
\node at (8.2,0) {\includegraphics[width=0.48\textwidth]{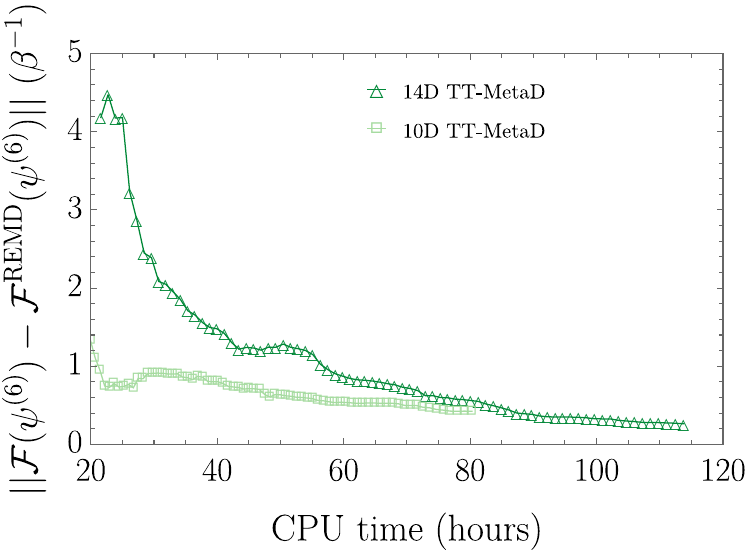}};
\end{tikzpicture}
\caption{1D free energy convergence and root mean square deviations (RMSD) of $\psi^{(6)}$ for AIB$_9$.}
\label{fig:aib9psi6}
\end{figure}

\begin{figure}
\centering
\begin{tikzpicture}
\node at (0,0) {\includegraphics[width=0.45\textwidth]{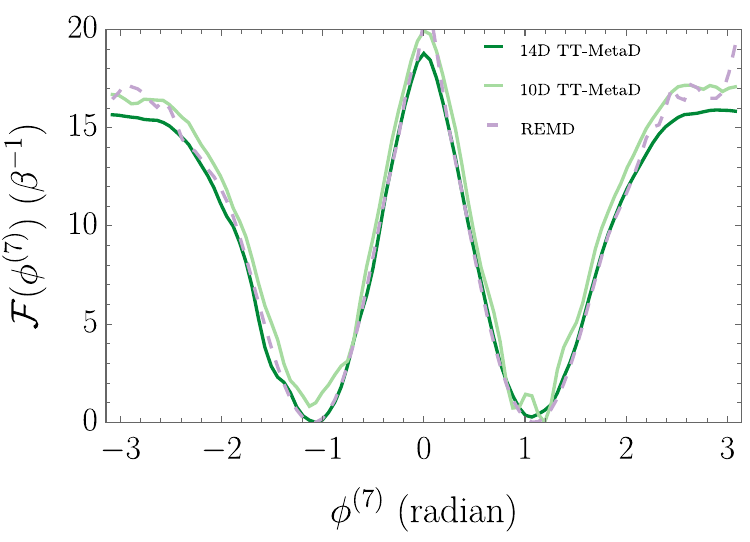}};
\node at (8.2,0) {\includegraphics[width=0.48\textwidth]{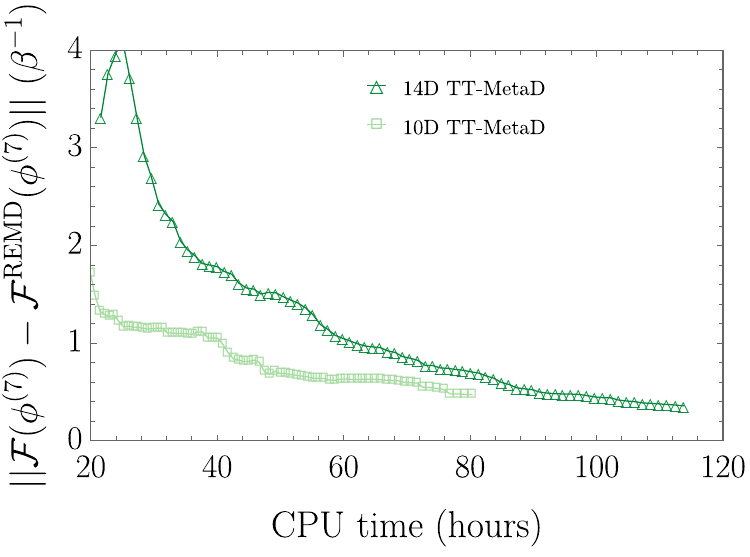}};
\end{tikzpicture}
\caption{1D free energy convergence and root mean square deviations (RMSD) of $\phi^{(7)}$ for AIB$_9$.}
\label{fig:aib9phi7}
\end{figure}

\begin{figure}
\centering
\begin{tikzpicture}
\node at (0,0) {\includegraphics[width=0.45\textwidth]{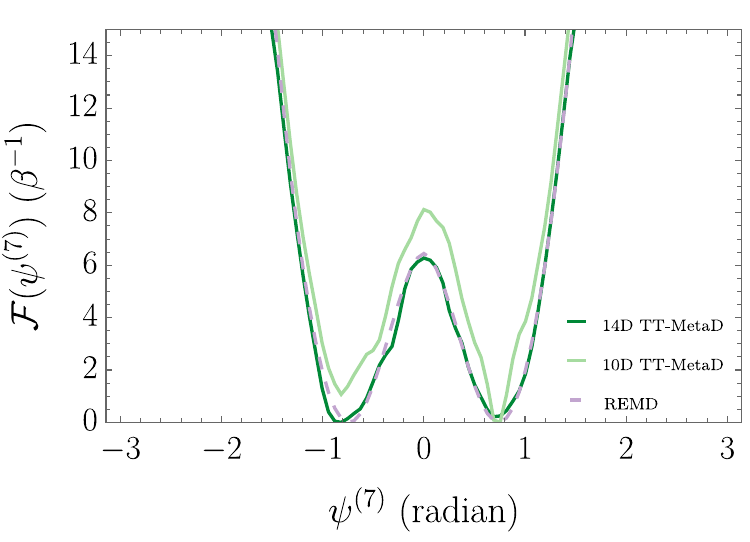}};
\node at (8.2,0) {\includegraphics[width=0.48\textwidth]{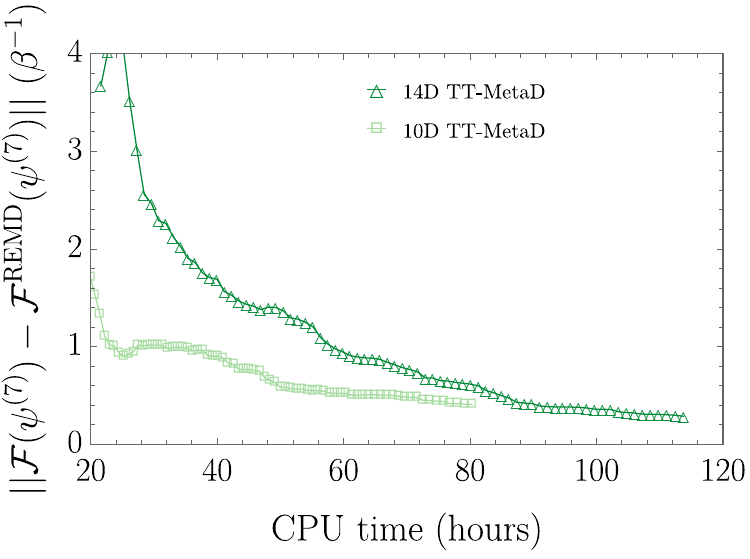}};
\end{tikzpicture}
\caption{1D free energy convergence and root mean square deviations (RMSD) of $\psi^{(7)}$ for AIB$_9$.}
\label{fig:aib9psi7}
\end{figure}

\begin{figure}
\centering
\begin{tikzpicture}
\node at (0,0) {\includegraphics[width=0.45\textwidth]{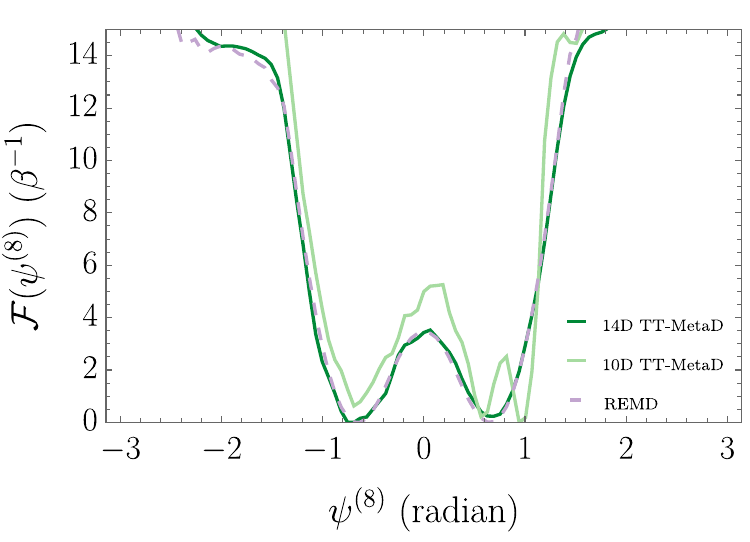}};
\node at (8.2,0) {\includegraphics[width=0.48\textwidth]{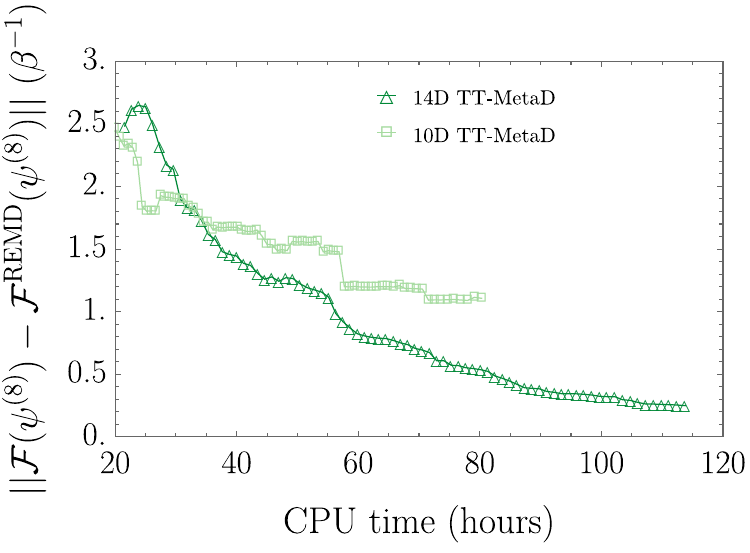}};
\end{tikzpicture}
\caption{1D free energy convergence and root mean square deviations (RMSD) of $\psi^{(8)}$ for AIB$_9$.}
\label{fig:aib9psi8}
\end{figure}

\begin{figure}
\centering
\begin{tikzpicture}
\node at (0,0) {\includegraphics[width=0.45\textwidth]{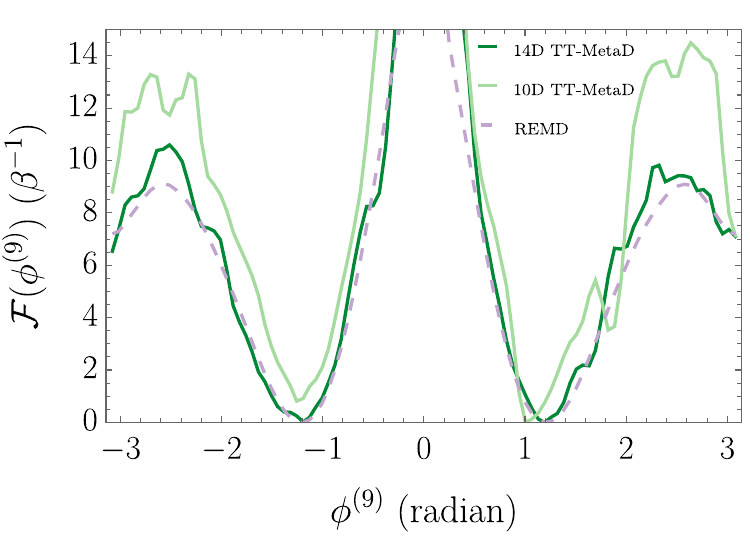}};
\node at (8.2,0) {\includegraphics[width=0.48\textwidth]{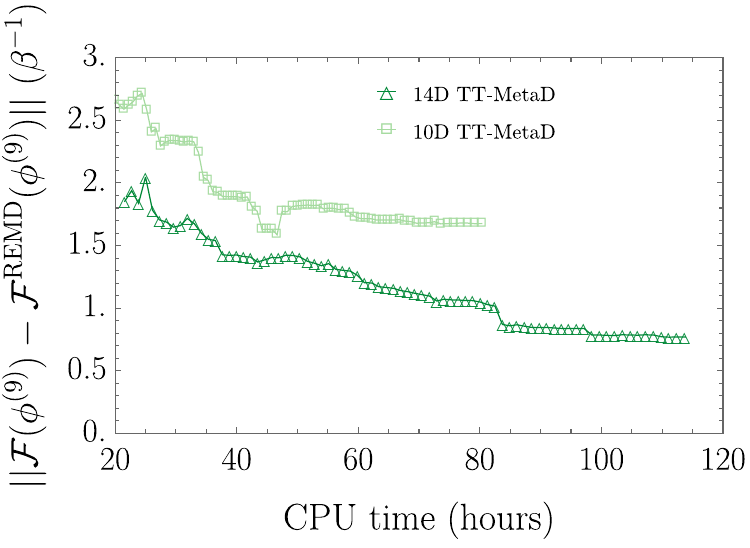}};
\end{tikzpicture}
\caption{1D free energy convergence and root mean square deviations (RMSD) of $\phi^{(9)}$ for AIB$_9$.}
\label{fig:aib9phi9}
\end{figure}

\begin{figure}
\centering
\begin{tikzpicture}
\node at (0,0) {\includegraphics[width=0.45\textwidth]{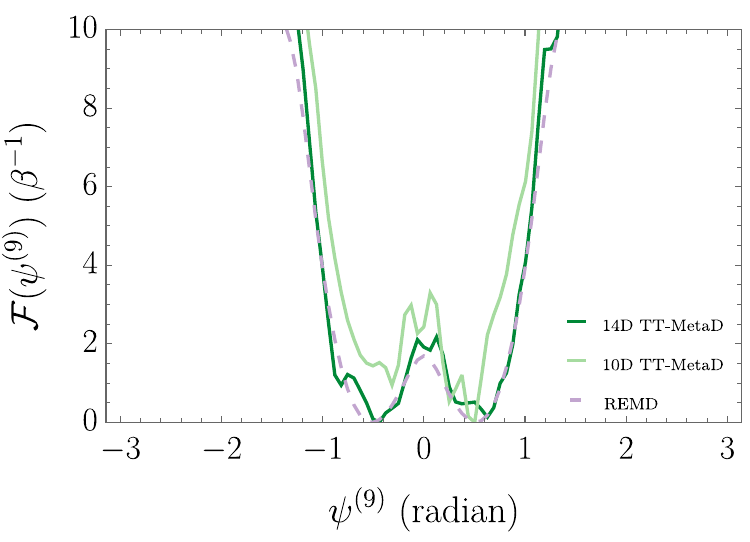}};
\node at (8.2,0) {\includegraphics[width=0.48\textwidth]{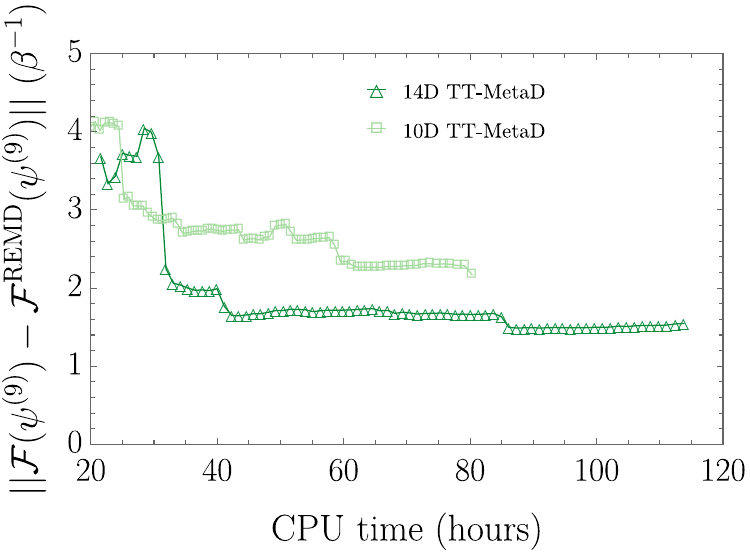}};
\end{tikzpicture}
\caption{1D free energy convergence and root mean square deviations (RMSD) of $\psi^{(9)}$ for AIB$_9$.}
\label{fig:aib9psi9}
\end{figure}

\begin{figure}
\centering
\begin{tikzpicture}
\node at (0,0) {\includegraphics[width=0.32\textwidth]{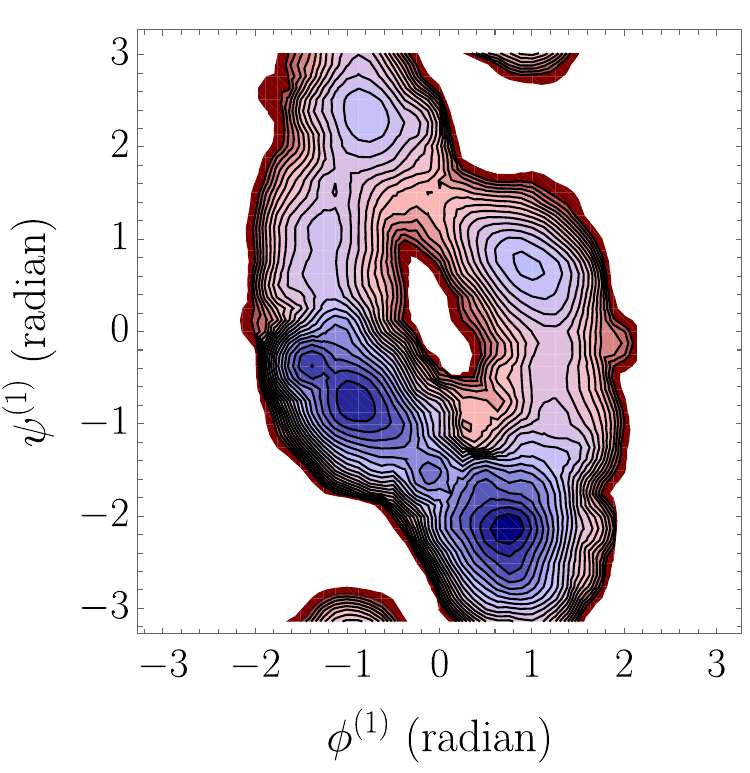}};
\node at (5.5,0) {\includegraphics[width=0.32\textwidth]{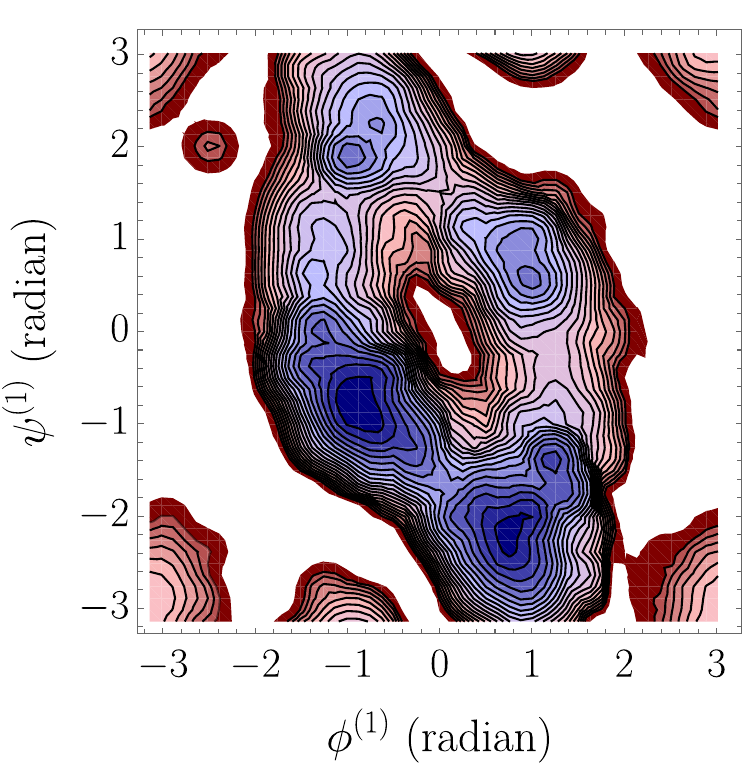}};
\node at (11,0) {\includegraphics[width=0.32\textwidth]{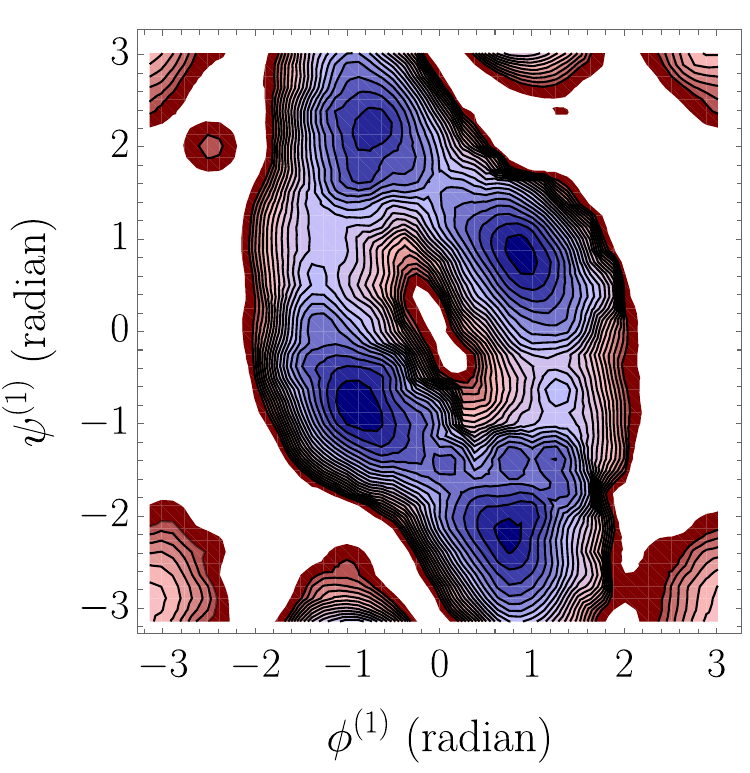}};
\end{tikzpicture}
\caption{2D free energy convergence of $(\phi^{(1)},\psi^{(1)})$ for AIB$_9$. Left: 100~ns; middle: 300~ns; right: 1~$\upmu$s.}
\label{fig:aib9_2d_1}
\end{figure}

\begin{figure}
\centering
\begin{tikzpicture}
\node at (0,0) {\includegraphics[width=0.32\textwidth]{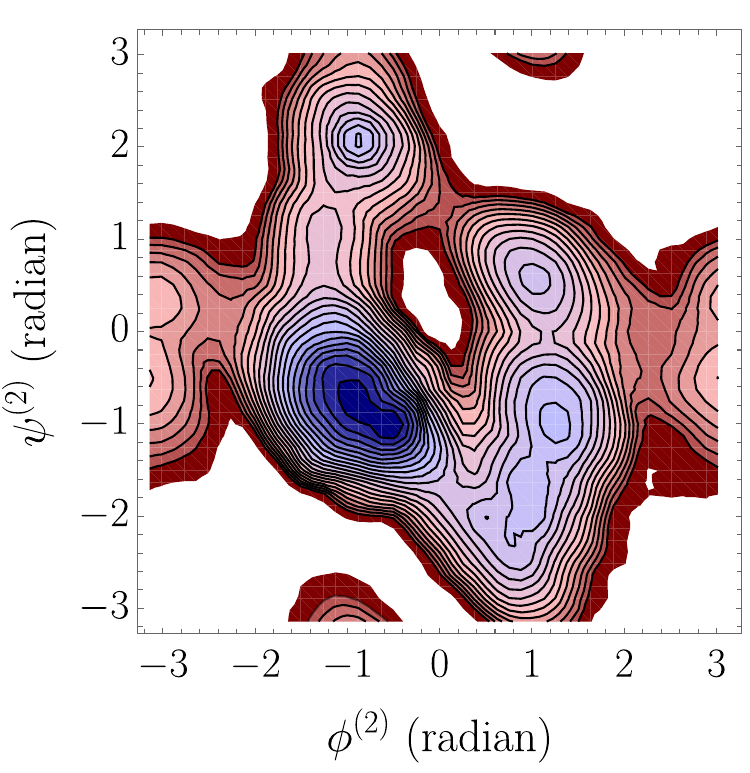}};
\node at (5.5,0) {\includegraphics[width=0.32\textwidth]{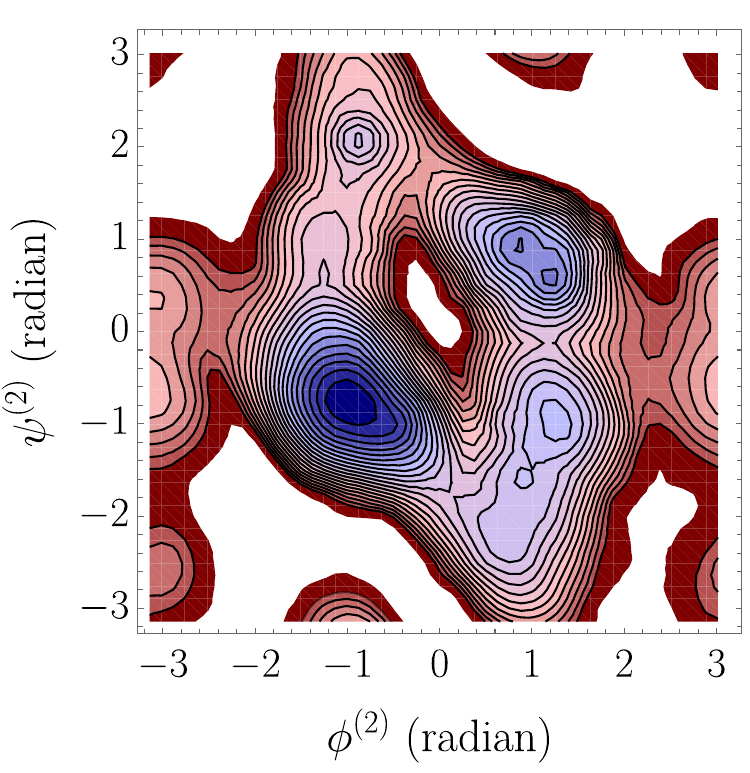}};
\node at (11,0) {\includegraphics[width=0.32\textwidth]{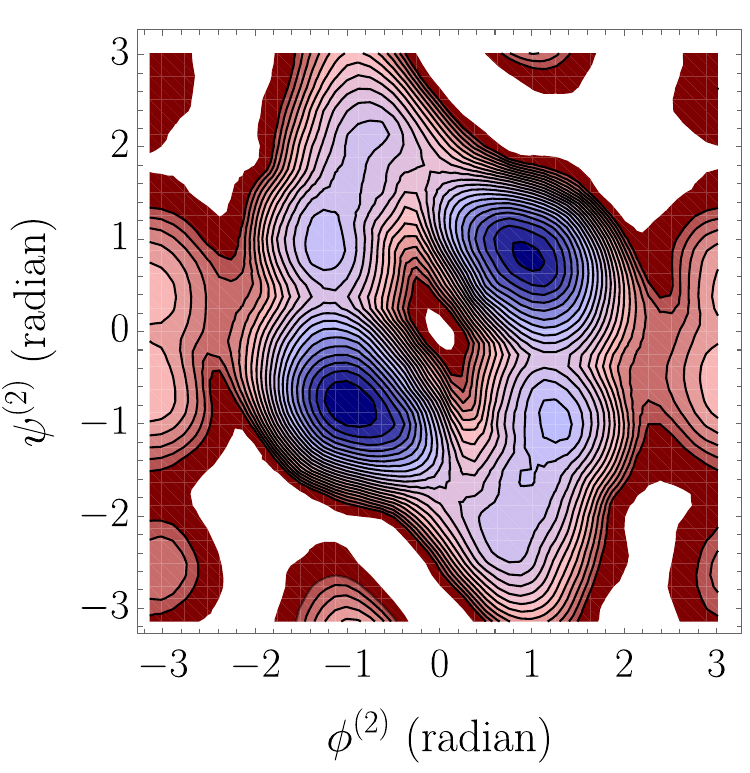}};
\end{tikzpicture}
\caption{2D free energy convergence of $(\phi^{(2)},\psi^{(2)})$ for AIB$_9$. Left: 100~ns; middle: 300~ns; right: 1~$\upmu$s.}
\label{fig:aib9_2d_2}
\end{figure}

\begin{figure}
\centering
\begin{tikzpicture}
\node at (0,0) {\includegraphics[width=0.32\textwidth]{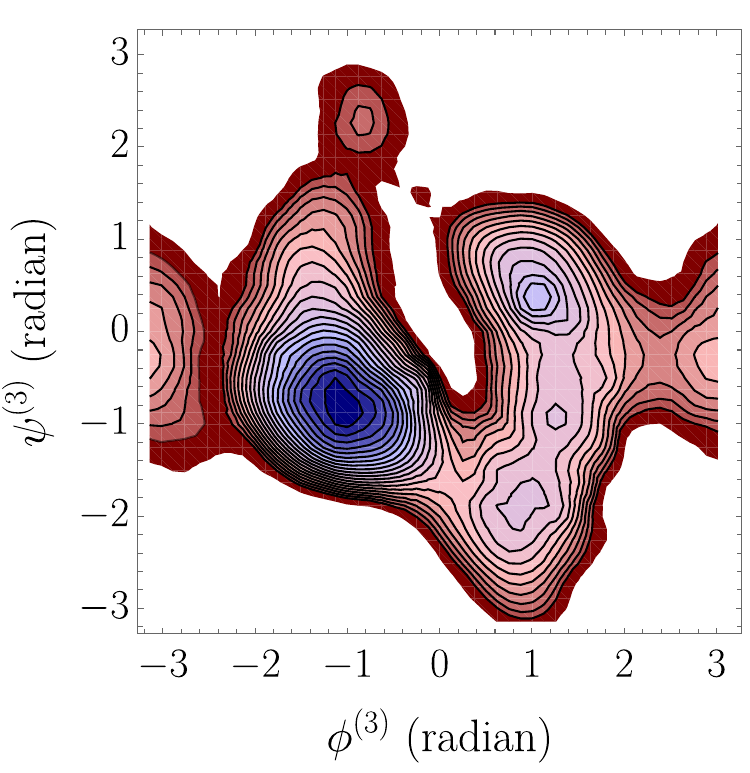}};
\node at (5.5,0) {\includegraphics[width=0.32\textwidth]{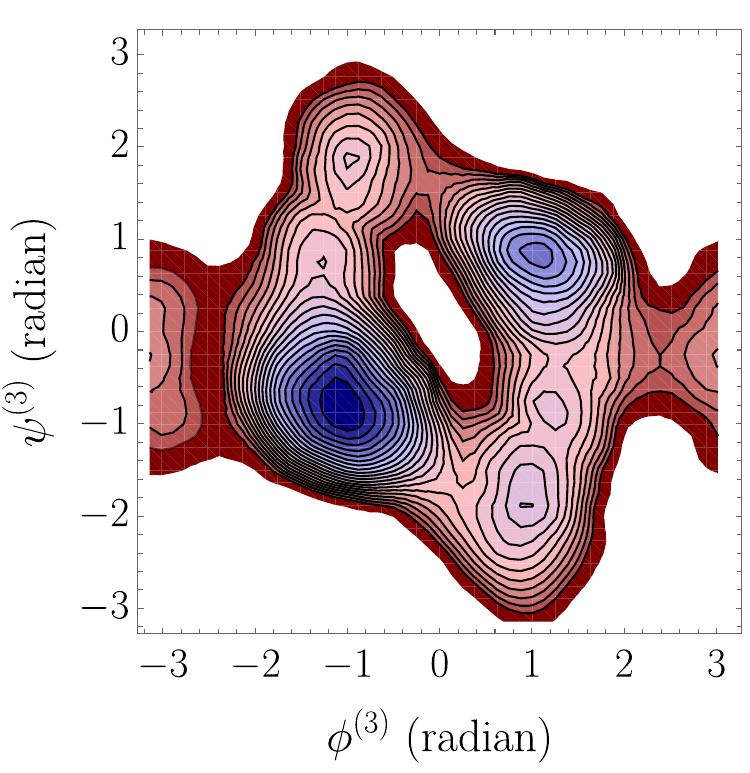}};
\node at (11,0) {\includegraphics[width=0.32\textwidth]{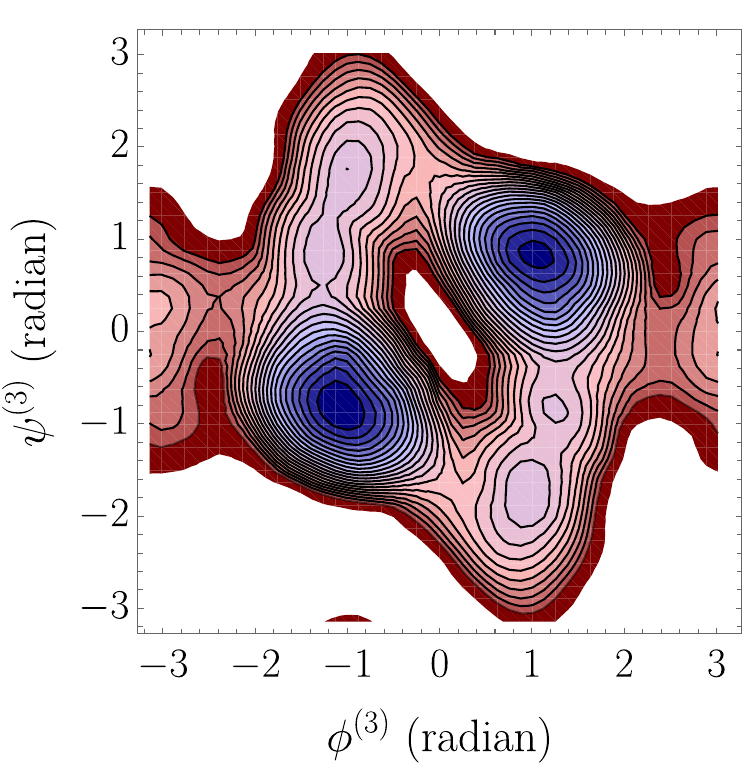}};
\end{tikzpicture}
\caption{2D free energy convergence of $(\phi^{(3)},\psi^{(3)})$ for AIB$_9$. Left: 100~ns; middle: 300~ns; right: 1~$\upmu$s.}
\label{fig:aib9_2d_3}
\end{figure}

\begin{figure}
\centering
\begin{tikzpicture}
\node at (0,0) {\includegraphics[width=0.32\textwidth]{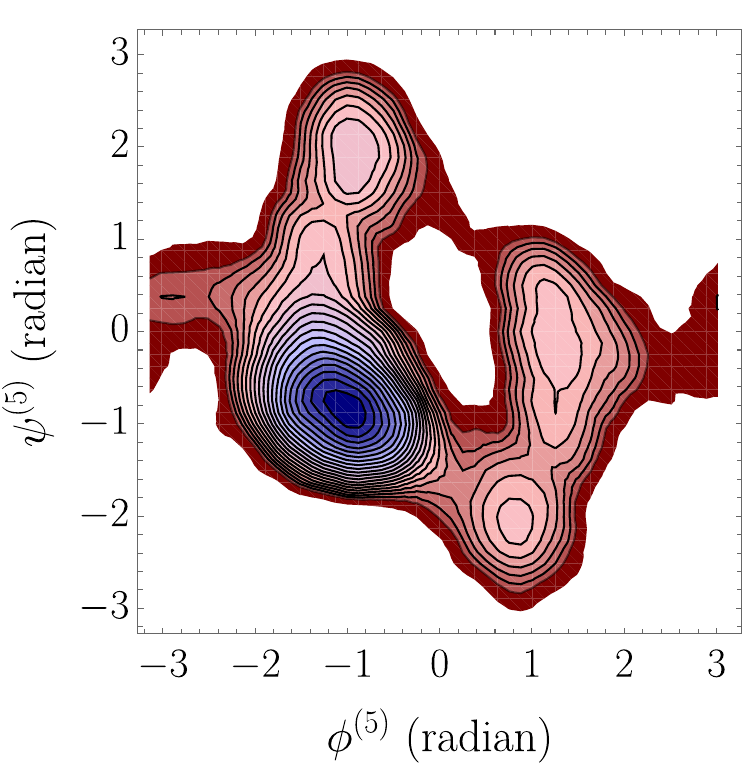}};
\node at (5.5,0) {\includegraphics[width=0.32\textwidth]{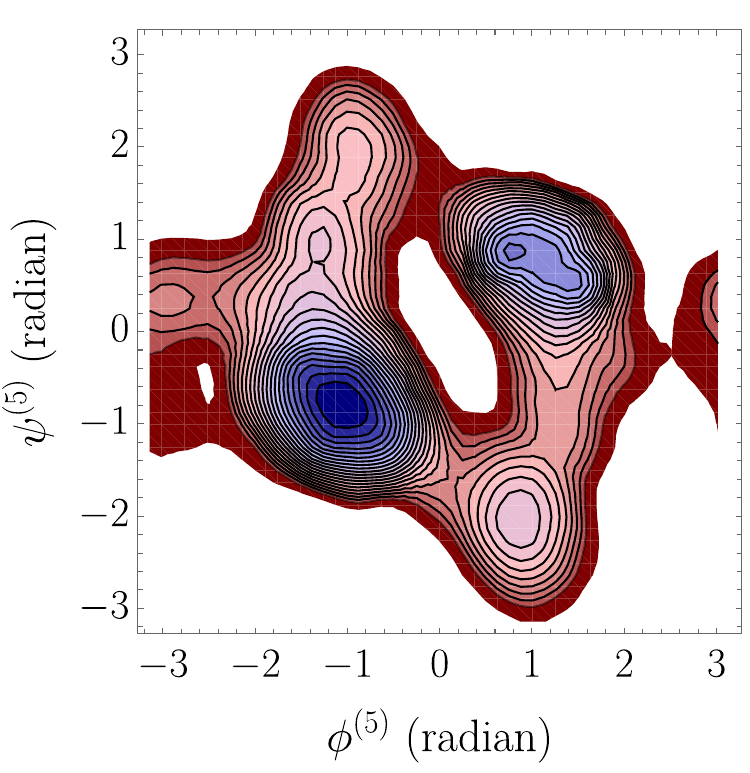}};
\node at (11,0) {\includegraphics[width=0.32\textwidth]{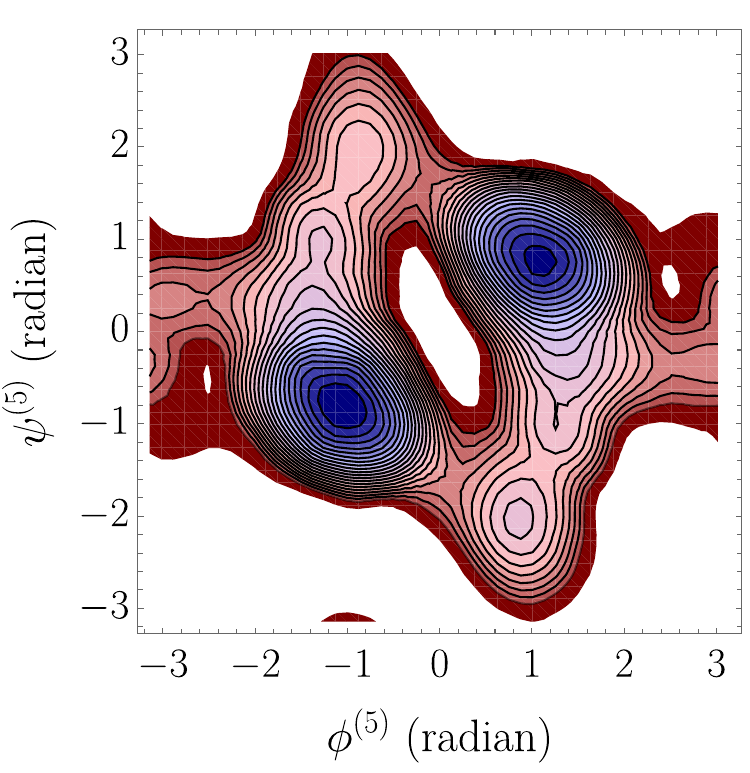}};
\end{tikzpicture}
\caption{2D free energy convergence of $(\phi^{(5)},\psi^{(5)})$ for AIB$_9$. Left: 100~ns; middle: 300~ns; right: 1~$\upmu$s.}
\label{fig:aib9_2d_5}
\end{figure}

\begin{figure}
\centering
\begin{tikzpicture}
\node at (0,0) {\includegraphics[width=0.32\textwidth]{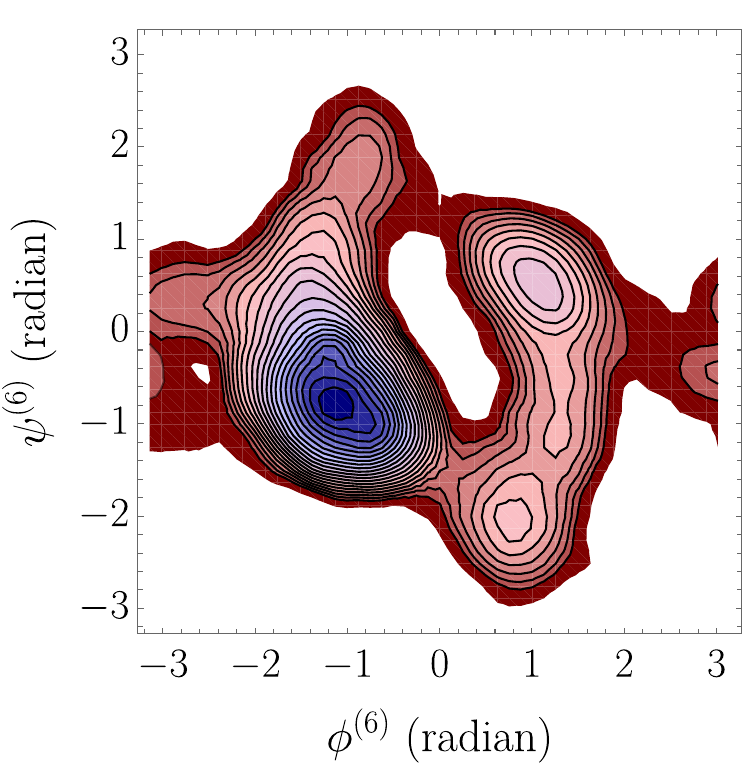}};
\node at (5.5,0) {\includegraphics[width=0.32\textwidth]{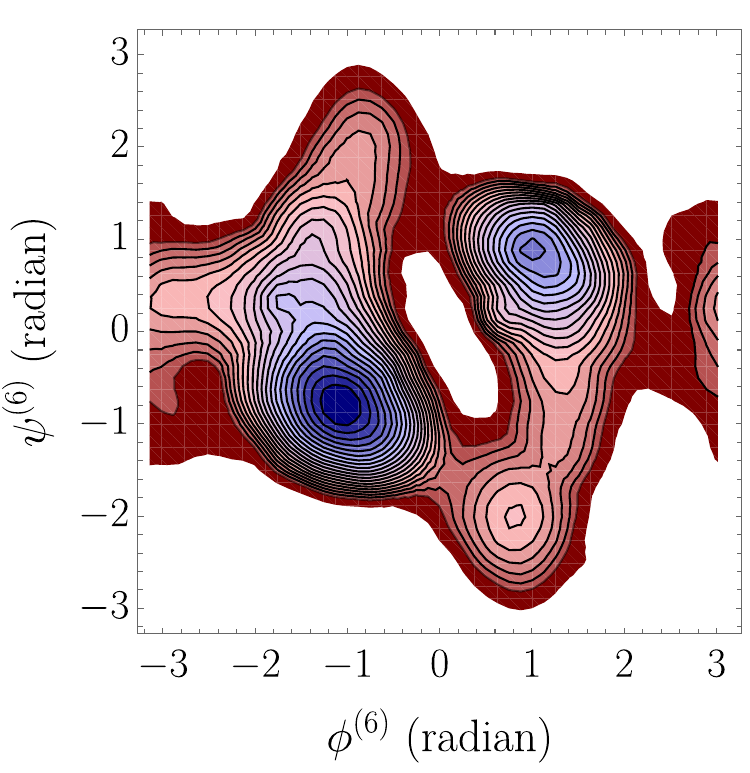}};
\node at (11,0) {\includegraphics[width=0.32\textwidth]{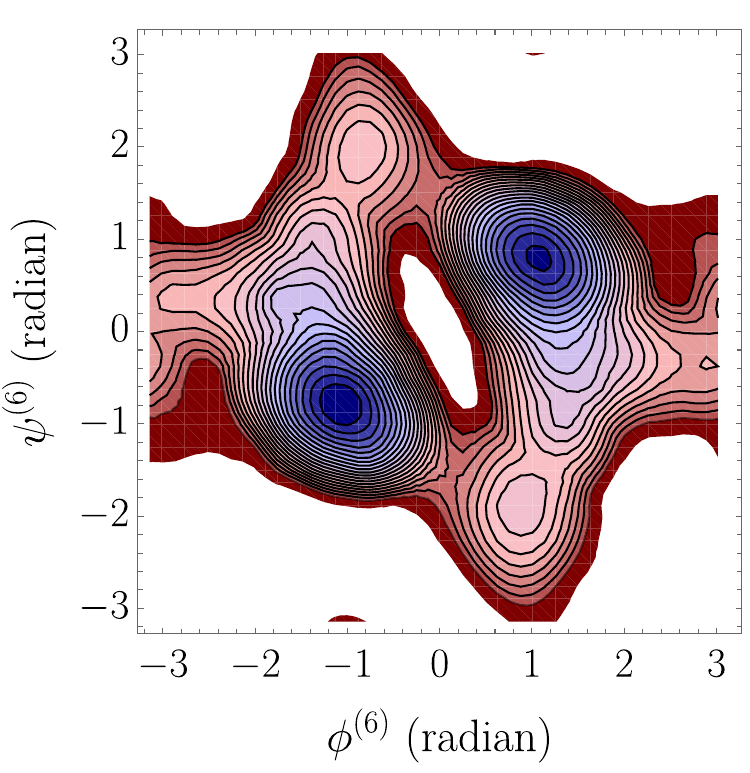}};
\end{tikzpicture}
\caption{2D free energy convergence of $(\phi^{(6)},\psi^{(6)})$ for AIB$_9$. Left: 100~ns; middle: 300~ns; right: 1~$\upmu$s.}
\label{fig:aib9_2d_6}
\end{figure}

\begin{figure}
\centering
\begin{tikzpicture}
\node at (0,0) {\includegraphics[width=0.32\textwidth]{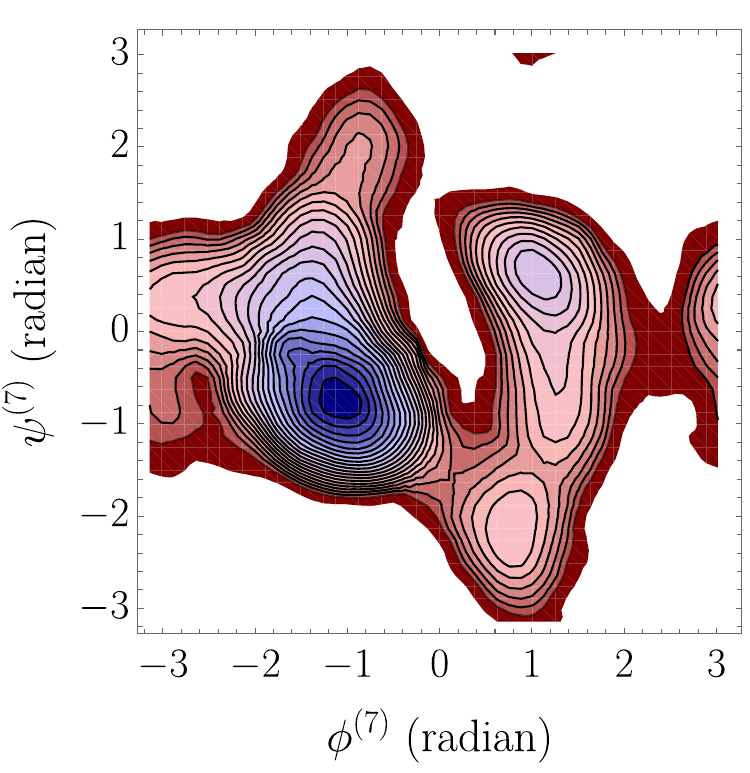}};
\node at (5.5,0) {\includegraphics[width=0.32\textwidth]{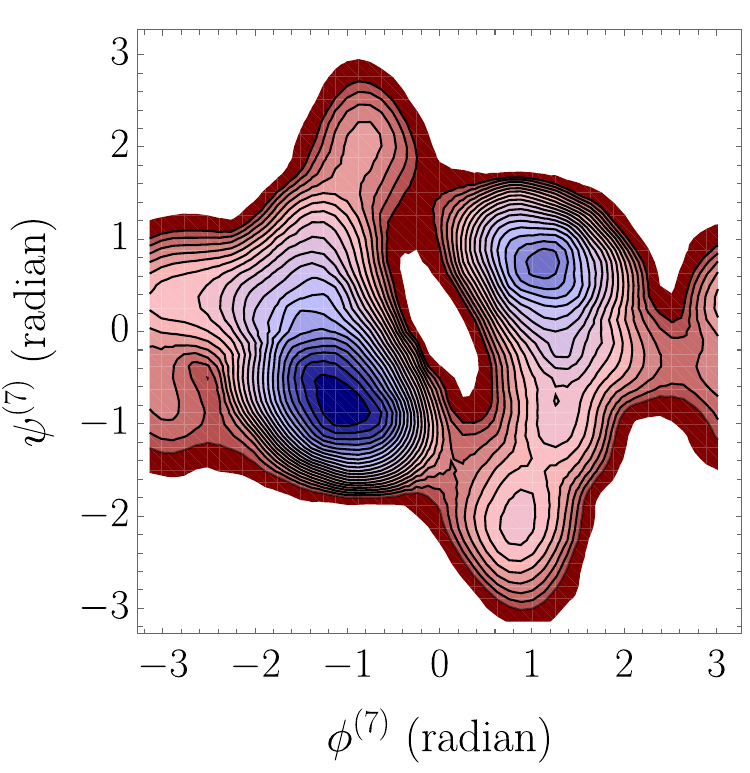}};
\node at (11,0) {\includegraphics[width=0.32\textwidth]{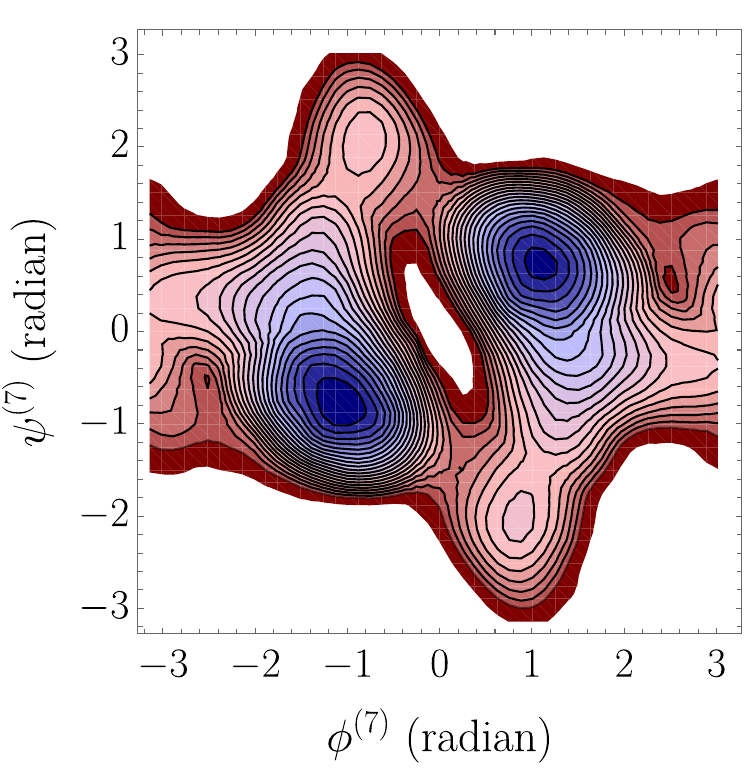}};
\end{tikzpicture}
\caption{2D free energy convergence of $(\phi^{(7)},\psi^{(7)})$ for AIB$_9$. Left: 100~ns; middle: 300~ns; right: 1~$\upmu$s.}
\label{fig:aib9_2d_7}
\end{figure}

\begin{figure}
\centering
\begin{tikzpicture}
\node at (0,0) {\includegraphics[width=0.32\textwidth]{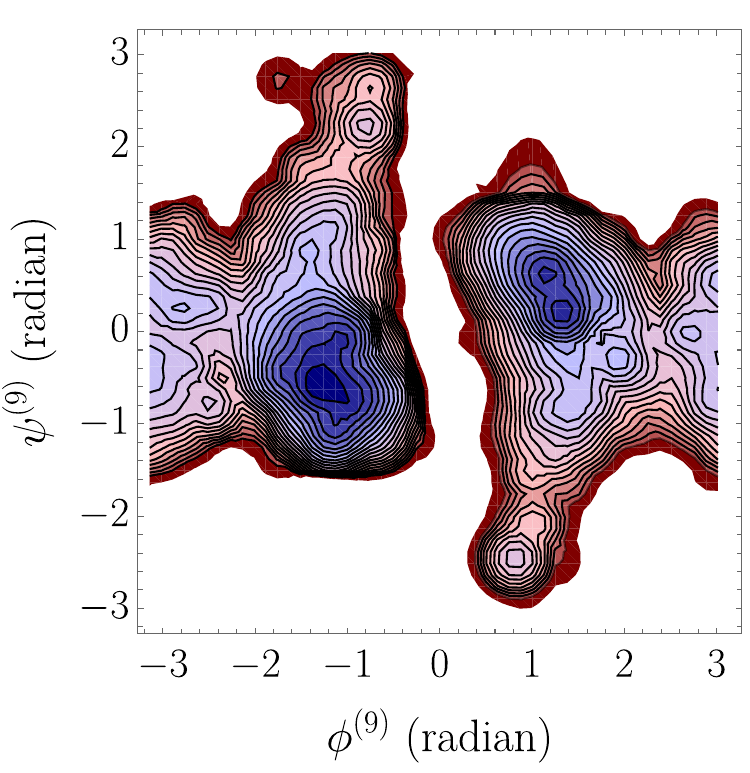}};
\node at (5.5,0) {\includegraphics[width=0.32\textwidth]{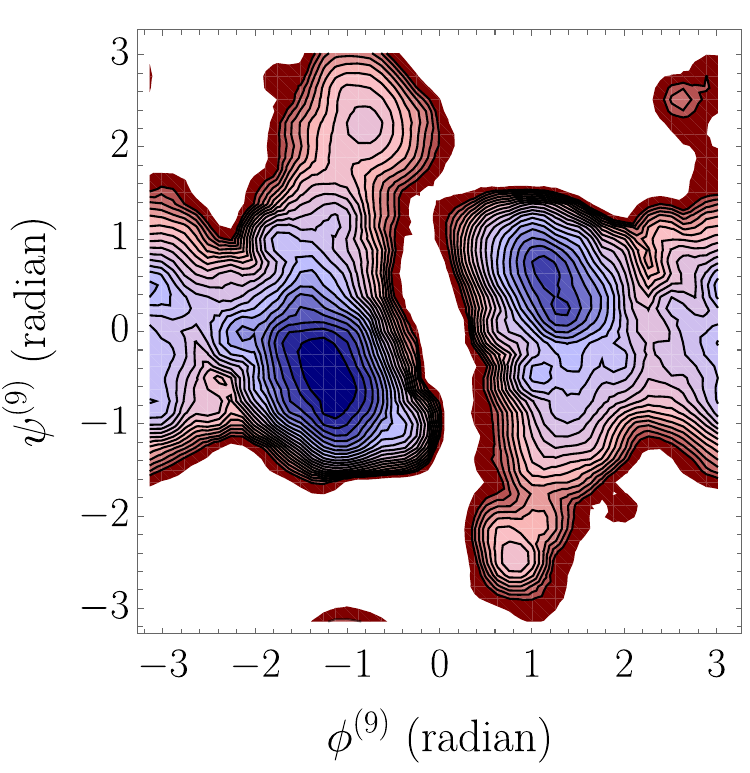}};
\node at (11,0) {\includegraphics[width=0.32\textwidth]{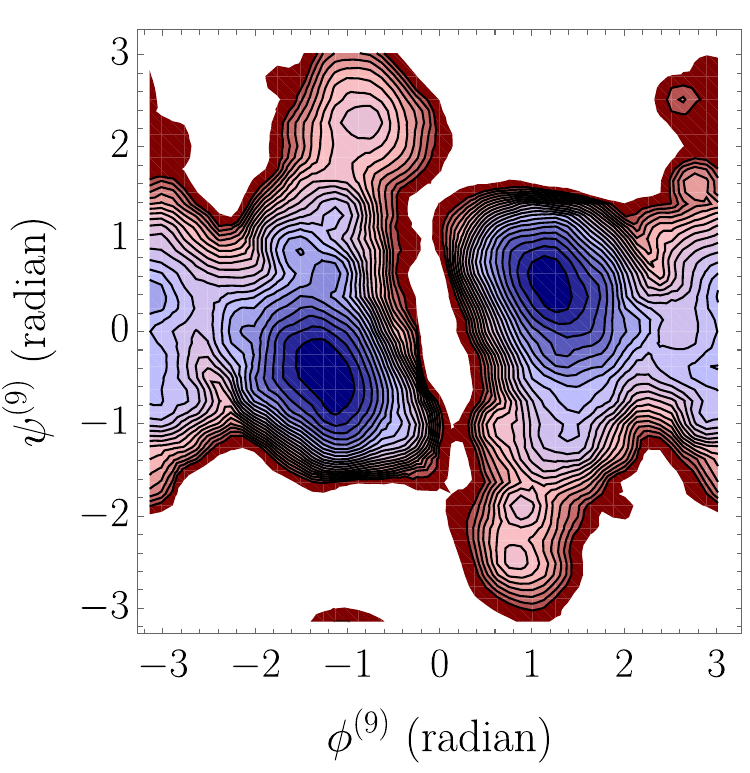}};
\end{tikzpicture}
\caption{2D free energy convergence of $(\phi^{(9)},\psi^{(9)})$ for AIB$_9$. Left: 100~ns; middle: 300~ns; right: 1~$\upmu$s.}
\label{fig:aib9_2d_9}
\end{figure}

\begin{figure}
\centering
\begin{tikzpicture}
\node at (0,0) {\includegraphics[width=0.45\textwidth]{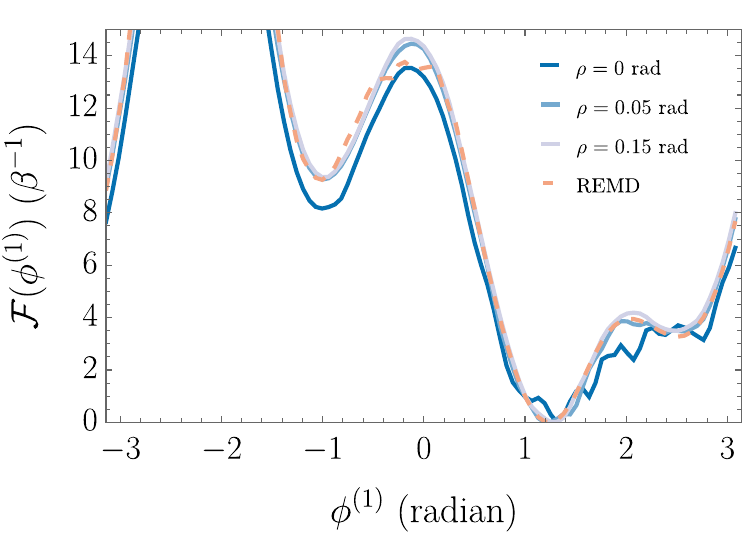}};
\node at (8.2,0) {\includegraphics[width=0.48\textwidth]{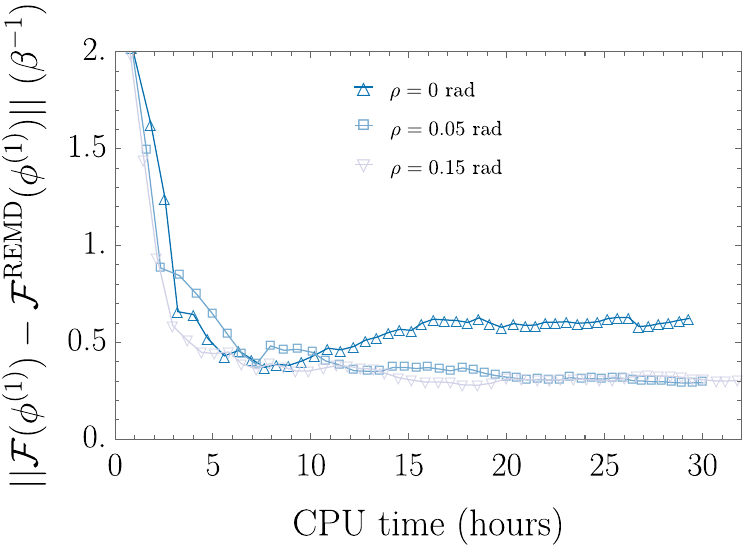}};
\end{tikzpicture}
\caption{1D free energy convergence and root mean square deviations (RMSD) of $\phi^{(1)}$ for ditrpyptophan as a function of the kernel smoothing bandwidth $\rho$.}
\label{fig:ditryptophanphi1_rho}
\end{figure}

\begin{figure}
\centering
\begin{tikzpicture}
\node at (0,0) {\includegraphics[width=0.45\textwidth]{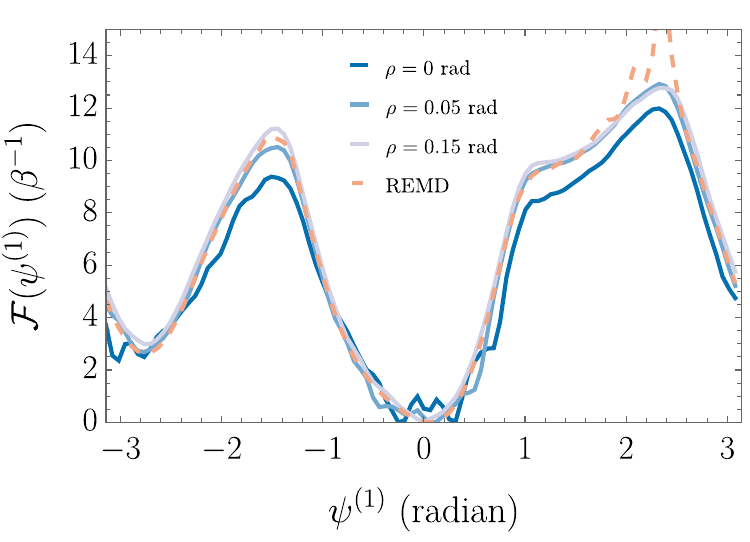}};
\node at (8.2,0) {\includegraphics[width=0.48\textwidth]{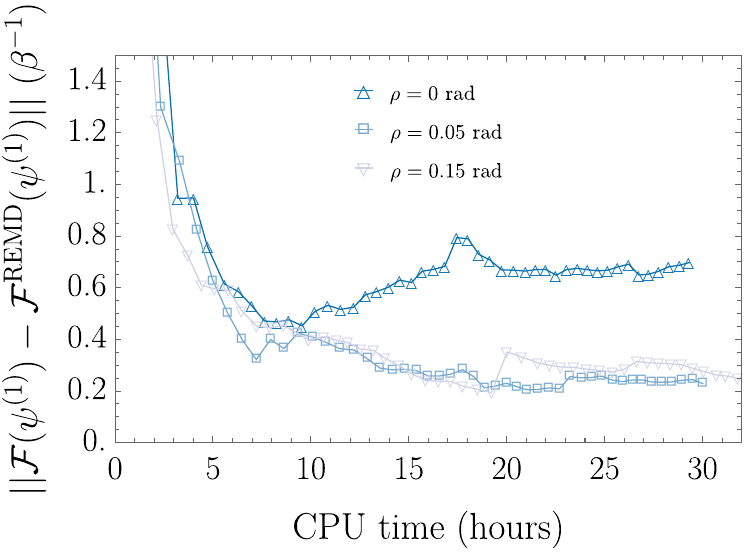}};
\end{tikzpicture}
\caption{1D free energy convergence and root mean square deviations (RMSD) of $\psi^{(1)}$ for ditrpyptophan as a function of the kernel smoothing bandwidth $\rho$.}
\label{fig:ditryptophanpsi1_rho}
\end{figure}

\begin{figure}
\centering
\begin{tikzpicture}
\node at (0,0) {\includegraphics[width=0.45\textwidth]{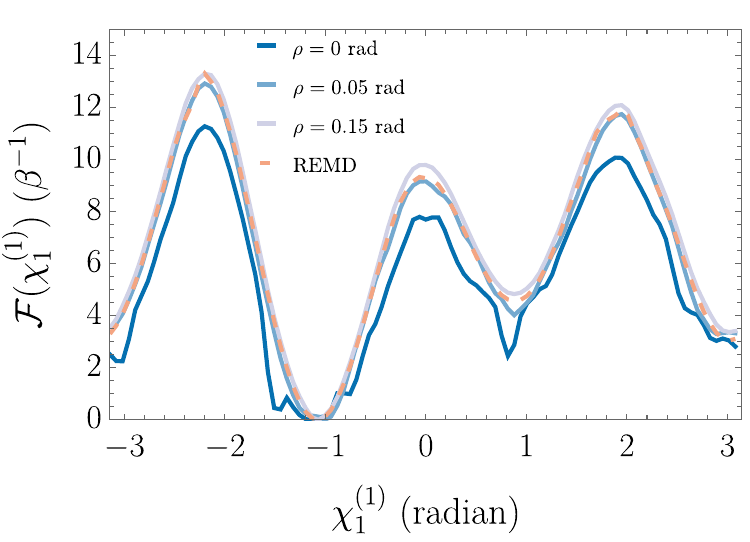}};
\node at (8.2,0) {\includegraphics[width=0.48\textwidth]{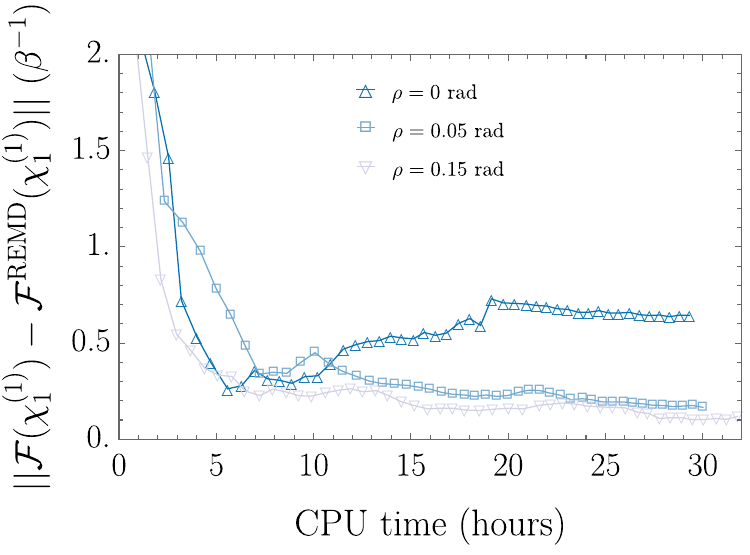}};
\end{tikzpicture}
\caption{1D free energy convergence and root mean square deviations (RMSD) of $\chi_1^{(1)}$ for ditrpyptophan as a function of the kernel smoothing bandwidth $\rho$.}
\label{fig:ditryptophanchi11_rho}
\end{figure}

\begin{figure}
\centering
\begin{tikzpicture}
\node at (0,0) {\includegraphics[width=0.45\textwidth]{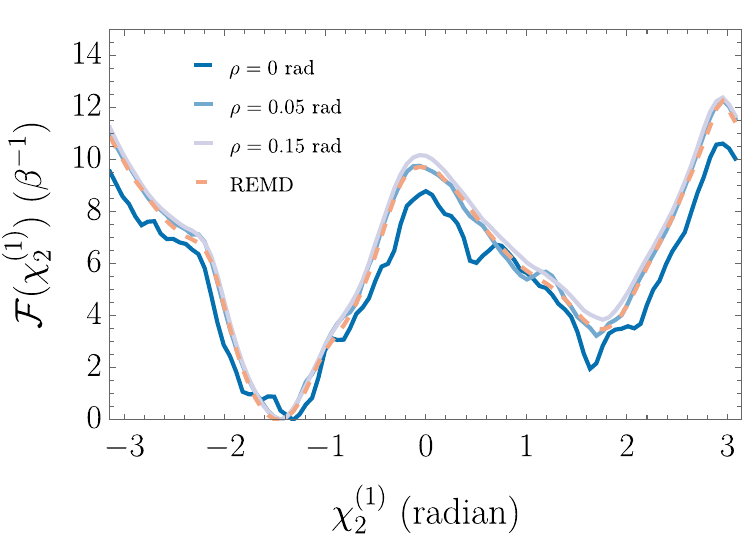}};
\node at (8.2,0) {\includegraphics[width=0.48\textwidth]{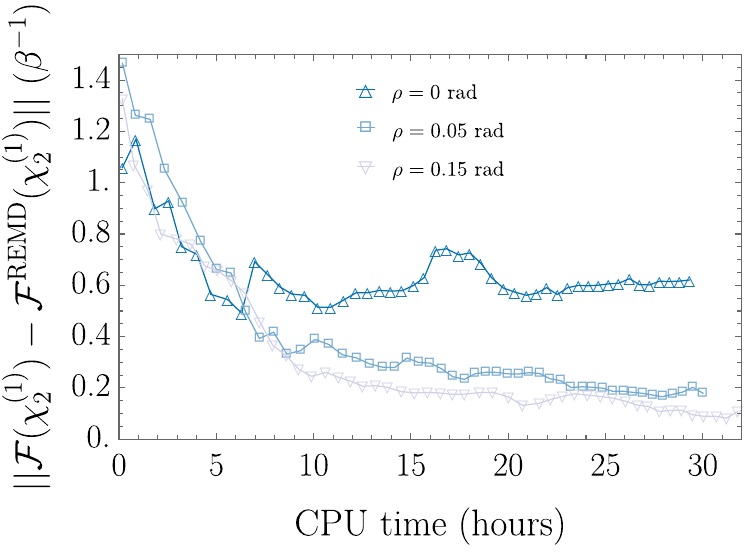}};
\end{tikzpicture}
\caption{1D free energy convergence and root mean square deviations (RMSD) of $\chi_2^{(1)}$ for ditrpyptophan as a function of the kernel smoothing bandwidth $\rho$.}
\label{fig:ditryptophanchi12_rho}
\end{figure}

\begin{figure}
\centering
\begin{tikzpicture}
\node at (0,0) {\includegraphics[width=0.45\textwidth]{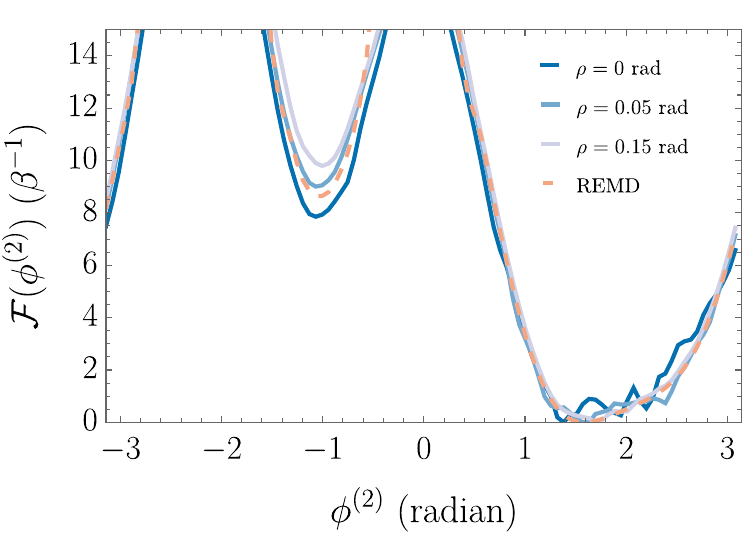}};
\node at (8.2,0) {\includegraphics[width=0.48\textwidth]{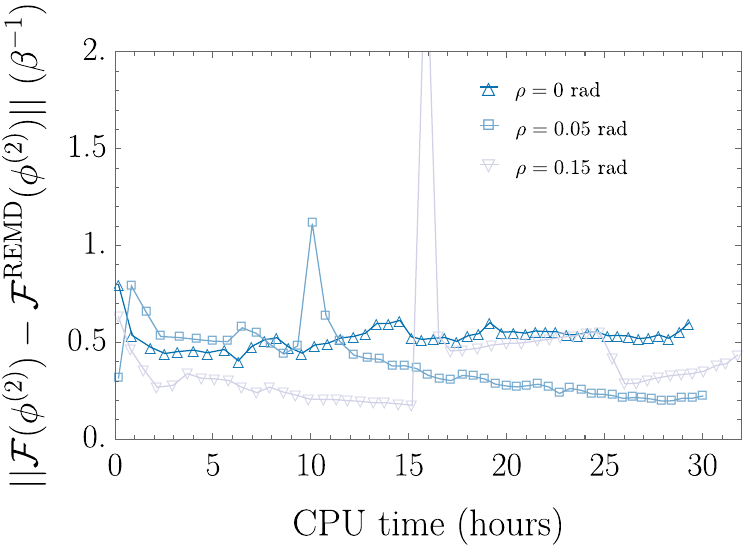}};
\end{tikzpicture}
\caption{1D free energy convergence and root mean square deviations (RMSD) of $\phi^{(2)}$ for ditrpyptophan as a function of the kernel smoothing bandwidth $\rho$.}
\label{fig:ditryptophanphi2_rho}
\end{figure}

\begin{figure}
\centering
\begin{tikzpicture}
\node at (0,0) {\includegraphics[width=0.45\textwidth]{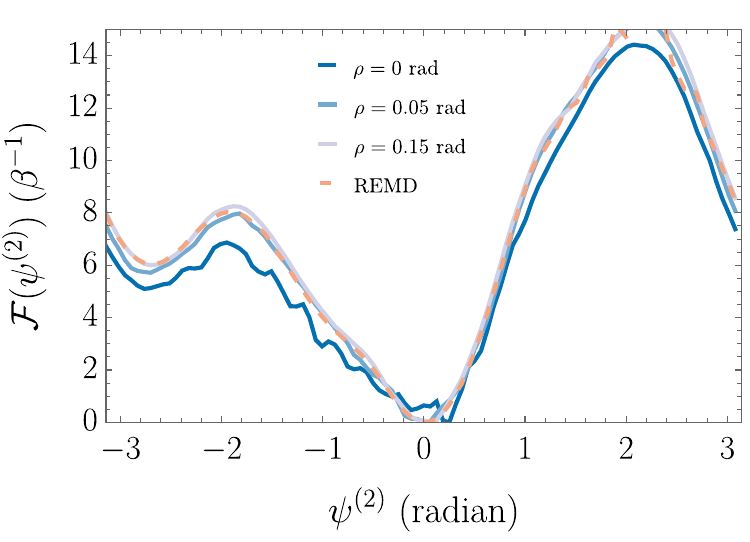}};
\node at (8.2,0) {\includegraphics[width=0.48\textwidth]{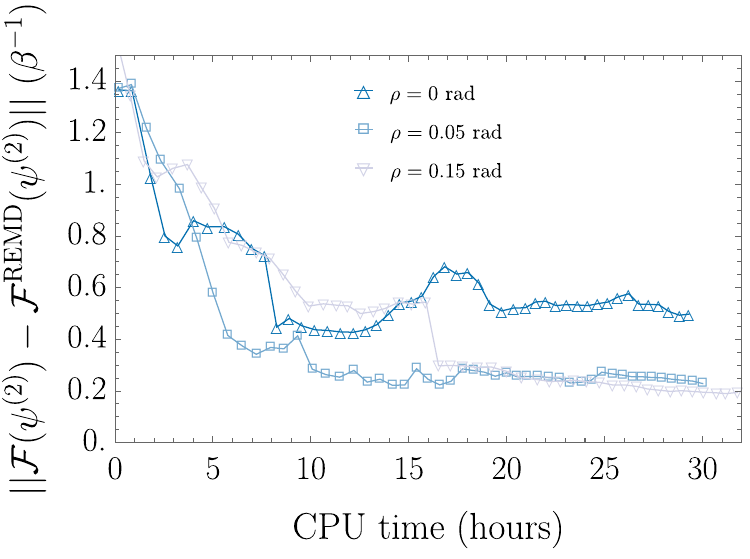}};
\end{tikzpicture}
\caption{1D free energy convergence and root mean square deviations (RMSD) of $\psi^{(2)}$ for ditrpyptophan as a function of the kernel smoothing bandwidth $\rho$.}
\label{fig:ditryptophanpsi2_rho}
\end{figure}

\begin{figure}
\centering
\begin{tikzpicture}
\node at (0,0) {\includegraphics[width=0.45\textwidth]{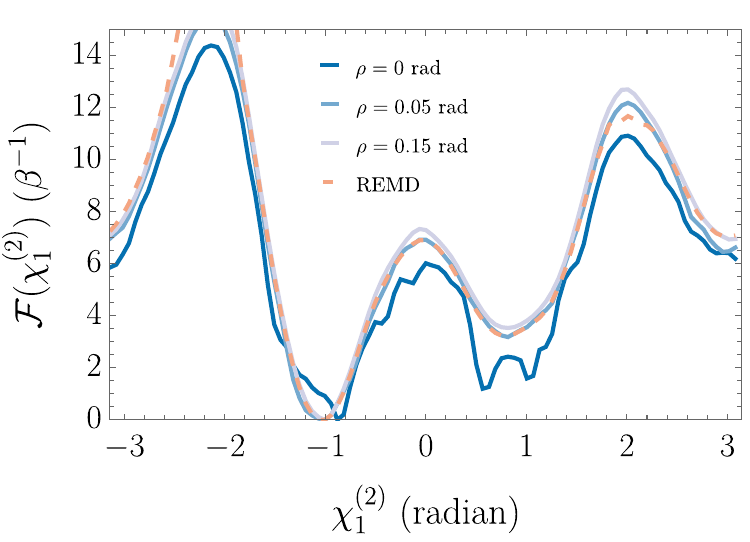}};
\node at (8.2,0) {\includegraphics[width=0.48\textwidth]{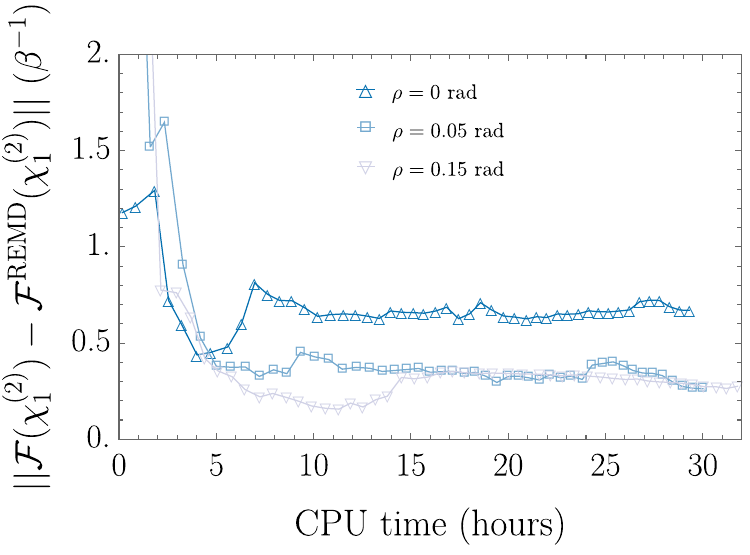}};
\end{tikzpicture}
\caption{1D free energy convergence and root mean square deviations (RMSD) of $\chi_1^{(2)}$ for ditrpyptophan as a function of the kernel smoothing bandwidth $\rho$.}
\label{fig:ditryptophanchi21_rho}
\end{figure}

\begin{figure}
\centering
\begin{tikzpicture}
\node at (0,0) {\includegraphics[width=0.45\textwidth]{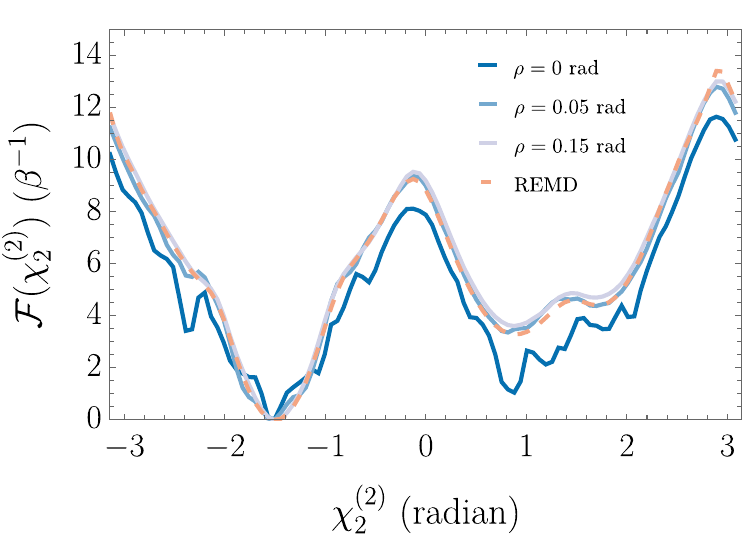}};
\node at (8.2,0) {\includegraphics[width=0.48\textwidth]{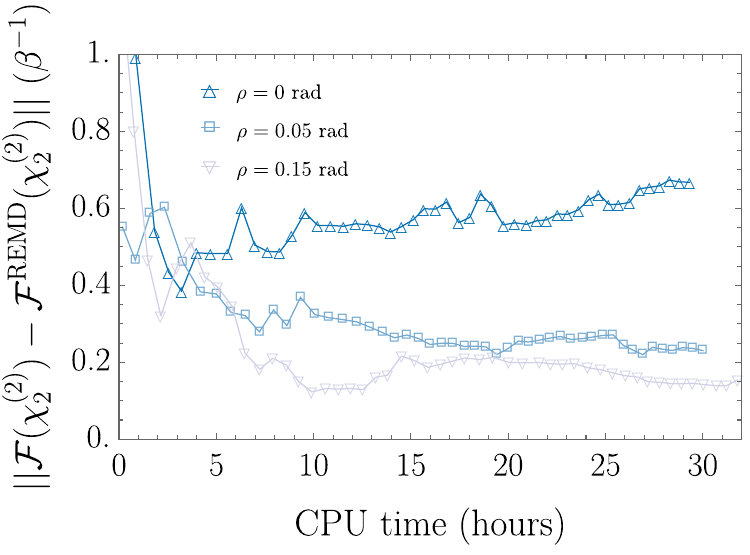}};
\end{tikzpicture}
\caption{1D free energy convergence and root mean square deviations (RMSD) of $\chi_2^{(2)}$ for ditrpyptophan as a function of the kernel smoothing bandwidth $\rho$.}
\label{fig:ditryptophanchi22_rho}
\end{figure}

\begin{figure}
\centering
\includegraphics[width=0.5\textwidth]{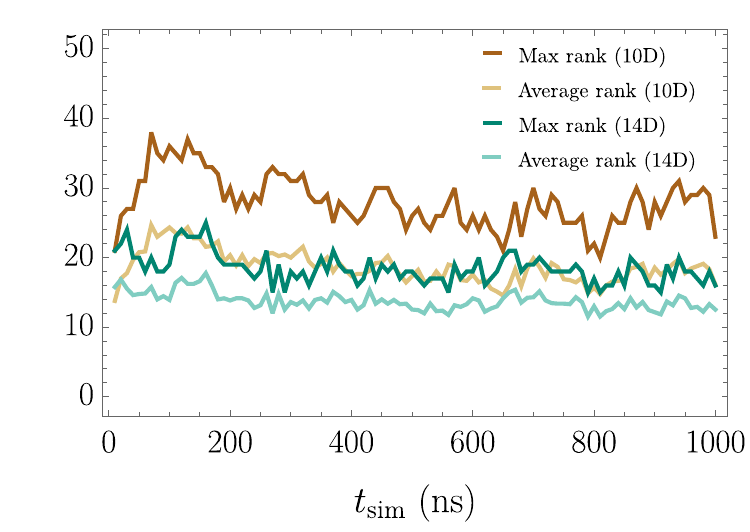}
\caption{Evolution of the TT rank (max, average) over the course of simulations of AIB$_9$ biased along 10 and 14 CVs.
Ranks shown are collected every 10 ns.}
\label{fig:aib9evolution}
\end{figure}

\end{document}